\pdfoutput=1
\documentclass[12pt]{article}
\usepackage{amssymb}
\usepackage{graphicx,epstopdf}
\usepackage{amsthm}
\usepackage{tikz}
\usepackage{subfig}
\usepackage{mathtools}
\usepackage{hyperref}
\usepackage[authoryear]{natbib}
\usepackage{multicol}
\usepackage{paralist}
\usepackage{url}
\usepackage{setspace}
\usepackage{amsmath}
\usepackage{amsfonts}
\usepackage{enumitem}
\usepackage{color}
\usepackage{float}
\usepackage{algorithm}
\usepackage{algpseudocode}
\usepackage{multirow}

\allowdisplaybreaks
\usepackage{setspace}
\theoremstyle{remark}
\newtheorem*{lemma*}{Lemma}
\theoremstyle{definition}

\newtheorem{theorem}{\n{Theorem}}[section]

\newtheorem{remark}{\n{Remark}}[section]
\newtheorem{lemma}{\n{Lemma}}[section]

\newcommand{\real}{\mathbb{R}}

\font\n=cmcsc10
\newcommand{\umod}[1]{ \left\lVert #1 \right\rVert}

\newcommand{\PP}{\mathbb{P}}
\newcommand{\EE}{\mathbb{E}}
\newcommand{\ER}{Erd\"{o}s-Ren\'{y}i}
\DeclareFontFamily{U}{mathx}{}
\DeclareFontShape{U}{mathx}{m}{n}{ <-> mathx10 }{}
\DeclareSymbolFont{mathx}{U}{mathx}{m}{n}
\DeclareFontSubstitution{U}{mathx}{m}{n}

\DeclareMathAccent{\widecheck}{0}{mathx}{"71}
\begin{document}
\title{{\Large\bf Detecting and Localizing Anomalous Cliques in Inhomogenous Networks using Egonets}}
\author{Subhankar Bhadra\hspace{.2cm}\\
    Department of Statistics, North Carolina State University\\
    Srijan Sengupta\\
    Department of Statistics, North Carolina State University
    }
\date{}
\maketitle

\textbf{Abstract:}
Cliques, or fully connected subgraphs, are among the most important and well-studied graph motifs in network science.
{We consider the problem of finding a \textit{statistically anomalous} clique  hidden in a large network. There are two 
parts to this problem: (1) detection, i.e., determining whether an anomalous clique is present, and (2) localization, i.e., determining which vertices of the network constitute
the detected clique. 
While this problem has been extensively studied under the homogeneous Erdős–Rényi model, little progress has been made beyond this simple setting, and no existing method can perform detection and localization in inhomogeneous networks within finite time.
To address this gap, we first show that in homogeneous networks, the anomalousness of a clique depends solely on its size. This property does not carry over to inhomogeneous networks, where the identity of the vertices forming the clique plays a critical role, and a smaller clique can be more anomalous than a larger one. Building on this insight, we propose a unified methods for clique detection and localization based on a class of subgraphs called \textit{egonets}.
The proposed method generalizes to a wide variety of inhomogeneous network models and is naturally amenable to parallel computing. We establish the theoretical properties of the proposed method and demonstrate its empirical performance through simulation studies.
}

\clearpage

\section{Introduction}
\label{sec:intro}
A clique is a fully connected subgraph of a given network.
We are interested in two problems related to cliques.
First, we want to determine whether a given network contains an anomalous clique --- this is the \textit{detection} problem.
Second, if an anomalous motif is detected, we want to know which nodes constitute the clique --- this is the \textit{localization} problem.
Note that cliques can appear randomly in a network \citep{bollobas1998random}. We are interested in \textit{anomalous} cliques that are very unlikely to appear randomly.

Anomalous clique detection is an important and relevant problem in many networked systems.
In financial trading networks \citep{li2010detecting, pan2012decoding}, an anomalous clique can indicate a group of
traders trading among each other to manipulate the stock
market.
Cliques can have important scientific interpretation in biological networks, e.g., in brain networks, neural cliques \citep{lin2006organizing} are network-level memory coding units in the hippocampus.
In social networks and online social media, an anomalous clique may be indicative of mechanisms of opinion manipulation or ideological echo chambers.

The detection problem can be formulated as a hypothesis testing problem as follows. Let $A$ be the symmetric adjacency matrix of a network with $n$ nodes where $A_{ij} = 1$ if nodes $i, j$  are connected, and $A_{ij} = 0$ otherwise. We consider there are no self-loops, that is, $A_{ii}=0$ for all $i$. Now, under the null hypothesis, we assume that $A_{ij}$ follows $\text{Bernoulli}(P_{ij})$ independently for all $i<j$, where $P \in \mathcal{P}$ is the unknown model which belongs to some model class $\mathcal{P}$.
The model class $\mathcal{P}$ is assumed to be known, for example, $\mathcal{P}$ could be the class of stochastic blockmodels or random dot product graphs \cite{lorrain1971structural, young2007random}.
Under the alternative, there exists some set $S\subseteq [n]$, 
where $[n] = \{1, \ldots, n\}$, 
such that the nodes in $S$ form a clique. We can write the testing problem mathematically as,
\begin{equation}
\begin{split}
H_0:\ &A_{ij} \sim \text{Bernoulli} (P_{ij}) \text{ for all } 1 \le i < j \le n, \text{ for some }  P \in \mathcal{P}, \text{vs.}\\
 H_1:\ &A_{ij} \sim \text{Bernoulli} (Q_{ij}) \text{ for all } 1 \le i < j \le n,\text{such that }\\
        &Q_{ij} = \left\{
        \begin{array}{ll}
            1 & \text{ if } i,j \in S \text{ and } i < j, \\
            P_{ij}  & \text{ if}\ i \notin S \text{ or } j \notin S, \text{ and } i < j.
        \end{array}\right.
\end{split}
    \label{detection_hyp}
\end{equation}
If an anomalous clique is detected, the second task is localization, i.e., to estimate the set of nodes $S$ forming the anomalous clique.

The simplest version of this problem is when $\mathcal{P}$ is assumed to be the class of \ER\ (ER) networks \cite{Erdoes1959}, where $P_{ij}=p$ for all $i,j$. 
The problem of finding a hidden clique in ER networks (the so-called ``planted clique'' problem) has been extensively studied in the computer science literature, some notable contributions being \citet{alon1998finding}, \citet{kuvcera1993expected}, \citet{dekel2014finding}, \citet{feige2010finding}. and \citet{deshpande2015finding}, to name a few.
The goal is similar to the \textit{localization} task we listed, i.e., to {identify} the nodes constituting the hidden clique, rather than to detect whether a given network contains a hidden clique. 
In the statistics literature, the dense subgraph \textit{detection} problem for ER networks was studied by \citet{arias2014community} and \citet{verzelen2015community} under a dense and a sparse setting respectively. 
For both the dense and the sparse settings, the authors derived theoretical lower bounds and proposed nearly-optimal tests for detection. 
However, the proposed optimal tests that achieve the theoretical lower bounds are, in most cases, computationally not feasible to perform, unless the network is assumed to be very sparse (see \cite{verzelen2015community}). 
\citet{Hajek2015computational} addressed this issue and derived separate statistical and computational bounds for the detection problem. Following this work, \citet{chen2016statistical} derived statistical and computational thresholds for the localization problem as well.

\begin{figure}[ht]
	\centering
	\includegraphics[height=0.13\textheight]{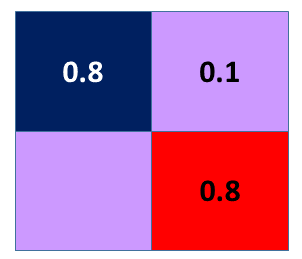}
	\hspace{1ex}
	\includegraphics[height=0.13\textheight]{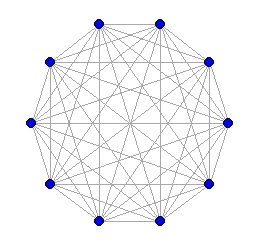}\\
	\vspace{1ex}
	\includegraphics[height=0.13\textheight]{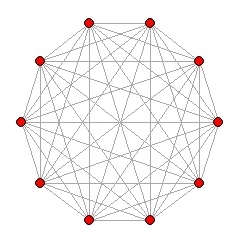}
	\hspace{2ex}
	\includegraphics[height=0.13\textheight]{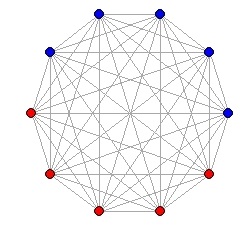}
	\caption{\small SBM with $n=25 \times 2$. All cliques have 10 nodes, but the ``balanced'' clique (bottom right) is almost impossible while the pure cliques are very likely (prob $\approx 87\%$).}
 \label{sbm_example}
\end{figure}

We consider the problem of finding anomalous cliques in a more general, inhomogeneous setting and we would like to point out a fundamental difference in the interpretation of this problem as we move to the inhomogeneous setting.
For ER networks, since the edge probabilities are all equal, the detection problem reduces to finding if there is an unusually large clique in the network. Then the localization problem is to find the largest clique in the network. 
Therefore, under the ER model, the `anomalousness' of a clique is defined solely by its size, and hence, the largest clique is also the most anomalous clique. 
However, this interpretation of anomalousness does not carry over to the inhomogeneous setting. 
In inhomogeneous networks, the anomalousness of a clique depends crucially on the members of the clique more than its size. 
To see this, consider a strongly homophilic SBM network with 50 nodes and two equal-sized communities, and suppose that the intra-community connection probability is 0.8 while the inter-community connection probability is 0.2 (see Figure \ref{sbm_example}). 
Using the classical result on largest cliques, it can be shown that a clique of size 10 consisting purely of community 1 nodes or consisting purely of community 2 nodes is very likely to occur randomly, whereas a mixed clique of size 10 consisting of 5 nodes from community 1 and 5 nodes from community 2 is very unlikely to occur randomly. 
This implies that the mixed clique of size 10 should be considered anomalous, but a pure clique of size 10 should be considered non-anomalous.
In such inhomogeneous settings, the hidden clique problem has received considerably less attention.  \citet{bogerd2021detecting} studied the dense subgraph detection problem in inhomogeneous networks as a hypothesis testing problem, formulated slightly differently from ours. 
For a general probability matrix $P$, the authors derived minimax lower bounds for the detection problem and proposed nearly-optimal tests when $P$ is known. 
However, despite its excellent theoretical guarantees, their proposed method is not computationally tractable, as they (at least) require scanning over all subgraphs in the network of some given clique size. 

In this paper, we propose an efficient solution to both the detection and localization problem by looking at a special class of subgraphs, called \textit{egonets} \citep{crossley2015social}.
The egonet of the $i^{th}$ node is defined as the subgraph spanned by all neighbors of the $i^{th}$ node.
Let $E_i$ denote the $i^{th}$ egonet degree, the number of edges in the $i^{th}$ egonet, i.e.,
$$
E_i = \underset{j<k}{\sum\sum} A_{ij} A_{ik} A_{jk}.
$$
Heuristically, in the presence of an anomalous clique, if the $i^{th}$ node is in the clique, then $E_i$ will be disproportionately larger than its expected value under the null given the neighbors of the $i^{th}$ node. On the other hand, the egonet degrees corresponding to the non-clique nodes will remain more concentrated around their expectations under the null. 
We formulate a statistical decision rule for the detection problem based on the centered egonet degrees falling above a certain threshold. If an anomalous clique is detected, we propose an algorithm based on ordering the centered egonet degrees to identify the nodes forming the clique. 
Since there is one egonet corresponding to each node, there are exactly $n$ subgraphs to monitor. Therefore, the egonet method is computationally easy to implement; in fact, we can compute all the egonets in cubic time.

Our main contribution in this paper is the formulation of a statistical inferential framework for anomalous clique detection under the inhomogeneous setting.
Using the egonet method, one can simultaneously perform two tasks --- first, detect whether there is a small anomalous clique in a network, and second, if the presence of an anomalous clique is detected, then identify the nodes forming the clique.
To the extent of our knowledge, this is one of the first proposed methods for detecting and localizing anomalous cliques in inhomogeneous networks that is both computationally feasible and has strong statistical guarantees.
The rest of the article is organized as follows.
In Section \ref{sec:egonet}, we formally describe the egonet method, and in Section \ref{sec:theory}, we present its theoretical properties pertaining to statistical guarantees.
In Section \ref{sec:sim}, we present simulation results for the egonet method for synthetic networks generated from a variety of network models.
In Section \ref{sec:data}, we present case studies reporting the results of the egonet method applied to two well-known network datasets.
We conclude the paper with a brief discussion in Section \ref{sec:discussion}.
The technical appendix contains proofs of the main theorems.

\textit{Notations:} All limits are taken as $n\rightarrow\infty$. We use standard asymptotic notations, e.g. for sequences $\{a_n\}$ and $\{b_n\}$, $a_n=o(b_n)$ if $a_n/b_n\rightarrow 0$; $a_n=O(b_n)$ if $a_n/b_n$ is bounded; $a_n=\omega(b_n)$ if $b_n=o(a_n)$; $a_n=\Omega(b_n)$ if $b_n=O(a_n)$; $a_n=\Theta(b_n)$ if $a_n=O(b_n)$ and $b_n=O(a_n)$. 
We also use the notation $a_n\ll b_n$(resp. $a_n\gg b_n$) which is equivalent to $a_n=o(b_n)$(resp. $a_n=\omega(b_n)$).
For any square matrix $M\in\mathbb{R}^{d\times d}$, we use the notation $M^{-i}$ to denote the $(d-1)\times(d-1)$ submatrix of $M$ excluding its $i^{th}$ row and column.
We say that an event $F_n$ occurs `almost surely' if $\PP(F_n)\rightarrow 1$, and $F_n$ occurs `with high probability' (`whp' in short) if, for any $\tau>1$, there exists $C_0>0$ such that $\PP(F_n)\geq 1-\frac{C_0}{n^\tau}$. 
For an $m\times n$ matrix $T$, we define the operators $\umod{.}_1, \delta(.)$ and $\umod{.}$ as follows:
$$\umod{T}_{1,1}=\underset{i,j}{\sum\sum}\,|T_{ij}|,\ \delta(T)=\max_i\sum_j\, |T_{ij}|,\ \umod{T}=\text{largest singular value of }T.$$
Note that all three operators above satisfy the properties of a matrix norm.

\section{The egonet method}
\label{sec:egonet}
Consider a simple, undirected network of $n$ nodes with no self-loops, and let $A\in \real^{n\times n}$ be its adjacency matrix, where $A_{ij} \sim \text{Bernoulli} (P_{ij})$ for $1\leq i < j\leq n$,
where $P$ is a symmetric probability matrix whose diagonals are zero.
Let $a_i$ denote the $i^{th}$ row (or column) of $A$, and $\mathcal{N}_i$ be the neighborhood of the $i^{th}$ node, i.e., 
\begin{equation}
    \mathcal{N}_i = \{j \in [n]: A_{ij} =1 \}.
\end{equation}
Let $D_i$ be the degree of the $i^{th}$ node, defined as the total number of nodes that are connected to the $i^{th}$ node, i.e.,
\begin{equation}
    D_i = \sum_{j=1}^n A_{ij}, \text{ or, equivalently, }
D_i = |\mathcal{N}_i|.
\end{equation}
The \textit{egonet} (short version of \textit{ego-centric network}) of the $i^{th}$ node is the subgraph spanned by its neighbors, given by $\{A_{jk}: j,k \in \mathcal{N}_i \}$.
Let $E_i$ be the \textit{egonet degree} of the $i^{th}$ node, defined as the number of edges in the $i^{th}$ egonet, i.e., 
\begin{equation}
    E_i = \underset{j<k:\,j,k \in \mathcal{N}_i}{\sum\sum} A_{jk}=\underset{j<k}{\sum\sum} A_{ij}A_{ik}A_{jk}.
\end{equation}
Observe that $A_{ij}A_{ik}=1$ if and only if $j,k \in \mathcal{N}_i$.

Egonets are going to be our primary focus for the detection and localization of anomalous cliques.
First, we provide a heuristic explanation of why it makes sense to use egonets for this problem.
Recall that our goal is to answer the two following questions.
\begin{itemize}
    \item[(a)] Detection: does the network $A$ contain an anomalous clique?
    \item[(b)] Localization: if the answer to (a) is yes, then which nodes are members of the anomalous clique?
\end{itemize}

We claim that we can give an answer to both questions (a) and (b) by examining the egonet degrees. To see this, suppose that the network $A$ contains an anomalous clique consisting of nodes in some set $S$. Now,
\begin{itemize}
    \item If the $i^{th}$ node belongs to $S$, then all the other clique members, being its neighbor, will belong to its egonet. Thus, the $i^{th}$ egonet will contain almost the entire anomalous clique.
    \item If the $i^{th}$ node is not in $S$, then only some of the clique members will be its neighbor. Thus, the $i^{th}$ egonet will contain only a small part of the anomalous clique formed by them.
\end{itemize}
Therefore, there is a structural connection between cliques and egonets. It is this connection that justifies the use of egonets as a means for the detection and localization of anomalous cliques. 

The next step is, how to operationalize this connection between cliques and egonets to establish a statistical method for the two inferential tasks. The basic idea is to simply use the centered egonet degrees.
When the network $A$ does not contain any anomalous clique, the \textit{conditional} expectation of the egonet degrees, conditioned on the neighborhoods, is given by
\begin{equation}
    \mathbb{E}_{H_0}(E_i \mid a_i) = \underset{j<k}{\sum\sum} A_{ij}A_{ik}P_{jk},
\label{e0ei}
\end{equation}
for $i=1, \ldots, n$.
On the other hand, consider the case when the network $A$ contains an anomalous clique consisting of nodes in $S$.
In this case, for the nodes that are members of the clique, i.e., for $i\in S$,
\begin{align*}
    \mathbb{E}_{H_1}(E_i \mid a_i) &=\underset{j<k}{\sum\sum} A_{ij}A_{ik}P_{jk} + \underset{j<k}{\sum\sum}(1 - P_{jk})1\{j,k\in S\}\\
    &= \mathbb{E}_{H_0}(E_i \mid a_i) + \underset{j<k}{\sum\sum}(1 - P_{jk})1\{j,k\in S\}
\end{align*}
 and for $i\in S^c$,
\begin{align*}
    \mathbb{E}_{H_1}(E_i \mid a_i) &=\underset{j<k}{\sum\sum} A_{ij}A_{ik}P_{jk} + \underset{j<k}{\sum\sum}A_{ij}A_{ik}(1 - P_{jk})1\{j,k\in S\}\\
    &= \mathbb{E}_{H_0}(E_i \mid a_i) + \underset{j<k}{\sum\sum}A_{ij}A_{ik}(1 - P_{jk})1\{j,k\in S\}.
\end{align*}
We note that when there is an anomalous clique, the conditional expectation of the egonet degrees of all the nodes increases from $\mathbb{E}_{H_0}(E_i \mid a_i)$ by some amount. If we have some suitable estimator of $P$, say $\widehat{P}$, then $\mathbb{E}_{H_0}(E_i \mid a_i)$ can be estimated simply by replacing ${P}_{jk}$ by $\widehat{P}_{jk}$ in \eqref{e0ei}. 
Then, we can define the $i^{th}$ centered egonet degree as follows:
\begin{equation}
    T_i=E_i - \widehat{\mathbb{E}_{H_0}(E_i \mid a_i)} = \underset{j<k}{\sum\sum} A_{ij}A_{ik}(A_{jk} - \widehat{P}_{jk}).
    \label{egodef}
\end{equation}
If there is an anomalous clique in the network, we would expect these centered egonet degrees, or at least one of them, to be `unusually' large. Therefore, to answer (a), we can simply find if at least one of the $T_i$'s falls above a certain threshold. 
Formally, we can use the $T_i$'s as test statistics for the testing problem in $\eqref{detection_hyp}$ and reject the null hypothesis if 
\begin{equation}
   \ T_i > C_n,
   \label{egodetect1}
\end{equation}
{for at least one } $i \in [n]$,
where $C_{n}$ is an appropriate threshold to be determined theoretically. 

If the answer to (a) is yes, our next task (b) is to estimate the set of nodes $S$ constituting the anomalous clique. We have already found that the egonet statistics $T_i$ have a positive bias, irrespective of whether $i$ is a member of the clique or not. For localization, we need to look more closely into this positive bias in the $T_i$'s. Consider a node $i$ that is a member of the clique, and another node $i'$ that is not a member of the clique. Assuming that $\widehat{P}$ estimates $P$ very accurately ($\widehat{P}\approx P$), the difference between the bias in $T_i$ and $T_{i'}$ is,
$$\EE_{H_1}(T_i\mid a_i) - \EE_{H_1}(T_{i'} \mid a_{i'}) \approx \underset{j<k}{\sum\sum}(1 - A_{i'j}A_{i'k})(1 - P_{jk})1\{j,k\in S\} > 0.$$
We see that the bias in the egonet statistics corresponding to the clique nodes are, in expectation, larger than the bias for the non-clique nodes.
We derive conditions under which the minimum of the egonet statistics for the clique nodes, $\min_{i\in S}T_{i}$, is larger than the maximum of the egonet statistics for the non-clique nodes, $\max\limits_{i\notin S}T_{i}$, in which case, we can estimate the nodes forming the anomalous clique by the egonet ranking algorithm, presented in Algorithm \ref{alg:cap}. 
Our theoretical results prove that this algorithm achieves \textit{exact recovery}, i.e., the estimate $\widehat{S}$ obtained from the algorithm is equal to $S$ almost surely.
\begin{algorithm}[ht]
    \begin{algorithmic}[1]
    \vspace{0.2cm}
    \Require adjacency matrix $A$, probability estimate $\widehat{P}$
    \vspace{0.2cm}
    \State for each $i\in[n]$, compute egonet statistic $T_i \gets {\sum\sum}_{j<k}\,A_{ij}A_{ik}(A_{jk}-\widehat{P}_{jk})$
    \vspace{.2cm}
    \State arrange $T_i$'s in descending order, let $T_{(i)}$ be the $i^{th}$ ordered egonet statistic and $v_i$ be the corresponding node index, that is, $T_{v_i}=T_{(i)}$
    \vspace{0.2cm}
    \State set $\widehat{S} \gets\{v_1\},\,i\gets 1$
    \vspace{0.2cm}
    \While{$v_{i+1}$ is connected to all nodes in $S$}
    \vspace{0.2cm}
        \State $S\gets \widehat{S} \cup \{v_{i+1}\}$
        \vspace{0.2cm}
        \State $i\gets i+1$
        \vspace{0.2cm}
    \EndWhile\\
    \vspace{0.2cm}
    \Return $\widehat{S}$
    \end{algorithmic}
    \caption{ Egonet ranking algorithm}\label{alg:cap}
\end{algorithm}

\subsection{Estimation of \texorpdfstring{$P$}{TEXT}}
\label{subsec:egonet_unknown}
When $P$ is known, the egonet method is straightforward to implement. We can compute the egonet statistics in \eqref{egodef} with $\widehat{P}=P$, and perform the detection and localization steps as described above. In this case, we can do inference for any $P\in [0,1]^{{n\choose 2}}$ without making any assumptions on the structure of $P$. In practice, we do not expect to know $P$. 
When $P$ is unknown,  
it is obvious that we can not estimate $P$ in its full generality, but we can do so by assuming some underlying structure on $P$. 
In this paper, we studied the egonet method for three such underlying model structures: the \ER\ model, the Chung-Lu model, and the Random Dot-Product Graph model.
When $P$ is estimated, the detection and localization steps can be executed the same way as for the known $P$ case, only the detection threshold $C_n$ in \eqref{egodetect1} would be different depending on the underlying models.

The task of estimating $P$ is non-trivial.
Existing estimators of $P$ that are consistent under the null hypothesis do not guarantee an accurate estimation of $P$ under the alternative hypothesis, where an anomalous clique is superimposed on the underlying network. 
This inaccuracy in estimating $P$ can be managed if the effect of the anomalous clique is small enough. 
However, if the anomaly is too strong, $\widehat{P}$ can become so much corrupted that the egonet statistics $T_{i}$ between the clique and non-clique nodes are no longer distinguishable. 
\begin{figure}[ht]
	\centering
	\includegraphics[height=0.3\textheight]{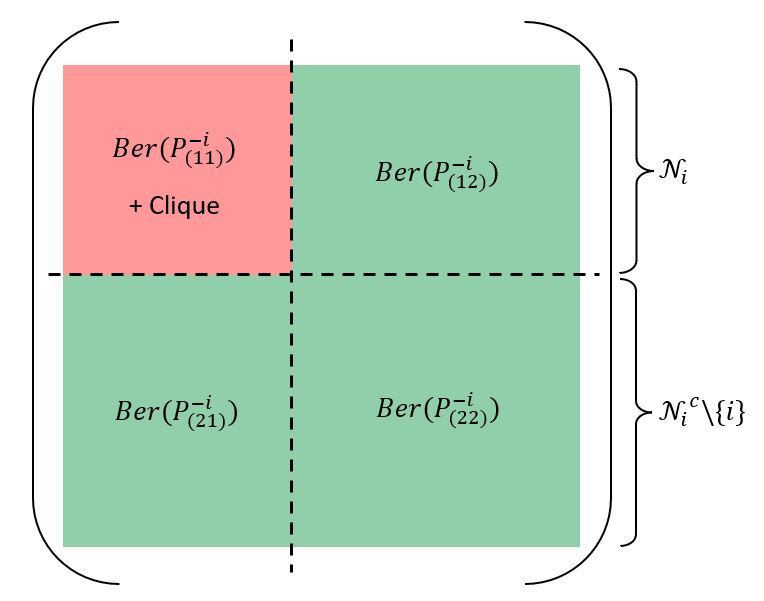}
	\caption{ \small Distribution of the adjacency matrix $A^{-i}$ under $H_1$ (partitioned based on the neighbors of a node $i$ in the clique).}	
	\label{probsec}
\end{figure}

We propose two alternatives to estimate $P$.
First, we can naively estimate $P$ by an estimator $\widehat{P}$ that is consistent under the null hypothesis, but does not guarantee consistency in the presence of an anomalous clique.
We theoretically prove that with a naive estimator $\widehat{P}$, the egonet method can detect and localize anomalous cliques provided that the effect of the anomaly is not too strong, and we derive theoretical lower and upper bounds on the strength of the clique within which detection and localization are possible.
Alternatively, we propose a strategy to estimate $P$ separately for each node $i$, which can produce consistent estimators of $P$ under both the null hypothesis, as well as the alternative hypothesis if the node $i$ belongs to the anomalous clique.
Note that for any node $i$, we only need to estimate $P^{-i}$ to compute the $i^{th}$ statistic.
Now, according to the neighbors $\mathcal{N}_i$ of $i$, $A^{-i}$ can then be partitioned into four parts: $A^{-i}_{(11)}$, $A^{-i}_{(12)}$, $A^{-i}_{(21)}$ and $A^{-i}_{(22)}$, as demonstrated in Figure \ref{probsec}.
In the presence of an anomalous clique $S$, if $i$ is in $S$, then only the submatrix spanned by the nodes $\mathcal{N}_i$, $A^{-i}_{(11)}$, is corrupted by the anomalous clique, whereas the rest of $A^{-i}$ remains uncorrupted.
Based on this idea, if an estimator $\widehat{P}^{-i}$ is constructed excluding the corresponding part of the adjacency matrix, $A^{-i}_{(11)}$, then $\widehat{P}^{-i}$ should consistently estimate $P^{-i}$, under both $H_0$ and $H_1$.
Of course, when $i$ is not in $S$, all parts of $A$ are corrupted, and we can no longer make such a claim.

Note that $\widehat{P}^{-i}$ is a $(N-1)\times (N-1)$ matrix, so that the indices of $A$ and $\widehat{P}^{-i}$ are not comparable. For notational convenience, we introduce a new $N\times N$ matrix $\widehat{P}^{(i)}$ such that its $(N-1)\times (N-1)$ submatrix excluding the $i^{th}$ row and column is exactly equal to $\widehat{P}^{-i}$, and the remaining elements are all zero.
The specific choice of $\widehat{P}^{(i)}$ depends on the underlying true model.
We explicitly provide a choice of $\widehat{P}^{(i)}$ for each of the three models: ER, Chung-Lu, and RDPG, below.

\subsubsection{ER model}
Under the ER model, $P_{jk}=p$ for all $j<k$. Naively,  we can estimate $p$ by $\widehat{p}=(1/n(n-1))\sum\sum_{j\neq k}A_{jk}$.
Alternatively, for the $i^{th}$ egonet statistic, we propose to estimate $p$ by the average connectivity in the network outside $A^{-i}_{(11)}$ (see Figure \ref{probsec}), that is, the edges $\{A_{jk}:A_{ij}A_{ik}=0,\,j,k\neq i\}$. 
Mathematically, for all $j\neq k$ such that $j,k\neq i$, we define
\begin{equation}
    \widehat{P}^{(i)}_{jk}=\frac{1}{{{n-1}\choose 2}-{D_i\choose 2}}\underset{j<k:\,j,k\neq i}{\sum\sum} (1-A_{ij}A_{ik})A_{jk}=\widehat{p}^{\,(i)},\text{ say,}
    \label{phat_er}
\end{equation}
and $\widehat{P}^{(i)}_{jk}=0$ otherwise.
\begin{figure}[ht]
	\centering
	\includegraphics[height=0.3\textheight]{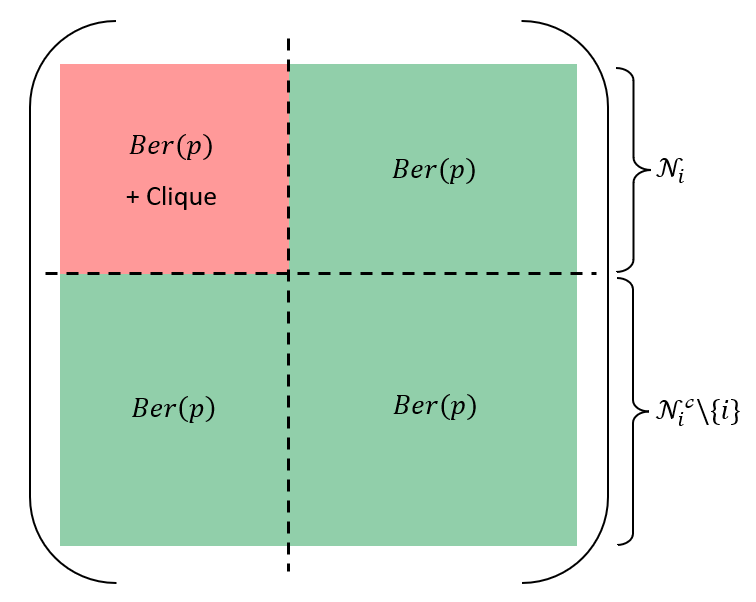}
	\caption{ \small Distribution of the adjacency matrix $A^{-i}$ under $H_1$ when the null model is ER (partitioned based on the neighbors of a node $i$ in the clique).}	
	\label{ersec}
\end{figure}

Then, the egonet statistic is,
\begin{equation}
    T_{i,ER}= \underset{j<k}{\sum\sum} A_{ij}A_{ik}(A_{jk}-\widehat{p}^{\,(i)})=\underset{j<k}{\sum\sum} A_{ij}A_{ik}A_{jk}-{D_i\choose 2}\widehat{p}^{\,(i)}.
\end{equation}

If $i\in S$, the entire anomalous clique is contained inside the egonet, and hence no part of the anomalous clique is present outside the egonet (see Figure \ref{ersec} for demonstration). 
Therefore, $\widehat{p}^{\,(i)}$ is the mean of i.i.d. $\text{Bernoulli}(p)$ random variables, an unbiased estimator of $p$. 
Hence, for nodes in $S$, we do not lose any contribution of the clique in the egonet statistics, which would not happen if we use the naive estimator $\widehat{p}=(1/n(n-1))\sum\sum_{j\neq k}A_{jk}$. 
This gives us an advantage both in the detection and localization step of our method. 

\subsubsection{Chung-Lu model}
Under the Chung-Lu model \cite{chung2002average}, $P_{jk}=\theta_j\theta_k$ for all $j<k$, where $\theta_i\in\mathbb{R}^+,i\in [n]$ are the degree parameters. 

\subsubsection*{Naive estimator}
We can estimate $P$ as follows:
\begin{equation}
    \widehat{P}_{jk}=\left\{
    \begin{array}{cc}
       \dfrac{D_j\,D_k}{2M}  &  j\neq k\\[.2cm]
       0  & \text{otherwise}
    \end{array},\right.
    \label{phat_cl}
\end{equation}
where $D_j=\sum\limits_{\ell=1}^n A_{j\ell}$ and $M=\underset{\ell<m}{\sum\sum}A_{\ell,m}$.
\subsubsection*{Node-based estimator}
To construct the $i^{th}$ egonet statistic, we propose to estimate $P$ as
\begin{equation}
    \widehat{P}^{(i)}_{jk}=\left\{
    \begin{array}{cc}
       \dfrac{\sum_{\ell\in{\mathcal{N}_i^c}} A_{j\ell} \sum_{m\in{\mathcal{N}_i^c}}A_{km}}{\sum_{\ell\in{\mathcal{N}_i^c}}\sum_{m\in{\mathcal{N}_i^c}}A_{\ell m}}  &  j\neq k;\ j,k\neq i\\[.2cm]
       0  & \text{otherwise}
    \end{array}\right.
    \label{phat_cl_node_based}
\end{equation}
Then, the egonet statistic is given by
\begin{equation*}
    T_{i,CL}= \underset{j<k}{\sum\sum} A_{ij}A_{ik}\left(A_{jk} - \widehat{P}^{(i)}_{jk}\right).
\end{equation*}
As before, it is evident that $\widehat{P}^{(i)}$ only depends on the edges $\{A_{jk}:A_{ij}A_{ik}=0,\,j,k\neq i\}$, which are unaffected by the anomalous clique if $i$ belongs to $S$.

\subsubsection{RDPG model}
Under the RDPG model \cite{young2007random}, $P=XX^\top $ for some $X\in \real^{n\times d}$, such that $P$ is a positive semidefinite matrix of rank $d$, and $d$ is assumed to be known. 
The Chung-Lu model is a special case of the RDPG model.
Here, each node $i$  has a $d$-dimensional latent position vector $X_i$, the $i^{th}$ row of $X$, and the probability that nodes $j$ and $k$ are connected is given by, $P_{jk}=X_j^\top X_k$. 
When $d=1$, this reduces to the Chung-Lu model scenario. 
Many well-known network models, such as the stochastic blockmodel (SBM) \cite{lorrain1971structural} and the degree-corrected blockmodel (DCBM) \cite{karrer2011stochastic} can be expressed as a special case of the RDPG model.

\begin{figure}[h!]
	\centering
	\includegraphics[height=0.3\textheight]{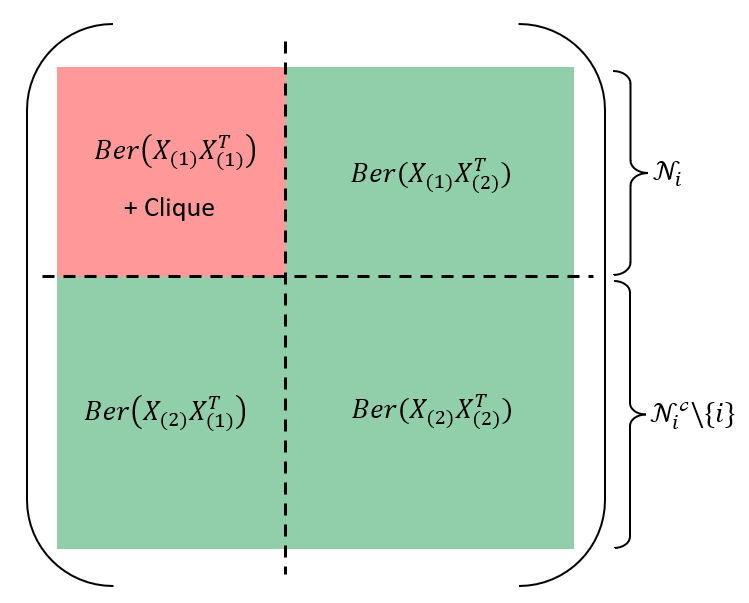}
	\caption{ \small Distribution of the adjacency matrix $A^{-i}$ under $H_1$ when the null model is RDPG (partitioned based on the neighbors of a node $i$ in the clique).}	
	\label{rdpgsec}
\end{figure}

\subsubsection*{Naive estimator}
We can estimate $P$ using the Adjacency Spectral Embedding (ASE) method, proposed in \cite{sussman2012consistent}. 
Let $P=U_PS_PU_P^\top $ be the spectral decomposition of $P$, where $$S_P=diag(\lambda_1(P),\ldots,\lambda_d(P)),\lambda_1(P)\geq\lambda_2(P)\geq\ldots\geq\lambda_d(P)>0.$$
And, let $$A=[U_A|U_A^{\perp}][S_A\oplus S_A^{\perp}][U_A|U_A^{\perp}]^\top $$
be the spectral decomposition of $A$, where
\begin{align*}
    &S_A=diag(\lambda_1(A),\ldots,\lambda_d(A)),\ S_A^{\perp}=diag(\lambda_{d+1}(A),\ldots,\lambda_n(A)),\\
    &|\lambda_1(A)|\geq|\lambda_2(A)|\geq\ldots\geq|\lambda_d(A)|\geq\ldots\geq |\lambda_n(A)|.
\end{align*}

We can define $X=U_PS_P^{\frac{1}{2}}$, noting that this choice of $X$ is not unique. 
The ASE of $A$ into $\real^d$ is given by, $\widehat{X}=U_AS_A^{\frac{1}{2}}$, and $P$ can be estimated as $\widehat{P}=\widehat{X}\widehat{X}^\top $. Then, the egonet statistic is given by,
\begin{equation}
    T_{i,RDPG}= \underset{j<k}{\sum\sum} A_{ij}A_{ik}(A_{jk}-\widehat{X}_j^\top \widehat{X}_k).
\end{equation}
Under the null, $\widehat{P}$ can be shown to be a consistent estimator of $P$ following the theory from \cite{xie2021entrywise}. 

\subsubsection*{Node-based estimator}
To construct the $i^{th}$ egonet statistic, we estimate $P^{-i}$ as follows. 
Let $X_{(1)}$ and $X_{(2)}$ denote the submatrices $X_{\mathcal{N}_i,.}$ and $X_{\mathcal{N}_i^c\backslash\{i\},.}$ respectively. 
Then, the subgraph $A^{-i}_{(22)}$ follows the RDPG model with latent positions $X_{(2)}$. So, we first estimate $X_{(2)}$ by applying Adjacency Spectral Embedding on the submatrix $A^{-i}_{(22)}$, let's call the estimator $\widehat{X}_{(2)}$. 
It is well-known from the RDPG theory that $\widehat{X}_{(2)}$ is stochastically close to $X_{(2)}W$ for some orthogonal matrix $W$. 
Now to estimate $X_{(1)}$, we observe the following relationship,
$$X_{(1)}X_{(2)}^\top \widehat{X}_{(2)}(\widehat{X}_{(2)}^\top \widehat{X}_{(2)})^{-1}\approx X_{(1)}W\widehat{X}_{(2)}^\top \widehat{X}_{(2)}(\widehat{X}_{(2)}^\top \widehat{X}_{(2)})^{-1}=X_{(1)}W.$$
The sample version of $X_{(1)}X_{(2)}^\top $ is given by $A^{-i}_{(12)}$. Therefore, we can estimate $X_{(1)}$ as
\begin{equation}
    \widehat{X}_{(1)} = A^{-i}_{(12)}\widehat{X}_{(2)}(\widehat{X}_{(2)}^\top \widehat{X}_{(2)})^{-1}.
\end{equation}
$\widehat{X}_{(1)}$ should be stochastically close to $X_{(1)}W$ for the same orthogonal matrix $W$. Therefore, the matrix $\widehat{X}^{-i}$ obtained by augmenting $\widehat{X}_{(1)}$ and $\widehat{X}_{(2)}$ should be close to $X^{-i}W$. 
Then, 
an estimate of $P^{-i}$ is given by $\widehat{P}^{-i}=\widehat{X}^{-i}({\widehat{X}^{-i}})^\top$.
The egonet statistic is given by
\begin{equation*}
    T_{i,RDPG}= \underset{j<k}{\sum\sum} A_{ij}A_{ik}\left(A_{jk} - \widehat{P}^{(i)}_{jk}\right),
\end{equation*}
where $\widehat{P}^{(i)}$ is constructed from $\widehat{P}^{-i}$ as described earlier.

\section{Theoretical results}
\label{sec:theory}
In this section, we provide the theoretical properties of the egonet method under the four following cases: when $P$ is known, or when $P$ is estimated assuming one of the three models considered in this paper, namely, \ER, Chung-Lu, and RDPG. 
For the Chung-Lu and the RDPG model, we derive results when $P$ is estimated by the naive estimators defined in Section \ref{subsec:egonet_unknown}.
We follow the definitions from Sections \ref{sec:intro} and \ref{sec:egonet}. 
In addition, we introduce some new notations to describe the results. Let us recall that we want to test
\begin{equation*}
\begin{split}
H_0:\ &A_{ij} \sim \text{Bernoulli} (P_{ij}) \text{ for all } 1 \le i < j \le n, \text{vs.}\\
 H_1:\ &A_{ij} \sim \text{Bernoulli} (Q_{ij}) \text{ for all } 1 \le i < j \le n,\text{such that }\\
        &Q_{ij} = \left\{
        \begin{array}{ll}
            1 & \text{ if } i,j \in S \text{ and } i < j, \\
            P_{ij}  & \text{ if}\ i \notin S \text{ or } j \notin S, \text{ and } i < j.
        \end{array}\right.
\end{split}
\end{equation*}

Let $\delta(P)=\max_{i}\sum_{j=1}^nP_{ij}$ be the maximum expected degree.
Throughout our theory, we are going to assume that 
\begin{equation}
    \delta(P)\gg \log n.
    \label{sparsity2}
\end{equation}

This assumption is often referred to as the \textit{dense regime} in networks (\cite{arias2014community}). For certain parts of our theory, we are also going to need slightly different conditions on the sparsity of the network, which we will specify as required.
Next, we define the difference 
\begin{equation}
    B_S=Q-P
    \label{signal}
\end{equation} as the \textit{signal} of the anomalous clique. 
Moreover, for each node $i$, define $B_S^{-i}$ to be the $(N-1)\times (N-1)$ matrix obtained by excluding the $i^{th}$ row and column of $B_S$. 
Note thst if $i$ is in the anomalous clique $S$, 
the contribution of the $S$ in the egonet statistic is captured by $B_S^{-i}$.
We provide theoretical guarantees for the egonet method that generally rely on conditions concerning the \textit{strength} of the anomalous clique, expressed by the norms of $B_S$ and $B_S^{-i}$-s.
Specifically, for each of the four cases, we derive conditions under which the detection test is asymptotically powerful, and under the alternative, the egonet ranking algorithm (Algorithm \ref{alg:cap}) finds the anomalous clique almost surely. 

\subsection{\texorpdfstring{$P$}{TEXT} is known}
We start with the case when $P$ is known. 

Define 
\begin{gather*}
    \sigma_{i,A}^2 \coloneqq \text{Var}_{H_0}(T_i\mid a_i) = \underset{j<k}{\sum\sum}\,A_{ij}A_{ik}P_{jk}(1-P_{jk}),\\
    \text{and},\sigma_{i,P}^2 \coloneqq \EE_{H_0}(\sigma_{i,A}^2) = \underset{j<k}{\sum\sum}\,P_{ij}P_{ik}P_{jk}(1-P_{jk}).
\end{gather*}
Define the matrix $V$ as
\begin{equation*}
    V_{jk} \coloneqq P_{jk}(1-P_{jk}),\ 1\leq j<k\leq n.
\end{equation*}
We have the following result.

\begin{theorem}\label{knownp_h0_alt}
{\em (Detection)
Under $H_0$, for all $i\in[n]$,
\begin{equation}
        \PP_{H_0}\left(T_i\leq \sqrt{2c\log n}\,\sigma_{i,A} + \frac{2c}{3}\log n\right)\geq 1-O(n^{-c}),
        \label{knownp_type1}
\end{equation}
for any constant $c>1$, depending on $c$.

Under $H_1$, suppose that
\begin{equation}
    \frac{1}{2}\min_{i\in S}\umod{B_S^{-i}}_{1,1} \gg C \left((\|V\|_F\log n)^{\frac{1}{2}} \max\{\|Q\|_{2\rightarrow\infty},\sqrt{\log n}\} + \sqrt{\delta(Q)}\log n\right).
    \label{knownp_detect_cond}
\end{equation}
Then, for all $i\in S$,
\begin{equation}
    \PP_{H_1}\left(T_i > \sqrt{2c\log n}\,\sigma_{i,A} + \frac{2c}{3}\log n\right)\geq 1-O(n^{-c}),
    \label{knownp_power}
\end{equation}
for any constant $c>1$, provided that $n$ is sufficiently large, depending on $c$.}
\end{theorem}

We propose the detection threshold as, $$C_n=\sqrt{2c\log n}\,\sigma_{i,A} + \frac{2c}{3}\log n.$$
Then, the type-I error of the detection test is given by,
\begin{equation*}
    \begin{split}
        & \PP_{H_0}\left(\bigcup\nolimits_{i\in [n]} \{T_i>C_n\}\right)= O(n^{c-1})\rightarrow 0\ \text{if}\ c>1.
    \end{split}
\end{equation*}
By Eq. \eqref{knownp_power}, the test is asymptotically of power 1.

For the next result, let us first define a matrix $P_S$ such that
\begin{equation*}
    (P_S)_{ij} = P_{ij}\,\mathbb{I}(i,j\in S),\ 1\leq i,j\leq n.
\end{equation*}
\begin{theorem}\label{knownp_ranking_alt}
{\em (Localization)
Under $H_1$, suppose Eq. \eqref{knownp_detect_cond} holds.
Moreover, assume that
\begin{equation}
    \begin{split}
        &\,\frac{1}{2}\min_{i\in S}\umod{B_S^{-i}}_{1,1} - \frac{1}{2}\max_{i\in S^c}\,p_i^\top B_S\,p_i \\[.2cm]
        \gg&\, C \left(\|B_S\|_F \sqrt{\log n}\max\{\|P_S\|_{2\rightarrow\infty},\sqrt{\log n}\} + \delta(P_S)\log n\right).
        \label{knownp_locate_cond1}
    \end{split}
\end{equation}
Then, for any constant $c>1$,
\begin{equation}
    \PP_{H_1}\left(\min_{i\in S} T_i > \max_{i'\in S^c} T_i'\right) > 1-O(n^{-(c-1)}),
    \label{knownp_loc}
\end{equation}
provided that $n$ is sufficiently large, depending on $c$.
In addition, if there exists $\epsilon > 0$ such that for all $i\in S^c$,
\begin{equation}
    \sum_{j\in S} P_{ij} \leq (1-\epsilon)\,|S|,\ \text{and}\ |S| \gg \log n,
    \label{knownp_locate_cond2}
\end{equation}
then the clique $\widehat{S}$ recovered by Algorithm \ref{alg:cap} is exactly equal to the anomalous clique $S$, that is, $\PP_{H_1}(\widehat{S}=S)\rightarrow 1$.}
\end{theorem}

\begin{remark}
    Eq. \eqref{knownp_detect_cond} guarantees detection, and Eq. \eqref{knownp_locate_cond1} is the additional condition required to ensure localization.
    To illustrate an example where detection can be achievable, but localization can be hard, consider an SBM network with 50 nodes and two equal-sized communities. 
    Suppose that the nodes in the $1^{st}$ community are popular, that is, they have a high connection probability, say 0.8, to all the nodes, while the nodes in the $2^{nd}$ community have a low connection probability, say 0.2, among themselves.
    If a mixed clique of size 10 consisting of 5 nodes from community 1 and 5 nodes from community 2 is embedded in the network, it can be easier to detect the clique using egonets, since there would be a significant increase in the egonet values for nodes in community 2.
    Mathematically, $\umod{B_S}_{1,1}$ will be large, thus Eq. \eqref{knownp_detect_cond} can be satisfied.
    On the other hand, since the non-clique nodes from community 1 are also strongly connected to all the clique nodes, it can be harder to localize the clique nodes.
    Mathematically, $p_i^\top B_S\,p_i$ will be large for non-clique nodes from community 1, and Eq. \eqref{knownp_detect_cond} may not be satisfied.
\end{remark}

\subsection{\texorpdfstring{$P$}{TEXT} is unknown}
\label{sec:theory_unknown}

Now, we consider the case when $P$ is unknown under three background model structures: the Erd\"{o}s-Ren\'{y}i model, the Chung-Lu model and the RDPG model. For the model parameters, we follow the notations from Section \ref{subsec:egonet_unknown}.

\subsubsection{ER model}
We assume that there exists a constant $p_1\in (0,1)$ such that $p < p_1$.

\begin{theorem}\label{unknownph0er}
{\em (Detection) Under $H_0$, for all $i\in[n]$,
\begin{align*}
    \PP_{H_0}\bigg(T_{i,ER}\,\leq\, C\max\left\{np^{3/2}\sqrt{(1-p)\log n},\log n\right\}\bigg)\geq 1-\frac{6}{n^c},
\end{align*}
for any constant $c>1$, provided that $n$ is sufficiently large, depending on $c$. 

Under $H_1$, uppose that there exists a constant $\alpha_1>0$ such that
\begin{equation}
    np\sqrt{(\log n)^{1+\alpha_1}} \ll k \ll n.
    \label{er_detectcond}
\end{equation}
Then, for all $i\in S$,
\begin{equation*}
    \PP_{H_1}\left(T_{i,ER}\geq \gamma_1 k^2\right)\geq 1-\frac{4}{n^c},
\end{equation*}
for some constant $\gamma_1>0$.}
\end{theorem}
The condition $k\ll n$ ensures that the egonet-complement part is sufficiently large so that $p$ is estimated accurately.
We propose the detection threshold, 
$$C_{n,ER}=n\widehat{p}^{\,(i)}\sqrt{(\log n)^{1+\beta}},\ \beta<\alpha_1.$$
As before, we can show that the type-I error of the detection test goes to 0 provided $c>1$, and the power of the detection test goes to 1 provided $c>0$.

\begin{theorem}\label{unknownprankinger}
{\em (Localization) Under $H_1$,
suppose that Eq. \eqref{er_detectcond} holds. Then
\begin{equation*}
    \PP_{H_1}\left(\min_{i\in S} T_{i,ER} > \max_{i'\in S^c} T_{i',ER}\right) \geq 1-O(n^{(c-1)}).
\end{equation*}
Consequently, the anomalous clique is exactly recovered, that is, $\PP_{H_1}(\widehat{S}=S) \geq 1-O(n^{(c-1)})$.}
\end{theorem}

\subsubsection{Chung-Lu model}
We assume the following condition on the degree parameters $\theta_i$'s:
\begin{equation}
     \sum_\ell \theta_{\ell}^2 \ll \left(\sum_\ell \theta_{\ell}\right)^2.
     \label{degree_cond}
\end{equation}
This condition ensures that $\theta_i$'s can not vary too much, thereby controlling the heterogeneity in the network.
The condition is satisfied, for example, if $\theta_{\min}/\theta_{\max}\gg 1/\sqrt{n}$.
We also assume that
$$\delta(P)\gg \log n.$$
We first state the result on detection.

\begin{theorem}\label{unknownph0cl} {\em (Detection)
Under $H_0$, for all $i\in[n]$,
\begin{equation*}
    \PP_{H_0}(T_{i,CL} \leq C\,\delta(P)\sqrt{\log n}) \geq 1-O(n^{-c}),
\end{equation*}
for any constant $c>1$, provided that $n$ is sufficiently large, depending on $c$.

Under $H_1$, suppose that there exists a constant $\alpha_1>0$ such that
\begin{gather}
    \umod{B_S}_{1,1} \geq \delta(P)\sqrt{(\log n)^{1 + \alpha_1}}, \label{cl_detectcond1}\\[.2cm]
    \delta(B_S)\ll \delta(P),\,\umod{B_S}_{1,1} \ll \left(\sum_\ell \theta_{\ell}\right)^2,\ \sum_{\ell\in S}\theta_\ell \ll \sum_\ell \theta_{\ell},\label{cl_detectcond2}\\[.2cm]
    \text{there exists } i\in S\ \text{such that } \max_{j\in S^c}P_{ij}<\sqrt{5}-2. \label{cl_detectcond3}
\end{gather}
Then,
\begin{equation*}
    \PP_{H_1}(\max_{i\in S} T_{i,CL}\geq \gamma_1\umod{B_S}_{1,1})\geq 1-O(n^{-c}),
\end{equation*}
for some constant $\gamma_1>0$.}
\end{theorem}

Conditions \eqref{cl_detectcond1} and \eqref{cl_detectcond2} are lower and upper bound requirements on the strength of the clique, respectively. 
The lower bounds in \eqref{cl_detectcond2} are required to control the bias of $\widehat{P}$ under $H_1$.

We propose the detection threshold, 
$$C_{n,CL}=\delta(\widehat{P})\sqrt{(\log n)^{1+\beta}},\ \beta<\alpha_1.$$
As before, we can show that the type-I error of the detection test goes to 0 provided $c>1$. and the power of the detection test goes to 1 provided $c>0$.

\begin{theorem}\label{unknownprankingcl}
{\em (Localization) Under $H_1$, suppose that Eq. \eqref{cl_detectcond1}-\eqref{cl_detectcond3} hold. Then,
\begin{equation*}
    \PP_{H_1}\left(\min_{i\in S} T_{i,CL} > \max_{i'\in S^c} T_{i',CL}\right) \geq 1-O(n^{(c-1)}).
\end{equation*}
Consequently, the anomalous clique is exactly recovered, that is, $\PP_{H_1}(\widehat{S}=S) \geq 1-O(n^{(c-1)})$.}
\end{theorem}

\subsubsection{RDPG model}
Before discussing our results in this case, we need to introduce some additional notations related to the probability matrix under the alternative, $Q$. Let $q_i$ be the $i^{th}$ row of $Q$. Let
\begin{equation*}
    Q=[U_Q\mid U_Q^{\perp}]\, [S_Q\oplus S_Q^{\perp}]\, [U_Q\mid U_Q^{\perp}]^\top    
\end{equation*}
be the spectral decomposition of $Q$, and
\begin{align*}
    &S_Q=diag(\lambda_1(Q),\ldots,\lambda_d(Q)),\ S_Q^{\perp}=diag(\lambda_{d+1}(Q),\ldots,\lambda_n(Q)),\\
    &|\lambda_1(Q)|\geq|\lambda_2(Q)|\geq\ldots\geq|\lambda_d(Q)|\geq\ldots\geq |\lambda_n(Q)|.
\end{align*}
Let $u_i(Q)$ be the $i^{th}$ eigenvector of $Q$, that is, $Q=\sum\limits_{i=1}^n \lambda_i(Q)u_i(Q)u_i(Q)^\top $. 

Next, we define the matrix $\widecheck{P}$ as the best rank-$d$ approximation of $Q$, that is,  
$$\widecheck{P}=\widecheck{X}\widecheck{X}^\top ,\ \text{where }\widecheck{X}=U_QS_Q^{\frac{1}{2}}.$$
$\widecheck{P}$ plays a crucial role in understanding the theoretical properties of the egonet method under the RDPG model.
Note that, when $Q=P$, that is under the null, $\widecheck{P}=P$ as well. 
Under the alternative, the ASE-based estimator $\widehat{P}$ is likely to be not close to $P$, and we do not want $\widehat{P}$ to be close to $Q$ either, since that would make our egonet statistics small even under the alternative. 
We show that under some suitable assumptions, $\widehat{P}$ is close to $\widecheck{P}$. Therefore, heuristically, the contribution of $(Q-\widecheck{P})$ would need to be significant enough for the egonet method to succeed. 
In our theoretical results, we are going to find the exact conditions involving $(Q-\widecheck{P})$ that are required for the detection and localization step.

We assume the following condition holds:
\begin{equation}
    \lambda_d(P)\geq c_1\delta(P),\text{ for some constant }c_1>0, \delta(P)\gg (\log n).
    \label{rdpg_cond}
\end{equation}
Similar sparsity assumptions are common in the RDPG model literature, e.g., \cite{cape2019signal,xie2021entrywise}.

\begin{theorem}\label{unknownph0rdpg} 
{\em (Detection) Under $H_0$, for all $i\in[n]$,
\begin{equation*}
    \PP_{H_0}\left(|T_{i,RDPG}|\leq C\,\delta(P)\sqrt{\log n}\max\left\{1,\sqrt{\dfrac{\delta(P)\log n}{n}}\right\}\right)\geq 1-O(n^{-c}),
\end{equation*}
for any constant $c>1$, provided that $n$ is sufficiently large, depending on $c$.

Under $H_1$, suppose that there exists constants $\alpha_1>0, \gamma_1 > 0$, such that
\begin{gather}
    \delta(B_S)\ll \delta(P) \ll \umod{B_S}^2/(\log n)^{1+\alpha_1}, \label{rdpg_detectcond1} \\[.2cm]
    \max_{\,i\in[n]}\ q_i^\top (Q-\widecheck{P})q_i \geq \gamma_1 \umod{B_S}^2,
    \label{rdpg_detectcond2}
\end{gather}
Then, 
\begin{equation}
    \PP_{H_1}\left(\max_{i\in[n]} T_{i,RDPG} \geq (\gamma_1/2)\umod{B_S}^2\right) \geq 1-O(n^{-c}).
    \label{unknownp_power_rdpg}
\end{equation}}
\end{theorem}

Eq. \eqref{rdpg_detectcond1} imposes a lower and upper bound restriction on the strength of the clique, where the lower bound is required to control the bias in $\widehat{P}$ under $H_1$.
To interpret the condition \eqref{rdpg_detectcond2}, observe that
\begin{align*}
    &q_i^\top (Q - \widecheck{P})q_i=e_i^\top Q\,U_Q^{\perp} S_Q^{\perp}{U_Q^{\perp}}^\top Q\,e_i=e_i^\top U_Q^{\perp}{S_Q^{\perp}}^3{U_Q^{\perp}}^\top e_i\\[.2cm]
    =& \sum_{\ell =d+1}^n \lambda_\ell(Q)^3(e_i^\top u_\ell(Q))^2=\sum_{\ell =d+1}^n \lambda_\ell(Q)^3(u_\ell(Q))_i^2=\sum_{\ell =d+1}^n \lambda_\ell(Q)^3(U_Q^\top )_{i\ell}^2.
\end{align*}
Therefore, we essentially need that 
$$\max_{i\in[n]} \sum_{\ell =d+1}^n \lambda_\ell(Q)^3(U_Q^\top)_{i\ell}^2 = \Omega(\umod{B_S}^2).$$
If $\lambda_{d+1}(Q)$ is very small (and hence, the subsequent eigenvalues), the left-hand side will also be very small and the condition will not be met. 
Now, $\lambda_{d+1}(Q)$ is the operator norm of $U_Q^{\perp}S_Q^{\perp}{U_Q^{\perp}}^\top =Q-U_QU_Q^\top Q$. 
So, if $Q$ is well-approximated by its rank-$d$ approximation $\widecheck{P}$, $\lambda_{d+1}(Q)$ will be small and this condition will not be met. 
In this case, $\widehat{P}$ will be a very good estimator of $Q$ and hence, $a_i^\top (A-\widehat{P})a_i\approx a_i(A-Q)a_i$ will not be large enough. 
As an example of when this situation might arise, consider the 2 community SBM we discussed in the introduction (Figure \ref{sbm_example}). 
If we plant a clique exactly on the subgraph of all nodes belonging to the first community, then we again have an SBM network with 2 communities, although with different connection probabilities. 
Naturally, $\widehat{P}$ will be a good estimator of $Q$, and the egonet method will fail to detect the anomalous clique. 
Therefore, for the egonet method to succeed, we need the effect of the anomalous clique to show up in the space projected by the non-top eigenvectors of $Q$.

We propose the detection threshold, 
$$C_{n,RDPG}=\delta(\widehat{P})(\log n)^{1+\beta},\ \beta<\alpha_1.$$

\begin{theorem}\label{unknownprankingrdpg}
{\em (Localization) Under $H_1$,
suppose that Eq. \eqref{rdpg_detectcond1} and \eqref{rdpg_detectcond2} hold. 
In addition, define $q_i^{(1)}$ such that $(q_i^{(1)})_j=(q_i)_j$ for all $j\in S$, and $(q_i^{(1)})_j=0$ otherwise. Let $q_i^{(2)}=q_i-q_i^{(1)}$. If there exists constants $\gamma_2>0$ and $\epsilon\in(0,1/8)$ such that

\begin{equation}
    {q_i^{(1)}}^\top (Q-\widecheck{P})q_i^{(1)}\left\{
    \begin{array}{ll}
        \geq & \gamma_2\umod{B_S}^2\ \text{for }i\in S,\\ [.2cm]
        \leq & \gamma_2\epsilon\umod{B_S}^2\ \text{for }i\in S^c,
    \end{array}\right.
    \label{rdpg_localizationcond1}
\end{equation}
and
 \begin{equation}
     \max({q_i^{(1)}}^\top (Q-\widecheck{P})q_i^{(2)},{q_i^{(2)}}^\top (Q-\widecheck{P})q_i^{(2)})\leq \gamma_2\epsilon\umod{B_S}^2\ \text{for all }i\in[n],
     \label{rdpg_localizationcond2}
 \end{equation} then, for any constant $c>1$,
\begin{equation*}
    \PP_{H_1}\left(\min_{i\in S} T_{i,RDPG} > \max_{i'\in S^c} T_{i',RDPG}\right) \geq 1-O(n^{-(c-1)}),
\end{equation*}
provided that $n$ is sufficiently large, depending on $c$.}
\end{theorem}

For localization, we need the bias in the quantity $q_i^\top (Q-\widecheck{P})q_i$ to not only be large but also be concentrated on the submatrix of $Q-\widecheck{P}$ induced by $S$. Let us expand the quantity $q_i^\top (Q-\widecheck{P})q_i$.
$$q_i^\top (Q-\widecheck{P})q_i={q_i^{(1)}}^\top (Q-\widecheck{P})q_i^{(1)} + 2{q_i^{(1)}}^\top (Q-\widecheck{P})q_i^{(2)}+{q_i^{(2)}}^\top (Q-\widecheck{P})q_i^{(2)}.$$
The first quantity, ${q_i^{(1)}}^\top (Q-\widecheck{P})q_i^{(1)}$, captures the contribution from the submatrix of $Q-\widecheck{P}$ induced by $S$. The conditions \eqref{rdpg_localizationcond1} and \eqref{rdpg_localizationcond2} imply that for $i\in S$, the contribution from the submatrix of $Q-\widecheck{P}$ induced by $S$ dominates the contribution from the other parts of $Q-\widecheck{P}$. For $i\in S^c$, the entire quantity $q_i^\top (Q-\widecheck{P})q_i$ needs to be small enough. In simpler models such as ER and Chung-Lu, these conditions translate to requiring the sparsity of the network to be small enough.

\subsection{Summary}
We conclude this section with a summary of the results we obtained above.\\
[1em]
1. \textbf{(Detection thresholds)}
In Table \ref{tab:detection_thresholds}, we report the detection thresholds derived for the four cases: when $P$ is known, and when $P$ is estimated under the \ER, Chung-Lu and RDPG model.
\begin{table}[ht]
    \centering
    \footnotesize
    \begin{tabular}{|c|c|c|c|c|}
        \hline
        Model & $P$ is known & \ER & Chung-Lu & RDPG \rule{0pt}{3ex}\\
        [.2cm]
        \hline
        Detection threshold & $\sqrt{2c\log n}\,\sigma_{i,A} + (2c/3)\log n$ & $n\widehat{p}^{\,(i)}\sqrt{(\log n)^{1+\beta}}$ & $\delta(\widehat{P})\sqrt{(\log n)^{1+\beta}}$ & $\delta(\widehat{P}){(\log n)}^{1+\beta}$ \rule{0pt}{3ex}\\
        [.2cm]
        \hline
    \end{tabular}
    \caption{Detection thresholds for the egonet method under different models.}
    \label{tab:detection_thresholds}
\end{table}

For the known-$P$ case, the detection threshold is nuanced and is based on the conditional variances, ${\sigma}_{i,A}^2$. It can be shown that the threshold for the known-$P$ case is bounded above by $\delta(P)\sqrt{(\log n)^{1+\beta}}$ for any $\beta>0$ with high probability. 
When $P$ is estimated assuming the \ER model, we achieve a threshold of the same order, with $\delta(P)\approx np$ replaced by its estimate $n\widehat{p}^{\,(i)}$. 
For the Chung-Lu Model as well, we achieve the same threshold, $\delta(\widehat{P})\sqrt{(\log n)^{1+\beta}}$. 
Finally, for the RDPG model, we require a slighly stronger threshold $\delta(\widehat{P}){(\log n)}^{1+\beta}$, which can be attributed to the additional complexity of the model.

\noindent 2. \textbf{(Theoretical bounds on the clique's strength)}
A key part of our assumptions is requiring bounds on the strength of the clique, quantified by the norms of $B_S$. 
It is expected that we would need the anomalous clique to be strong enough for our method to find the clique. 
In the first row of Table \ref{tab:bounds}, we report the lower bound requirements on the clique's strength for the four cases: when $P$ is known, and when $P$ is estimated under the \ER, Chung-Lu and RDPG model. 
When $P$ is estimated, we also need an upper bound on the clique's strength to identify the clique (second row).  
Recall that in Section \ref{subsec:egonet_unknown}, we discussed how the estimation of $P$ can be problematic under the alternative when the anomalous clique is `too strong'. 
Here, we have quantified that strength through norms of $B_S$, and we find that when $P$ is estimated, we indeed need an upper bound on the clique strength for our method to succeed.
\begin{table}[ht]
    \centering
    \footnotesize
    \begin{tabular}{|c|c|c|c|}
        \hline
        $P$ is known & \ER & Chung-Lu & RDPG \rule{0pt}{3ex}\\
        [.2cm]
        \hline
        $\umod{B_S}_{1,1}\gg \delta(P)\sqrt{\log n}$ & $\umod{B_S}_{1,1}\gg np\sqrt{\log n}$ & $\umod{B_S}_{1,1}\gg n\sqrt{\rho_n\log n}$ & $\umod{B_S}^2\gg \delta(P)\log n$ \rule{0pt}{3ex}\\
        [.2cm]
        \hline
        - & $\sqrt{\umod{B_S}_{1,1}} = o(n)$ & 
        $\delta(B_S)\ll\sum_{\ell=1}^n \theta_{\ell}$, & $\delta(B_S)\ll \delta(P)$\rule{0pt}{3ex}\\
         & & $\umod{B_S}_{1,1}\ll (\sum_{\ell=1}^n \theta_{\ell})^2$ & \rule{0pt}{3ex}\\
        \hline
    \end{tabular}
    \caption{Theoretical lower (first row) and upper (second row) bounds on the clique signal for the egonet method under different models.}
    \label{tab:bounds}
\end{table}

\section{Simulation study}
\label{sec:sim}
In this Section, we present results for the egonet method for synthetic networks generated from the Chung-Lu model, the stochastic block model, and the RDPG model. We evaluate the performance of the method using the following metrics:
\begin{enumerate}
	\item False alarm rate (Size): The false alarm rate is the percentage of non-anomalous networks where the null hypothesis is rejected, i.e., the method wrongly detects the presence of an anomalous clique although there isn't one. 
    This should be small, close to zero.
	\item Detection rate (Power): The detection rate is the percentage of anomalous networks where the null hypothesis is rejected, i.e., the method correctly detects the presence of a clique. This should be close to 100.
    \item Node detection rate (Localization): in networks with an embedded clique, the node detection rate is the percentage of anomalous nodes whose egonet statistic value is higher than the threshold, i.e., nodes that are correctly flagged by the method. This should ideally be close to 100.
    \item Node false alarm rate (False alarm): the node false alarm rate is the percentage of non-anomalous nodes whose egonet statistic value lies above the threshold, i.e., nodes that are wrongly flagged by the method. This should be close to zero.
\end{enumerate}

\subsection{Chung-Lu model}
We generated networks from the Chung-Lu model with $n=500, 1000$ nodes, and of three different values of density parameter, $\delta=0.05, 0.1$ and $0.2$. The model parameters $\{\theta_i\}$ were sampled from the $\text{Uniform}(0,\rho_n)$ distribution, so that ${\rho}_n$ is the maximum edge probability and $\delta={\rho}_n/4$ is the expected network density. To generate networks under the alternative, we randomly selected a set of $k$ nodes and embedded a clique on the subgraph spanned by them. We explored three different ways to select the clique nodes,
\begin{enumerate}
    \item selecting the nodes uniformly, i.e., with equal probability
    \item selecting the $k$ nodes with the largest expected degree in the underlying network
    \item selecting the $k$ nodes with the smallest expected degree in the underlying network
\end{enumerate}
We developed our theory under the first scenario, where the anomalous clique is formed by any uniformly random $k$ nodes in the network. Here, we also examine the second and third scenario where the nodes are selected based on their degrees. This will help us understand if the degrees of nodes forming the clique have any effect on the performance of our method.

The egonet method is implemented on both the original non-anomalous networks and the networks with an embedded clique. For the anomalous networks, we chose $k=c\sqrt{2\,\delta(P)\sqrt{\log n}}$ so that the contribution of the anomalous clique in the egonet degree, approximately $ k^2/2$, is equal to $c^2$ times the detection threshold. The results are tabulated in Tables \ref{simclh0}-\ref{simclh1maxdeg}.

\begin{table}[ht]
\centering
\footnotesize
\begin{tabular}{|c|cc|cc|}
\hline
   & \multicolumn{2}{c|}{$n=500$} & \multicolumn{2}{c|}{$n=1000$}\\
  \hline
   $\delta$ & Size & False alarm & Size & False alarm\\
  \hline
    0.05 & 0.00 & - & 0.00 & - \\
    0.1 & 0.00 & - & 0.00 & - \\
    0.2 & 0.00 & - & 0.00 & - \\
  \hline
  \end{tabular}
  \captionof{table}{Results for the Chung-Lu model with no anomalous clique}
  \label{simclh0}
  \end{table}

\begin{table}[ht]
\centering
\fontsize{9pt}{9pt}\selectfont
\begin{tabular}{|c|c|rrcc|rrcc|}
\hline
   & & \multicolumn{4}{c|}{$n=500$} & \multicolumn{4}{c|}{$n=1000$}\\
  \hline
   $\delta$ & $c$ & $\sqrt{\umod{B_S}_{1,1}}$ &
  Power & Localization & False alarm & $\sqrt{\umod{B_S}_{1,1}}$ & Power & Localization & False alarm\\
  \hline
    0.05 & 1 & 20.96 & 0 & - & - & 30.67 & 0 & - & -  \\ 
    0.05 & 2 & 42.37 & 3 & 100.00 & 0.00 & 61.91 & 99 & 100.00 & 0.00\\ 
    0.05 & 4 & 86.23 & 100 & 100.00 & 0.00 & 125.31 & 100 & 100.00 & 0.00\\ 
    \hline
    0.1 & 1 & 28.95 & 0 & - & - & 42.21 & 0 & - & -\\ 
  0.1 & 2 & 59.32 & 92 & 100.00 & 0.00 & 85.84 & 100 & 100.00 & 0.00 \\ 
  0.1 & 4 & 118.92 & 100 & 100.00 & 0.00 & 173.12 & 100 & 100.00 & 0.00\\ 
  \hline
  0.2 & 1 & 38.82 & 0 & - & - & 56.98 &0 & - & -\\ 
  0.2 & 2 & 78.69 & 100 & 97.55 & 0.00 & 114.77 &100 & 100.00 & 0.00 \\ 
  0.2 & 4 & 159.03 & 100 & 100.00 & 0.00 & 230.98 &100 & 100.00 & 0.00 \\ 
  \hline
  \end{tabular}
  \captionof{table}{\small Results for the Chung-Lu model when clique nodes are selected uniformly}
  \label{simclh1random}
\vspace{\baselineskip}
\begin{tabular}{|c|c|rrcc|rrcc|}
\hline
   & & \multicolumn{4}{c|}{$n=500$} & \multicolumn{4}{c|}{$n=1000$}\\
  \hline
   $\delta$ & $c$ & $\sqrt{\umod{B_S}_{1,1}}$ &
  Power & Localization & False alarm & $\sqrt{\umod{B_S}_{1,1}}$ & Power & Localization & False alarm\\
  \hline
0.05 & 1 & 21.49 &0 & - & - & 31.50 &0 & - & - \\ 
0.05 & 2 & 43.48 & 100 & 100.00 & 0.00 & 63.49 & 100 & 100.00 & 0.00 \\ 
0.05 & 4 & 88.42 & 100 & 100.00 & 0.00 & 128.44 & 100 & 100.00 & 0.00 \\ 
 \hline
0.1 & 1 & 30.49 &0 & - & - &  44.49 &0 & - & - \\ 
0.1 & 2 & 62.45 &100 & 100.00 & 0.00 & 90.46 & 100 & 100.00 & 0.00 \\ 
0.1 & 4 & 125.10 &100 & 100.00 & 0.00 & 182.18 &100 & 100.00 & 0.00 \\ 
\hline
0.2 & 1 & 43.46&0 & - & - & 63.47 &0 & - & - \\ 
 0.2 & 2 &88.22 & 100 & 100.00 & 0.00 & 128.29 & 100 & 100.00 & 0.00 \\ 
 0.2 & 4 & 175.28 & 100 & 100.00 & 0.00 & 256.74 & 100 & 100.00 & 0.00\\ 
\hline
\end{tabular}
\captionof{table}{\small Results for the Chung-Lu model when nodes with the lowest degrees are selected in the clique}
\label{simclh1mindeg}
\vspace{\baselineskip}
\begin{tabular}{|c|c|rrcc|rrcc|}
\hline
   & & \multicolumn{4}{c|}{$n=500$} & \multicolumn{4}{c|}{$n=1000$}\\
  \hline
   $\delta$ & $c$ & $\sqrt{\umod{B_S}_{1,1}}$ &
  Power & Localization & False alarm & $\sqrt{\umod{B_S}_{1,1}}$ & Power & Localization & False alarm\\
  \hline
  0.05 & 1 & 19.33 & 0 & - & - & 28.29 & 0 & - & - \\ 
  0.05 & 2 & 39.34 & 0 & - & - &  57.24 &0 & - & - \\ 
  0.05 & 4 & 80.84 & 0 & - & - & 116.72 & 100 & 100.00 & 0.00 \\
  \hline
  0.1 & 1& 24.10 & 0 & - & - & 34.98 &0 & - & - \\ 
  0.1 & 2 & 50.37 & 0 & - & - &72.19 & 0 & - & - \\ 
  0.1 & 4 & 104.60 & 0 & - & - & 149.41 & 100 & 100.00 & 0.00 \\
  \hline
  0.2 & 1& 22.61 & 0 & - & - & 31.65 & 0 & - & - \\ 
  0.2 & 2 & 51.57 & 0 & - & - & 70.59 & 0 & - & - \\ 
  0.2 & 4 & 120.49 & 0 & - & - & 162.48 & 0 & - & -\\ 
  \hline
  \end{tabular}
  \captionof{table}{\small Results for the Chung-Lu model when nodes with the highest degrees are selected in the clique}
  \label{simclh1maxdeg}
\end{table}
In Table \ref{simclh0}, we find that for non-anomalous networks generated from the Chung-Lu model, the false alarm is 0\% for all choices of $n$ and $\delta$, that is, the detection test correctly does not reject the null hypothesis.

In Table \ref{simclh1random}, we have networks from the Chung-Lu model with an anomalous clique formed by uniformly randomly selected nodes. For $n=500$, we find that when $\umod{B_S}_{1,1}$ is small ($c=1$), the detection rate is 0\%. So the detection test fails to reject the null hypothesis when $\umod{B_S}_{1,1}$ is small. As $\umod{B_S}_{1,1}$ gets larger, the detection rate improves, and for $c=4$, we have a 100\% detection rate for all the choices of $\delta$. For $n=1000$, the detection rates further improve, as expected. Whenever a clique is detected, the node detection rate is almost 100\% and the node false alarm rate is almost 0\%, that is, the egonet ranking algorithm almost always recovers the clique exactly.

In Tables \ref{simclh1mindeg} and \ref{simclh1maxdeg}, we have anomalous networks with the same choices of $n, \delta$, and $c$, but the clique nodes are selected based on their expected degrees. When the nodes with the lowest degrees are selected (Table \ref{simclh1mindeg}), we find that the performance of the egonet method improves significantly. The detection rate is 100\% for all the cases except when $c=1$, and then, the egonet ranking algorithm almost always recovers the clique exactly. On the other hand, when the high-degree nodes are selected (Table \ref{simclh1maxdeg}), the detection test almost always fails to detect the presence of the clique for all choices of $n, \delta$, and $k$, except when $n=1000$ and $\delta=0.05$ and 0.1, where the detection rate is still 100\%.

\subsection{Stochastic block model}
We generated networks from the stochastic block model with $n=1000,2000$ nodes and 2 equal-sized communities such that
the block probability matrix,
$$\Omega\propto \begin{pmatrix}
    4 & 1\\
    1 & 4
\end{pmatrix}.$$
We varied the network density over $\delta=0.1,0.2$ and $0.3$. To generate networks under the alternative, we embedded a clique of size $k$ on a network from the true model. we explored two ways of selecting the anomalous clique nodes,
\begin{enumerate}
    \item selecting half of the nodes from the first community and the remaining half from the second community
    \item selecting all the nodes from the same community
\end{enumerate}
\begin{table}[ht]
\footnotesize
\begin{tabular}{|c|c|rrcc|rrcc|}
\hline
   & & \multicolumn{4}{c|}{$n=1000$} & \multicolumn{4}{c|}{$n=2000$}\\
  \hline
   $\delta$ & $c$ & $\umod{B_S}$ &
  Power & Localization & False alarm & $\umod{B_S}$ & Power & Localization & False alarm\\
  \hline
  0.1 & 1 & 18.96 & 0 & - & - & 28.86 & 0 & - & - \\ 
  0.1 & 2 & 39.65 & 100 & 100.00 & 0.00 & 58.55 & 100 & 100.00 & 0.00 \\ 
  0.1 & 3 & 60.35 & 100 & 100.00 & 0.00 & 88.25 & 100 & 100.00 & 0.00 \\ 
  0.1 & 4 & 81.05 & 0 & - & - & 117.95 & 100 & 100.00 & 0.00 \\ 
  \hline
  0.2 & 1 & 24.91 & 0 & - & - & 36.11 & 0 & - & - \\ 
  0.2 & 2 & 50.50 & 59 & 100.00 & 0.00 & 73.70 & 100 & 100.00 & 0.00 \\ 
  0.2 & 3 & 76.89 & 100 & 100.00 & 0.00 & 111.30 & 100 & 100.00 & 0.00 \\ 
  0.2 & 4 & 102.48 & 100 & 100.00 & 0.00 & 148.89 & 100 & 100.00 & 0.00 \\ 
  \hline
  0.3 & 1 & 26.76 & 0 & - & - & 39.36 & 0 & - & - \\ 
  0.3 & 2 & 54.74 & 0 & - & - & 79.95 & 0 & - & - \\ 
  0.3 & 3 & 82.72 & 100 & 100.00 & 0.00 & 119.84 & 100 & 100.00 & 0.00 \\ 
  0.3 & 4 & 110.00 & 100 & 100.00 & 0.00 & 160.43 & 100 & 100.00 & 0.00 \\ 
   \hline
\end{tabular}
\captionof{table}{\small Results for SBM when nodes are selected from both communities}
\label{simsbmh1mix}
\vspace{\baselineskip}
\begin{tabular}{|c|c|rrcc|rrcc|}
\hline
   & & \multicolumn{4}{c|}{$n=1000$} & \multicolumn{4}{c|}{$n=2000$}\\
  \hline
   $\delta$ & $c$ & $\umod{B_S}$ &
  Power & Localization & False alarm & $\umod{B_S}$ & Power & Localization & False alarm\\
  \hline
  0.1 & 1 & 17.63 & 0 & - & - & 26.88 & 0 & - & - \\ 
  0.1 & 2 & 36.95 & 0 & - & - & 54.59 & 85 & 100.00 & 0.00 \\ 
  0.1 & 3 & 56.26 & 0 & - & - & 82.31 & 100 & 100.00 & 0.00 \\ 
  0.1 & 4 & 75.58 & 0 & - & - & 110.02 & 40 & 75.13 & 0.01 \\ 
  0.2 & 1 & 21.06 & 0 & - & - & 30.59 & 0 & - & - \\ 
  0.2 & 2 & 42.81 & 0 & - & - & 62.54 & 0 & - & -\\ 
  0.2 & 3 & 65.23 & 0 & - & - & 94.48 & 13 & 100.00 & 0.00 \\ 
  0.2 & 4 & 86.97 & 0 & - & - & 126.43 & 51 & 92.16 & 0.00 \\ 
  0.3 & 1 & 19.73 & 0 & - & - & 29.10 & 0 & - & - \\ 
  0.3 & 2 & 40.50 & 0 & - & - & 59.24 & 0 & - & - \\ 
  0.3 & 3 & 61.27 & 0 & - & - & 88.85 & 0 & - & - \\ 
  0.3 & 4 & 81.52 & 0 & - & - & 118.99 & 0 & - & - \\ 
   \hline
\end{tabular}
\captionof{table}{\small Results for SBM when nodes are selected from only one community}
\label{simsbmh1pure}
\end{table}
We chose $k=c\sqrt{2\delta(P)\sqrt{\log n}}$ so that the contribution of the anomalous clique in the egonet degree, approximately $ k^2/2$, is equal to $c^2$ times the detection threshold. 
Since the SBM is a special case of the RDPG model, we applied the egonet method proposed for the RDPG case. The false alarm rate was 0\% for all the replications, that is, the non-anomalous networks were never rejected by the detection test.
So, we only report the results of the egonet method for the anomalous networks. The results are presented in Tables \ref{simsbmh1mix} and \ref{simsbmh1pure}.

In Table \ref{simsbmh1mix}, we have networks from the SBM with an anomalous clique formed by $k$ nodes of which, about half were randomly selected from one community and the remaining from the other community. 
For both $n=1000$ and 2000, we find that when $\umod{B_S}$ is small ($c=1$), the detection rate is 0\%. 
As $\umod{B_S}$ gets larger, the detection rate improves, and for $c=4$, we have a 100\% detection rate for all the choices of $n,\delta$ except one. For $n=1000$ and $\delta=0.1$, we find that the detection rate is again 0\% for $c=4$, despite $\umod{B_S}$ being large. 
This case is actually an example where the anomaly of the clique has affected the estimation of $P$ so much that the detection test is not able to detect the anomalous clique anymore. 
Whenever a clique is detected, the node detection rate is almost 100\% and the node false alarm rate is almost 0\%, that is, the egonet ranking algorithm almost always recovers the clique exactly.

In Table \ref{simsbmh1pure}, we have networks from the SBM with an anomalous clique formed by $k$ nodes all of which were randomly selected from one community. 
We find that for $n=1000$, the detection test always fails to detect the anomalous clique. For $n=2000$, the detection test is able to detect the anomalous clique for some of the cases, when $\delta$ is small. 
Comparing with Table \ref{simsbmh1mix}, we see that overall, the performance of the egonet method dropped significantly in this one-community scenario, although cliques of the same size were embedded for both scenarios. 
We first note that for the same clique size, $\umod{B_S}$ is slightly smaller in Table \ref{simsbmh1pure}, but that does not seem to justify this massive drop in performance. To explain this, let us recall our discussion after Theorem \ref{unknownph0rdpg}. 
To estimate $P$ accurately, we need the effect of the anomalous clique to mostly appear in the space of the non-top eigenvectors of $Q$. 
In this case, since the clique nodes belong from the same community, the clique's effect is `sucked up' by the top eigenvectors of $Q$, and this is the reason why the egonet method fails in this scenario.

\subsection{RDPG model}
\begin{table}[ht]
\centering
\fontsize{10pt}{9pt}\selectfont
\begin{tabular}{|c|c|rcc|rcc|}
\hline
   \multicolumn{8}{|c|}{$n=2000,\delta=0.1$}\\
   \hline
   & & \multicolumn{3}{|c|}{Naive estimator} & \multicolumn{3}{|c|}{Node-based estimator}\\
  \hline
  $k$ &  $\frac{\umod{B_S}}{\lambda_d(P)}$ &  Power & Localization & False alarm &  Power & Localization & False alarm \\ 
  \hline 
  35 & 0.63 & 0 & - & - & 4 & 100.00 & 0.00 \\ 
  40 & 0.73 & 0 & - & - & 100 & 100.00 & 0.00 \\ 
  45 & 0.82 & 2 & 100.00 & 0.00 & 100 & 100.00 & 0.00 \\ 
  50 & 0.92 & 84 & 100.00 & 0.00 & 100 & 100.00 & 0.00 \\ 
  55 & 0.99 & 92 & 91.60 & 0.00 & 100 & 100.00 & 0.00 \\ 
  60 & 1.10 & 4 & 5.00 & 0.00 & 100 & 100.00 & 0.00 \\ 
  65 & 1.19 & 1 & 3.08 & 0.00 & 100 & 100.00 & 0.00 \\ 
  70 & 1.28 & 0 & - & - & 100 & 100.00 & 0.00 \\ 
  75 & 1.37 & 0 & - & - & 100 & 100.00 & 0.00 \\
   \hline
\end{tabular}
\captionof{table}{\small Results for the RDPG model with network density $\delta=0.1$.}
\label{simrdpgh1delta1}
\vspace{\baselineskip}
\begin{tabular}{|c|c|rcc|rcc|}
\hline
  \multicolumn{8}{|c|}{$n=2000,\delta=0.2$}\\
   \hline
   & & \multicolumn{3}{|c|}{Naive estimator} & \multicolumn{3}{|c|}{Node-based estimator}\\
  \hline
  $k$ &  $\frac{\umod{B_S}}{\lambda_d(P)}$ &  Power & Localization & False alarm &  Power & Localization & False alarm \\ 
  \hline 
  60 & 0.49 & 0 & - & - & 100 & 100.00 & 0.00 \\ 
   70 & 0.57 & 0 & - & - & 100 & 100.00 & 0.00 \\ 
   80  & 0.65 & 73 & 100.00 & 0.00 & 100 & 100.00 & 0.00 \\ 
   90  & 0.73 & 100 & 100.00 & 0.00 & 100 & 100.00 & 0.00 \\ 
  100  & 0.81 & 100 & 100.00 & 0.00 & 100 & 100.00 & 0.00 \\ 
  110  & 0.90 & 100 & 100.00 & 0.00 & 100 & 99.64 & 0.00 \\ 
  120  & 0.98 & 100 & 97.04 & 0.00 & 100 & 100.00 & 0.00 \\ 
  130  & 1.06 & 100 & 48.72 & 0.00 & 100 & 100.00 & 0.00 \\ 
  140 & 1.14 & 93 & 3.41 & 2.05 & 100 & 100.00 & 0.00 \\ 
  150  & 1.23 & 92 & 1.82 & 1.11 & 100 & 100.00 & 0.00 \\ 
  160  & 1.31 & 92 & 0.26 & 0.13 & 100 & 100.00 & 0.00 \\ 
  170  & 1.39 & 84 & 2.18 & 2.13 & 100 & 100.00 & 0.00 \\ 
  180 & 1.47 & 89 & 3.16 & 3.11 & 100 & 100.00 & 0.00 \\ 
  190  & 1.56 & 76 & 0.09 & 0.12 & 100 & 100.00 & 0.00  \\ 
  200  & 1.64 & 75 & 1.12 & 1.13 & 100 & 100.00 & 0.00  \\ 
   \hline
\end{tabular}
\captionof{table}{\small Results for the RDPG model with network density $\delta=0.2$.}
\label{simrdpgh1delta2}
\end{table}

We generated networks from a 3-dimensional RDPG model with $n=2000$ nodes, where the latent positions $\{X_i\}$ were sampled from the Dirichlet distribution with parameter $(1,1,1)^\top $. 
We scaled the probability matrix so that the average network density is equal to $\delta$ and varied $\delta$ over $(0.1,0.2)$. 
To generate networks under the alternative, we randomly selected a set of $k$ nodes from the network and embedded a clique on the subgraph spanned by the selected nodes.
As before, the false alarm rate was 0\% for all the replications, that is, the non-anomalous networks were never rejected.
For the simulation results under the alternative, we estimated $P$ both using the naive estimator and using the node-based estimator.
We report the results in Tables \ref{simrdpgh1delta1} and \ref{simrdpgh1delta2}.

For every choice of $(n,\delta)$, we started with a sequence of clique sizes $k$'s at regular intervals to observe when the egonet method is able to find the anomalous clique.
The node-based estimator performed well in all scenarios: the anomalous clique was successfully detected and localized for nearly all values of $k$.
In case of the naive estimator, We discovered a common pattern: there appears to be a range $(B_0,B_1)$ such that when $\umod{B_S}$ is less than $B_0$, the detection test is unable to detect the anomalous clique; 
between $B_0$ and $B_1$, the egonet method works almost perfectly, that is, the detection test finds the anomalous clique and the egonet ranking algorithm recovers the anomalous clique perfectly; 
above $B_1$, the performance of both detection and localization starts to drop, eventually leading to 0\% detection. Both of these bounds also seem to vary with the network density $\delta$. 
This observation is consistent with our theoretical findings for RDPG, which shows that we require both a lower and upper bound on $\umod{B_S}$ for the egonet method to succeed. 

\section{Case studies with real-world network data}
\label{sec:data}
We implemented the egonet method on two well-known networks. In real-world applications, an important consideration is which null model to use for fitting the data and computing the egonet statistics. A practitioner can choose a null model based on heuristic considerations about the structure of the network, or take the cautious route of carrying out multiple versions of the test using different null models. In the latter case, the inference can vary from one null model to another. 
Another option is to use the most general model (e.g., among the models used in this paper, the RDPG model is the most general model and all other models are its special cases), although this can potentially lead to overfitting the data. 
In the following well-studied examples, we used null models from recent studies where a certain model was shown to work well for a certain network.

Our first dataset is the British MP's Twitter dataset which is a network of twitter interactions (follows, mentions, and retweets) between 419 Members of Parliament (MP) in the United Kingdom belonging to five different political parties \cite{greene2013producing}. For our analysis, we considered the network of retweets. We picked a subset of this network consisting of 360 MPs belonging to the Conservative Party and the Labour Party, the two parties with the highest number of members in the dataset. We extracted the largest connected component of this subnetwork, which had 329 nodes and 5720 edges. Following \citet{senguptapabm}, we used the 2-community popularity adjusted blockmodel (RDPG with dimension 4) as the null model. The null hypothesis was not rejected, i.e., no anomalous clique was detected.

Our second dataset is the political blogs network \cite{adamic2005political}, representing 16,714 hyperlinks between 1222 political weblogs before the 2004 US Presidential election. This network has been studied by many papers \cite{karrer2011stochastic,zhao2012consistency,amini2013, Bickel2016estimating} in the statistics literature pertaining to community structure. A prominent feature of this network is its strong community structure with two communities representing liberals.
and conservatives. Following \citet{karrer2011stochastic}, we used the 2-community degree corrected block model as the
null model. Here as well, the null hypothesis was not rejected, i.e., no anomalous clique was detected.

While no anomalous clique was detected for both datasets, we wanted to check if there was any large clique at all, even if not anomalous, in these datasets. 
For that, we applied a largest clique finding algorithm from the \texttt{igraph} package in R on the two networks. 
We found that the British MP network contained a clique of 19 nodes and the political blogs network also contained a clique of 20 nodes. 
Note that, the two networks have an average edge probability of 0.106 and 0.022 respectively. 
From the theory of clique numbers, we know that the largest clique size in an ER$(n,p)$ network lies on two consecutive integers around $2\log_{\frac{1}{p}} n$ almost surely \cite{matula1976largest}. 
This quantity evaluated for the two networks are approximately equal to 5.2 and 3.7 respectively. 
So, if these networks were modeled as homogeneous, the corresponding largest cliques would have been identified as anomalous. 
Looking into the true communities of the nodes forming the largest cliques, we found that in both cases, the nodes belonged to the same community.
Therefore, we can say that the egonet method correctly did not identify the two large cliques to be anomalous as their presence is explained by the underlying block structure in the networks.

\begin{table}[ht]
\centering
\footnotesize
\begin{tabular}{|c|rcc|}
  \hline
  $k$ &  Power & Localization & False alarm \\ 
  \hline 
40  & 83 & 98.32 & 0.00\\
50  & 100 & 96.86 & 0.00\\ 
60  & 100 & 84.68 & 0.00 \\ 
\hline
\end{tabular}
\captionof{table}{\small Results for the political blogs network with a planted clique.}
\label{polblogs}
\end{table}

What if the networks actually had an anomalous clique? Would the egonet method be able to find the clique in that case? 
To check that, we planted cliques of three different sizes, 40, 50, and 60 on the political blogs network in such a way that half of the clique nodes are from the first true community and the remaining are from the second true community. 
Then, we applied the egonet method to the network. We repeated this procedure 100 times, we have the results in Table \ref{polblogs}. 
We find that when $k=50$ and 60, the egonet method was successful to detect the presence of the anomalous clique 100\% of the time, and when $k=40$, the anomalous clique was detected 83\% of the time. 
When a clique was detected, our method was able to identify more than 95\% of the clique nodes for $k=40$ and 50, and about 85\% of the clique nodes when $k=60$. 
In all the cases, almost no non-clique nodes were wrongly flagged.
From this analysis, we can conclude that our method would be able to find if there was a clique anomalous with respect to the DCSBM model in this manner.

\section{Discussion}
\label{sec:discussion}
The goal of anomaly detection is to detect whether there is an anomaly, and if any anomaly is detected, to identify the part of the system that constitutes the anomaly.
In this paper, we have formulated a statistical inferential framework using egonets to address both these questions simultaneously in the context of clique detection in static networks.
The egonet method is naturally amenable to parallel computing, making it computationally scalable.
The flexible nature of the method makes it amenable to a wide variety of network models.
We the theoretical properties of the method under the ER model, the Chung-Lu model, and the RDPG model.
In our simulation study, we applied the method to synthetic networks generated from a variety of heterogeneous models, and the results support our theoretical findings.
 
In this paper, we focused on exact cliques where all pairs of anomalous nodes are connected. 
A closely related form of anomaly is a soft clique or an unusually dense subgraph, where a group of anomalous nodes are, maybe not fully, but densely connected to each other compared to other parts of the network.
As a next step, we plan to study the dense subgraph problem theoretically and drive the conditions under which the egonet method can detect and recover such dense subgraphs.
Also, there can be several other relevant forms of anomaly in a network, e.g., star subgraphs, long chains, etc.
The overarching idea of the egonet method is that if we fit a model for the entire network, and if there is some anomalous subgraph in the network, then trying to fit a class of subgraphs using the fitted network model will make the unusual substructures pop out, as they won't fit the empirical network model. 
It would be of great interest to try and extend this basic principle to other forms of anomaly.
The problem of anomaly detection in networks is an important and relevant problem that has not received much attention from the statistics community.
We envisage this work as one of the first steps towards addressing this methodological gap.

\clearpage

\section{Appendix}
We first introduce some standard inequalities that we are going to use to prove our results.

\begin{lemma}
    (Bernstein's inequality) Suppose $X_1, X_2,\ldots, X_n$ are mean zero, independent random variables, $|X_i|\leq M$. Then for any $t>0$,
\begin{equation*}
    \PP\left(\left|\sum_{i} X_i\right|\geq t\right)\leq 2\exp\left[-\frac{t^2/2}{\sum\limits_{i} \EE(X_i^2)+Mt/3}\right].
\end{equation*}
\end{lemma}

\begin{lemma}
    (McDiarmid's inequality) Suppose $X_1, X_2,\ldots, X_n$ are independent random variables, where $X_i\in\mathcal{X}_i$ for all $i$. Let $f:\mathcal{X}_1\times \mathcal{X}_2\times\ldots \times\mathcal{X}_n\mapsto \mathbb{R}$ be a function with $(c_1,c_2,\ldots,c_n)$-bounded differences property. Then for any $t>0$,
\begin{equation*}
    \PP\left(\left|f(X_1,X_2,\ldots,X_n)-\EE(f(X_1,X_2,\ldots,X_n))\right|\geq t\right)\leq 2\exp\left[-\frac{2t^2}{\sum\limits_{i} c_i^2}\right].
\end{equation*}
\label{lem:mcdiarmid}
\end{lemma}

\begin{lemma}
    (Inequality reversal lemma) Let $X$ be a random variable and $a,b>0,c,d\geq 0$ be constants such that, for all $t>0$, $\PP(|X|\geq t)\leq a\exp\left[-\dfrac{bt^2}{c+dt}\right]$. Then with probability at least $1-\delta$,
\begin{equation*}
    |X|\leq\sqrt{\frac{c}{b}\log\frac{a}{\delta}} + \frac{d}{b}\log\frac{a}{\delta}.
\end{equation*}
\label{lem:irl}
\end{lemma}

\begin{lemma}
    (Hanson-Wright inequality) \cite{rudelson2013hanson} Let $X = (X_1,\ldots,X_n)\in\mathbb{R}^n$ be a random vector with independent components $X_i$ which satisfy $\EE(X_i)=0$ and $\umod{X_i}_{\psi_2}\leq K$. Let A be an $n\times n$ matrix. Then, for any $t>0$,
    \begin{equation*}
        \PP(|X^\top AX - \EE(X^\top AX)|\geq t) \leq 2\exp\left[-c\min\left(\frac{t^2}{K^4\umod{A}_F^2}, \frac{t}{K^2\umod{A}} \right)\right].
    \end{equation*}
\end{lemma}

Here, for any random variable $\xi$, its subgaussian norm $\umod{.}_{\psi_2}$ is defined as
$$\umod{\xi}_{\psi_2}=\sup_{p\geq 1}p^{-1/2}(\EE(|\xi^p|))^{\frac{1}{p}}.$$

\subsection{Proof of Theorem \ref{knownp_h0_alt}}

\subsubsection*{Under $H_0$}
\begin{equation*}
    T_i= \underset{j<k}{\sum\sum} A_{ij}A_{ik}(A_{jk}-P_{jk})=S_{1i},\ \text{say.}
\end{equation*}
Given $a_i$, $S_{1i}$ is a sum of ${D_i \choose 2}$ independent centered Bernoulli random variables. Using Bernstein's inequality, we have
\begin{equation*}
    \PP_{H_0}(|S_{1i}| \geq t\mid a_i)\leq 2\exp\left[-\frac{t^2/2}{\sigma_{i,A}^2 + t/3}\right],
\end{equation*}
where
\begin{equation*}
    \sigma_{i,A}^2 \coloneqq \underset{j<k}{\sum\sum}\,A_{ij}A_{ik}V_{jk}.
\end{equation*}
Applying Lemma \ref{lem:irl}, we obtain that for any $c>1$,
\begin{equation}
    \PP_{H_0}\left(T_i \leq \sqrt{2c\log n}\,\sigma_{i,A} + \frac{2c}{3}\log n\right)\geq 1-\frac{2}{n^c}.
    \label{s1i_init_alt}
\end{equation}

\subsubsection*{Under $H_1$}
Consider $i\in S$.
\begin{equation*}
    \begin{split}
        T_i=&\, \underset{j<k}{\sum\sum} A_{ij}A_{ik}(A_{jk}-P_{jk})= \underset{j<k}{\sum\sum} A_{ij}A_{ik}(A_{jk}-Q_{jk}) + \underset{j<k:\,j,k\neq i}{\sum\sum} (Q_{jk}-P_{jk})\\[.2cm]
    =&\, S_{1i} + \frac{1}{2}\umod{B_S^{-i}}_{1,1},\ \text{say.}
    \end{split}
\end{equation*}

\textbf{Bounding $S_{1i}$:} Define a matrix $\widetilde{V}\in[0,1]^{n\times n}$ such that 
$$\widetilde{V}_{jk}=Q_{jk}(1-Q_{jk}),\ 1\leq j<k\leq n.$$

Given $a_i$, $S_{1i}$ is a sum of ${D_i \choose 2}$ independent centered Bernoulli random variables. Using Bernstein's inequality and Lemma \ref{lem:irl}, we obtain 
\begin{equation}
    \PP_{H_1}\left(|S_{1i}|\leq \sqrt{2c\log n}\,\widetilde\sigma_{i,A} + \frac{2c}{3}\log n\right)\geq 1-\frac{2}{n^c},
    \label{s1i}
\end{equation}
where
\begin{equation*}
    \widetilde\sigma_{i,A}^2 \coloneqq \underset{j<k}{\sum\sum}\,A_{ij}A_{ik}\widetilde{V}_{jk}.
\end{equation*}

\textbf{Concentration of $\widetilde\sigma_{i,A}^2$:}
Define
\begin{equation*}
    \widetilde\sigma_{i,Q}^2 \coloneqq \underset{j<k}{\sum\sum}\,Q_{ij}Q_{ik}\widetilde{V}_{jk}.
\end{equation*}

We derive
\begin{equation}
    \begin{split}
    &\,2\,(\widetilde\sigma_{i,A}^2 - \widetilde\sigma_{i,Q}^2) \,=\,\underset{j,k}{\sum\sum} (A_{ij}A_{ik}-Q_{ij}Q_{ik})\widetilde{V}_{jk} \\[.2cm]
    =&\, \underset{j,k}{\sum\sum} (A_{ij}-Q_{ij})(A_{ik}-Q_{ik})\widetilde{V}_{jk} + 2\,\sum_j Q_{ij}\sum_{k}(A_{ik}-Q_{ik})\widetilde{V}_{jk}.
\end{split}
\label{sia_sip}
\end{equation}

Let $y_i\coloneqq a_i-q_i$, where $a_i$ is the $i^{th}$ column of $A$, and $q_i$ is the $i^{th}$ column of $Q$. 
Then the first quantity above is $y_i^\top \widetilde{V} y_i$. 
Using the Hanson-Wright inequality, we obtain 
$$\PP_{H_1}(|y_i^\top \widetilde{V}y_i|\geq t) \leq 2\exp\left[-c'\min\left(\frac{t^2}{\|\widetilde{V}\|_F^2}, \frac{t}{\|\widetilde{V}\|} \right)\right].$$

Note that $\EE_{H_1}(y_i^\top \widetilde{V}y_i)=0$ since $\widetilde{V}_{jj}=0$ for all $j$. Equivalently, with probability at least $1-\dfrac{2}{n^c}$,
\begin{equation}
    |y_i^\top \widetilde{V} y_i|\leq C \max\{\|\widetilde{V}\|_F\sqrt{\log n}, \|\widetilde{V}\|\log n\}) \leq  C \|\widetilde{V}\|_F \log n.
    \label{sia_sip_pt1}
\end{equation}

For any $j\in [n]$, $\sum_{k}(A_{ik}-Q_{ik})Q_{jk}(1-Q_{jk})$ is a weighted sum of $(n-1)$ independent centered Bernoulli random variables. 
Using Bernstein's inequality and Lemma \ref{lem:irl}, we obtain that with probability at least $1-\dfrac{2}{n^{c+1}}$,
\begin{align*}
    \left|\sum_{k}(A_{ik}-Q_{ik})Q_{jk}(1-Q_{jk})\right|\leq& \sqrt{2c \sum_{k}Q_{ik}(1-Q_{ik})Q_{jk}^2(1-Q_{jk})^2\log n} + \frac{2c}{3}\log n\\[.2cm]
    \leq& \sqrt{2c \sum_{k}Q_{jk}^2(1-Q_{jk})^2\log n} + \frac{2c}{3}\log n\\[.2cm]
    =& \sqrt{2c \log n} \umod{\widetilde{v}_j} + \frac{2c}{3}\log n,
\end{align*}
where $\widetilde{v}_j$ is defined as the $j^{th}$ column of $\widetilde{V}$.

Hence, with probability at least $1-\dfrac{2}{n^c}$,
\begin{equation}
    \label{sia_sip_pt2}
    \begin{split}
        \left|\sum_j Q_{ij}\sum_{k}(A_{ik}-P_{ik})\widetilde{V}_{jk}\right|
    \leq & \sum_j Q_{ij}\left(\sqrt{2c\log n}\,\|\widetilde{v}_j\| + \frac{2c}{3}\log n\right)\\[.2cm]
    \leq & \sqrt{2c\log n}\sqrt{\sum_j Q_{ij}^2 \sum_j\|\widetilde{v}_j\|^2} + \frac{2c}{3}\log n \sum_j Q_{ij}\\[.2cm]
    \leq&\, C\sqrt{\log n}\left(\|q_i\|\|\widetilde{V}\|_F + d_i(Q)\sqrt{\log n}\right).
    \end{split}
\end{equation}

Combining \eqref{sia_sip}, \eqref{sia_sip_pt1} and \eqref{sia_sip_pt2}, we obtain that with probability at least $1-\dfrac{4}{n^c}$,
\begin{equation*}
    \widetilde\sigma_{i,A}^2 \leq \widetilde\sigma_{i,Q}^2 + C \|\widetilde{V}\|_F \log n + C\sqrt{\log n}\left(\|q_i\| \|\widetilde{V}\|_F + d_i(Q)\sqrt{\log n}\right).
\end{equation*}
Now, $\|\widetilde{V}\|_F\leq \|V\|_F$, since $\widetilde{V}_{jk}=V_{jk}$ if either $j\in S^c$ or $k\in S^c$, and $\widetilde{V}_{jk}=0$ if both $j,k\in S$. 
We also derive
\begin{align*}
    2\,\widetilde\sigma_{i,Q}^2 \,\coloneqq\, \underset{j,k}{\sum\sum}\,Q_{ij}Q_{ik}\widetilde{V}_{jk}
    \,\leq\, \sqrt{\underset{j,k}{\sum\sum}\,Q_{ij}^2Q_{ik}^2}\,\|\widetilde{V}\|_F \,\leq\, \|q_i\|^2\|\widetilde{V}\|_F.
\end{align*}
Hence, 
with probability at least $1-\dfrac{4}{n^c}$,
\begin{equation}
    \widetilde\sigma_{i,A}^2 \leq C\left(\|V\|_F \max\{\|q_i\|^2,\log n\} + d_i(Q)\log n\right).
    \label{sia_bound_2}
\end{equation}

\textbf{Lower bound on $T_i$:}
Note that $\|q_i\|\leq \|Q\|_{2\rightarrow\infty}$ and $d_i(Q)\leq \delta(Q)$ by definition. Combining \eqref{s1i} and \eqref{sia_bound_2}, we obtain that with probability at least $1-O(n^{-c})$,
\begin{equation}
    |S_{1i}| \leq C\left((\|V\|_F\log n)^{\frac{1}{2}} \max\{\|Q\|_{2\rightarrow\infty},\sqrt{\log n}\} + \sqrt{\delta(Q)}\log n\right).
    \label{s1i_bound_iinS}
\end{equation}
Therefore, with probability at least $1-O(n^{-c})$,
\begin{equation}
    T_i \,\geq\, \frac{1}{2}\umod{B_S^{-i}}_{1,1} - C\left((\|V\|_F\log n)^{\frac{1}{2}} \max\{\|Q\|_{2\rightarrow\infty},\sqrt{\log n}\} + \sqrt{\delta(Q)}\log n\right).
    \label{ti_bound}
\end{equation}

\textbf{Proving Eq, \eqref{knownp_power}:}
Note that $\sigma_{i,A}\coloneqq \sum\sum_{j,k}A_{ij}A_{ik}V_{jk}$ can be bounded the same way as $\widetilde\sigma_{i,A}\coloneqq \sum\sum_{j,k}A_{ij}A_{ik}\widetilde{V}_{jk}$, so that with probability at least $1-O(n^{-c})$,
\begin{align*}
    &\,\sqrt{2c\log n}\,\sigma_{i,A} + \frac{2c}{3}\log n \leq C \left((\|V\|_F\log n)^{\frac{1}{2}} \max\{ \|Q\|_{2\rightarrow\infty}, \sqrt{\log n}\} + \sqrt{\delta(Q)}\log n\right).
\end{align*}
Combining this with Eq. \eqref{ti_bound} and invoking Eq. \eqref{knownp_detect_cond}, we conclude that for any $i\in S$,
\begin{equation*}
    \PP_{H_1}\left(T_i > \sqrt{2c\log n}\,\sigma_{i,A} + \frac{2c}{3}\log n\right)\geq 1-O(n^{-c}).
\end{equation*} 

\subsection{Proof of Theorem \ref{knownp_ranking_alt}}
Fix any $i\in S^c$.
\begin{equation*}
    \begin{split}
        T_i=&\, \underset{j<k}{\sum\sum} A_{ij}A_{ik}(A_{jk}-P_{jk})= \underset{j<k}{\sum\sum} A_{ij}A_{ik}(A_{jk}-Q_{jk}) + \underset{j<k}{\sum\sum} A_{ij}A_{ik}(Q_{jk}-P_{jk})\\[.2cm]
        =&\,\underset{j<k}{\sum\sum} A_{ij}A_{ik}(A_{jk}-Q_{jk}) + \underset{j<k}{\sum\sum} (A_{ij}A_{ik} - P_{ij}P_{ik})(Q_{jk}-P_{jk})+ \underset{j<k}{\sum\sum} P_{ij}P_{ik}(Q_{jk}-P_{jk})\\[.2cm]
        =&\, S_{1i} + S_{2i} + \underset{j<k}{\sum\sum} P_{ij}P_{ik}(Q_{jk}-P_{jk}),\ \text{say}.
    \end{split}
\end{equation*}

\textbf{Bounding $S_{1i}$:} We can bound $S_{1i}$ similarly as in the proof of Theorem \ref{knownp_h0_alt}, so that 
\begin{equation*}
    \PP_{H_1}\left(|S_{1i}|\leq \sqrt{2c\log n}\,\widetilde\sigma_{i,A} + \frac{2c}{3}\log n\right)\geq 1-\frac{2}{n^c}.
\end{equation*}

Recall that we derived the following concentration bound for $\widetilde\sigma_{i,A}$ (see Eq. \eqref{sia_bound_2}):
\begin{equation*}
    \PP_{H_1}\left(\widetilde\sigma_{i,A}^2 \leq C\left(\|V\|_F \max\{\|q_i\|^2,\log n\} + d_i(Q)\log n\right)\right) \geq 1-\dfrac{4}{n^c}.
\end{equation*}

Note that for $i\in S^c$, $\|q_i\|=\|p_i\|\leq \|P\|_{2\rightarrow\infty}$ and $d_i(Q)=d_i(P)\leq\delta(P)$. Therefore, with probability at least $1-O(n^{-c})$,
\begin{equation}
    \label{s1i_bound_iinSc}
    |S_{1i}| \leq C\left((\|V\|_F\log n)^{\frac{1}{2}} \max\{\|P\|_{2\rightarrow\infty},\sqrt{\log n}\} + \sqrt{\delta(P)}\log n\right).
\end{equation}


\textbf{Bounding $S_{2i}$:}
\begin{equation}
    \begin{split}
        &\underset{j,k}{\sum\sum} (A_{ij}A_{ik}-P_{ij}P_{ik})(Q_{jk}-P_{jk})\\
        =&\, \underset{j,k}{\sum\sum} (A_{ij}-P_{ij})(A_{ik}-P_{ik})(Q_{jk}-P_{jk}) + 2\,\sum_j P_{ij}\sum_{k}(A_{ik}-P_{ik})(Q_{jk}-P_{jk}).
    \end{split}
\label{s2i}
\end{equation}

Define $y_i\coloneqq a_i-p_i$, so that the first quantity above is $y_i^\top B_S\,y_i$. Using the Hanson-Wright inequality, we obtain 
$$\PP(|y_i^\top B_S\,y_i|\geq t) \,\leq\, 2\exp\left[-c'\min\left(\frac{t^2}{\umod{B_S}_F^2}, \frac{t}{\umod{B_S}} \right)\right].$$
Note that $\EE_{H_1}(y_i^\top B_S\,y_i)=0$ since $B_S$ has diagonal entries all equal to zero. Equivalently, with probability at least 
$1- 2/n^c$,
\begin{equation}
    |y_i^\top B_S\,y_i| \,\leq\, C\,\max\{\umod{B_S}_F\sqrt{\log n}, \umod{B_S}\log n\} \,\leq\, C\umod{B_S}_F\log n.
    \label{s2i_pt1}
\end{equation}

For any $j\in [n]$, $\sum_{k}(A_{ik}-P_{ik})(Q_{jk}-P_{jk})$ is a weighted sum of $(n-1)$ independent centered Bernoulli random variables. Using Bernstein's inequality and Lemma \ref{lem:irl}, we obtain that with probability at least $1- 2/n^{(c+1)}$,
\begin{align*}
    \left|\sum_{k}(A_{ik}-P_{ik})(Q_{jk}-P_{jk})\right|\leq&\, \sqrt{2c \sum_{k}P_{ik}(1-P_{ik})(Q_{jk}-P_{jk})^2\log n} + \frac{2c}{3}\log n\\[.2cm]
    \leq&\, \sqrt{2c \sum_{k}(Q_{jk}-P_{jk})^2\log n} + \frac{2c}{3}\log n\\[.2cm]
    =&\, \sqrt{2c \log n} \umod{b_j} + \frac{2c}{3}\log n,
\end{align*}
where $b_j$ is defined as the $j^{th}$ column of $B_S$.

Hence, with probability at least $1-2/n^c$,
\begin{equation}
    \label{s2i_pt2}
    \begin{split}
    &\,\sum_j P_{ij}\sum_{k}(A_{ik}-P_{ik})(Q_{jk}-P_{jk})
    \,=\,\sum_j P_{ij}\,\mathbb{I}(j\in S)\,\sum_{k}(A_{ik}-P_{ik})(Q_{jk}-P_{jk}) \\[.2cm]
    \leq & \sum_j P_{ij}\,\mathbb{I}(j\in S)\left(\sqrt{2c\log n}\,\|b_j\| + \frac{2c}{3}\log n\right)\\[.2cm]
    \leq & \sqrt{2c\log n}\sqrt{\sum_j P_{ij}^2 \,\mathbb{I}(j\in S) \sum_j\|b_j\|^2} + \frac{2c}{3}\log n \sum_j P_{ij}\,\mathbb{I}(j\in S)\\[.2cm]
    \leq&\, C\sqrt{\log n}\left(\|P_S\|_{2\rightarrow\infty}\|B_S\|_F + \delta(P_S)\sqrt{\log n}\right).
    \end{split}
\end{equation}

Combining \eqref{s2i}, \eqref{s2i_pt1} and \eqref{s2i_pt2}, we obtain that with probability at least $1-\dfrac{4}{n^c}$,
\begin{equation}
    \begin{split}
        |S_{2i}| \leq&\, C \|B_S\|_F \log n + C\sqrt{\log n}\left(\|P_S\|_{2\rightarrow\infty}\|B_S\|_F + \delta(P_S)\sqrt{\log n}\right)\\[.2cm]
        \leq&\, C \left(\|B_S\|_F \sqrt{\log n}\max\{\|P_S\|_{2\rightarrow\infty},\sqrt{\log n}\} + \delta(P_S)\log n\right).
    \end{split}
    \label{s2i_bound}
\end{equation}

\textbf{Proving Eq. \eqref{knownp_loc}:}
Consider nodes $i\in S, i'\in S^c$.
\begin{align*}
    &\, T_{i}-T_{i'}=S_{1i} + \underset{j<k:\,j,k\neq i}{\sum\sum} (Q_{jk}-P_{jk}) -S_{1i'} - S_{2i'} -\underset{j<k:\,j,k\neq i}{\sum\sum} P_{ij}P_{ik}(Q_{jk}-P_{jk})\\[.2cm]
    =&\, \frac{1}{2}\umod{B_S^{-i}}_{1,1} -\frac{1}{2}p_i^\top B_S p_i - C\left((\|V\|_F\log n)^{\frac{1}{2}} \max\{\|Q\|_{2\rightarrow\infty},\sqrt{\log n}\} + \sqrt{\delta(Q)}\log n\right) \\[.2cm]
    -&\, C \left(\|B_S\|_F \sqrt{\log n}\max\{\|P_S\|_{2\rightarrow\infty},\sqrt{\log n}\} + \delta(P_S)\log n\right),
\end{align*}
with probability at least $1-O(n^{-c})$, invoking \eqref{s1i_bound_iinS},\eqref{s1i_bound_iinSc} and \eqref{s2i_bound}. 
Therefore, provided that Eq. \eqref{knownp_detect_cond} and \eqref{knownp_locate_cond1} holds, 
\begin{equation*}
    \PP_{H_1}\left(\min_{i\in S,\,i'\in S^c}(T_i-T_i')>0\right) > 1-O(n^{-(c-1)}).
\end{equation*}

The above result ensures that the set $\widehat{S}$ recovered by Algorithm \ref{alg:cap} will contain all the clique nodes, that is, $\widehat{S}\supseteq S$, with probability at least $1-O(n^{-(c-1)})$.
To ensure that $\widehat{S}$ is exactly equal to $S$, each node in $S^c$ must not be connected to all nodes in $S$.
The probability of the event is,
\begin{equation*}
    \PP_{H_1}\left(\bigcap_{i\in S^c}\left\{\sum_{j\in S}A_{i,j}\neq |S|\right\}\right) \geq 1-\sum_{i\in S^c}\PP_{H_1}\left(\sum_{j\in S}A_{i,j}= |S|\right)=1-\sum_{i\in S^c}\prod_{j\in S}P_{i,j}.
\end{equation*}
We derive
\begin{align*}
    \sum_{i\in S^c}\prod_{j\in S}P_{i,j} \leq&\, n\max_{i\in S^c}\prod_{j\in S}P_{i,j} = n\exp\left(\max_{i\in S^c}\sum_{j\in S}\log P_{i,j}\right)\\[.2cm]
     \leq&\, n\exp\left(\max_{i\in S^c}\sum_{j\in S} (P_{i,j} - 1)\right) \leq n\exp(-\epsilon |S|) \rightarrow 0,
\end{align*}
where the first inequality follows from the fact that $\log x \leq x-1$ for any $x>0$, while the second inequality and the convergence to 0 follows from Condition \eqref{knownp_locate_cond2}.
Therefore, $\PP_{H_1}(\widehat{S}=S)\rightarrow 1$.
\subsection{Proof of Theorems \ref{unknownph0er}}
\subsubsection*{Under $H_0$}
\begin{align*}
    T_{i,ER}&= \underset{j<k}{\sum\sum} A_{ij}A_{ik}(A_{jk}-\widehat{p}^{\,(i)})= \underset{j<k}{\sum\sum} A_{ij}A_{ik}(A_{jk}-p)+ \underset{j<k}{\sum\sum} A_{ij}A_{ik}(p- \widehat{p}^{\,(i)})\\
    &=\underset{j<k}{\sum\sum} A_{ij}A_{ik}(A_{jk}-p)+ {D_i\choose 2}(p- \widehat{p}^{\,(i)})\\
    &= \underset{j<k}{\sum\sum} A_{ij}A_{ik}(A_{jk}-p)- \frac{{D_i\choose 2}}{{n-1\choose 2}-{D_i\choose 2}}\underset{j<k:\, j,k\neq i}{\sum\sum} (1-A_{ij}A_{ik})(A_{jk}-p)\\
    &=S_{1i}-\frac{{D_i\choose 2}}{{n-1\choose 2}-{D_i\choose 2}}\widetilde{S}_{1i},\ \text{say.}
\end{align*}

From the proof of Theorem \ref{knownp_h0_alt}, we have
\begin{equation}
    \PP\left(|S_{1i}|\leq \sqrt{2c{D_i\choose 2} p(1-p)\log n} + \frac{2c}{3}\log n\right)\geq 1-\frac{2}{n^c},
    \label{erh0p1}
\end{equation}
since 
\begin{equation*}
    \sigma_{i,A}^2 \coloneqq \underset{j<k}{\sum\sum}\,A_{ij}A_{ik}V_{jk} = {D_i\choose 2} p(1-p).
\end{equation*}

Given $a_i$, $\widetilde{S}_{1i}$ is a sum of $\left({n-1 \choose 2} - {D_i \choose 2}\right)$ i.i.d. centered Bernoulli random variables. Using Bernstein's inequality, we obtain 
\begin{equation*}
    \PP(|\widetilde{S}_{1i}| \geq t\mid a_i)\leq 2\exp\left[-\frac{t^2/2}{\left({n-1 \choose 2} - {D_i \choose 2}\right)p(1-p) + t/3}\right].
\end{equation*}

From Lemma \ref{lem:irl}, we get that with probability at least $1-2/n^c$,
\begin{equation}
    |\widetilde{S}_{1i}|\leq \sqrt{2c \left({n-1 \choose 2} - {D_i \choose 2}\right)p(1-p)\log n} + \frac{2c}{3}\log n. 
    \label{erh0p3}
\end{equation}

From equations \eqref{erh0p1} and \eqref{erh0p3}, we have that with probability at least $1-4/n^c$,
\begin{align*}
    T_{i,ER}&\leq \sqrt{2c{D_i\choose 2} p(1-p)\log n} + \frac{{D_i\choose 2}}{\sqrt{{n-1\choose 2}-{D_i\choose 2}}} \sqrt{2c\,p(1-p)\log n} + \frac{2c}{3}\frac{{n-1\choose 2}}{{n-1\choose 2}-{D_i\choose 2}}\log n.
\end{align*}

\textbf{Concentration of $D_i$:} Using Bernstein's inequality and Lemma \ref{lem:irl}, we have that with probability at least $1-2/n^c$,
$$|D_i - \EE(D_i)|\leq \sqrt{2c\,(n-1)p\log n} + \frac{2c}{3}\log n.$$
Since $\EE(D_i)\leq (n-1)p$ and $(n-1)p\gg \log n$, we have that for any $\epsilon>0$,
\begin{equation}
    \PP_{H_0}\left(D_i\leq (n-1)p +C\sqrt{(n-1)p\log n}\right)\geq 1-\dfrac{2}{n^c},
    \label{di_bound_h0_er}
\end{equation}
for all sufficiently large $n$.

Hence, with probability at least $1-6/n^c$,
\begin{align*}
    T_{i,ER}&\leq C\max\left\{np^{3/2}\sqrt{(1-p)\log n},\log n\right\}.
\end{align*}
\subsubsection*{Under $H_1$}

For $i\in S$,
\begin{align*}
        T_{i,ER}=&\,\underset{j<k}{\sum\sum} A_{ij}A_{ik}(A_{jk}-\widehat{p}^{\,(i)}) = \underset{j<k}{\sum\sum} A_{ij}A_{ik}(A_{jk}-p)+ {D_i\choose 2}(p- \widehat{p}^{\,(i)})\\
        =&\,\underset{j<k}{\sum\sum} A_{ij}A_{ik}(A_{jk}-Q_{jk})+ {k-1\choose 2}(1-p) -\frac{{D_i\choose 2}}{{n-1\choose 2}-{D_i\choose 2}}\underset{j<k:\, j,k\neq i}{\sum\sum} (1-A_{ij}A_{ik})(A_{jk}-p)\\
        =&\,S_{1i} + {k-1\choose 2}(1-p) -\frac{{D_i\choose 2}}{{n-1\choose 2}-{D_i\choose 2}}\widetilde{S}_{1i},\ \text{say.}
\end{align*}

From the proof of Theorem \ref{knownp_h0_alt}, we have
\begin{equation}
    \PP\left(|S_{1i}|\leq \sqrt{2c\left({D_i\choose 2}-{{k-1}\choose 2}\right) p(1-p)\log n} + \frac{2c}{3}\log n\right)\geq 1-\frac{2}{n^c},
    \label{erh1p1s}
\end{equation}
since 
\begin{equation*}
    \widetilde\sigma_{i,A}^2 \coloneqq \underset{j<k}{\sum\sum}\,A_{ij}A_{ik}\widetilde{V}_{jk} = \left({D_i\choose 2}-{{k-1}\choose 2}\right) p(1-p).
\end{equation*}

Given $a_i$, $\widetilde{S}_{1i}$ is a sum of at most $\left({n-1 \choose 2} - {D_i \choose 2}\right)$ i.i.d. centered Bernoulli random variables. From Bernstein's inequality and Lemma \ref{lem:irl}, we obtain with probability at least $1-2/n^c$,
\begin{equation}
    |\widetilde{S}_{1i}|\leq \sqrt{2c \left({n-1 \choose 2} - {D_i \choose 2}\right)p(1-p)\log n} + \frac{2c}{3}\log n. 
    \label{erh1p2s}
\end{equation}

\textbf{Concentration of $D_i$:} Using Bernstein's inequality and Lemma \ref{lem:irl}, we have that with probability at least $1-2/n^c$,
$$|D_i - \EE(D_i)|\leq \sqrt{2c\,((n-1)p+k)\log n} + \frac{2c}{3}\log n.$$
Since $\EE(D_i)\leq (n-1)p + k$ and $(n-1)p\gg \log n$, we have that for any $\epsilon>0$,
\begin{equation}
    \PP_{H_1}\left(D_i\leq (n-1)p + k + C\sqrt{((n-1)p + k)\log n}\right)\geq 1-\dfrac{2}{n^c},
    \label{di_bound_h1_er}
\end{equation}
for all sufficiently large $n$.
Therefore, with probability at least $1-2/n^c$,
\begin{align*}
    &\frac{{D_i\choose 2}}{{n-1\choose 2}}\leq \frac{D_i^2}{(n-1)^2} \leq  \frac{((n-1)p+k+C\sqrt{((n-1)p + k)\log n})^2}{(n-1)^2}=p^2+\gamma^1_{n,k,p},
\end{align*}
where $\gamma^1_{n,k,p}=o(1)$, since $k=o(n)$. Hence, 
\begin{equation*}
    \frac{\frac{{D_i\choose 2}}{{n-1\choose 2}}}{1-\frac{{D_i\choose 2}}{{n-1\choose 2}}}\leq \frac{p^2+ \gamma^1_{n,k,p}}{1-p^2-\gamma^1_{n,k,p}}\implies \frac{{D_i\choose 2}}{{n-1\choose 2}-{D_i\choose 2}} -\frac{p^2}{1-p^2} \leq \frac{\gamma^1_{n,k,p}}{\left(1-p^2-\gamma^1_{n,k,p}\right)(1-p^2)}.
\end{equation*}

Since $p<p_1<1$, we have with probability at least $1-2/n^c$,
\begin{equation}
\frac{{D_i\choose 2}}{{n-1\choose 2}-{D_i\choose 2}} \leq \frac{p^2}{1-p^2}+C\,\gamma^1_{n,k,p}.
\label{erh1p3s}
\end{equation}

From Equations \eqref{erh1p1s},\eqref{erh1p2s}, \eqref{di_bound_h1_er} and \eqref{erh1p3s}, we have that for $i\in S$, with probability at least $1-6/n^c$,
\begin{align*}
        T_{i,ER} \geq&\,{k-1\choose 2}(1-p) - C\,(np + k)\sqrt{p(1-p)\log n} + \frac{2c}{3}\log n\\[.1cm]
        -&\,\left(\frac{p^2}{1-p^2}+O(\gamma^1_{n,k,p})\right)\sqrt{c(1+\epsilon)}\, n \sqrt{p\log n}\\[.1cm]
        =&\,{k-1\choose 2}(1-p) - O\left(np^{\frac{3}{2}} \max\left\{\sqrt{\frac{2k}{np}},\,1,\,\frac{p}{1-p^2}\right\} \sqrt{\log n}\right) - O(n\gamma^1_{n,k,p} \sqrt{p\log n}).
\end{align*}

By Eq. \eqref{er_detectcond}, we conclude that for $i\in S$, there exists a constant $\gamma_1>0$ such that
\begin{equation*}
    \PP_{H_1}\left(T_{i,ER}\geq \gamma_1 k^2\right)\geq 1-\frac{6}{n^c}.
\end{equation*}

\subsection{Proof of Theorem \ref{unknownprankinger}}
For $i\in S^c$, recall the quantity $R_i$ as the number of nodes in $S$ that are neighbors of $i$, i.e., $R_i=\sum_j A_{ij}\,\mathbb{I}(j\in S)$. We see that in this case,
\begin{equation*}
    \underset{j<k}{\sum\sum} A_{ij}A_{ik}\,\mathbb{I}(j,k\in S)={R_i\choose 2},
    \underset{j<k:\, j,k\neq i}{\sum\sum} (1-A_{ij}A_{ik})\,\mathbb{I}(j,k\in S)={k\choose 2}-{R_i\choose 2}.
\end{equation*}

Therefore,
\begin{align*}
    T_{i,ER}=& \underset{j<k}{\sum\sum} A_{ij}A_{ik}(A_{jk}-\widehat{p}^{\,(i)})= \underset{j<k}{\sum\sum} A_{ij}A_{ik}(A_{jk}-p)+ {D_i\choose 2}(p- \widehat{p}^{\,(i)})\\[.2cm]
        =& \underset{j<k}{\sum\sum} A_{ij}A_{ik}(A_{jk}-p)\mathbb{I}(j\in S^c\ \text{or}\ k\in S^c)+ (1-p)\underset{j<k}{\sum\sum} A_{ij}A_{ik}\mathbb{I}(j,k\in S) \\[.2cm]
        -& \frac{{D_i\choose 2}}{{n-1\choose 2}-{D_i\choose 2}}\underset{j<k:\, j,k\neq i}{\sum\sum} (1-A_{ij}A_{ik})(A_{jk}-p)\\
        =& \underset{j<k}{\sum\sum} A_{ij}A_{ik}(A_{jk}-p)\mathbb{I}(j\in S^c\ \text{or}\ k\in S^c)+ (1-p)\underset{j<k}{\sum\sum} A_{ij}A_{ik}\mathbb{I}(j,k\in S) \\[.2cm]
        -& \frac{{D_i\choose 2}}{{n-1\choose 2}-{D_i\choose 2}}\underset{j<k:\, j,k\neq i}{\sum\sum} (1-A_{ij}A_{ik})(A_{jk}-p)\mathbb{I}(j\in S^c\ \text{or}\ k\in S^c) \\[.2cm]
        -& \frac{{D_i\choose 2}(1-p)}{{n-1\choose 2}-{D_i\choose 2}}\underset{j<k:\, j,k\neq i}{\sum\sum} (1-A_{ij}A_{ik})\mathbb{I}(j,k\in S) \\[.2cm]
        =& S_{1i}  -\frac{{D_i\choose 2}}{{n-1\choose 2}-{D_i\choose 2}}\widetilde{S}_{1i} + (1-p)\left({R_i\choose 2}-\frac{{D_i\choose 2}}{{n-1\choose 2}-{D_i\choose 2}}\left({k\choose 2}-{R_i\choose 2}\right)\right).
\end{align*}

It is straightforward to show that the bounds for $S_{1i}$ and $\widetilde{S}_{1i}$ derived for $i\in S$ also hold for $i\in S^c$.
We focus on the last term on the expansion of $T_{i,ER}$ for $i\in S^c$, $${R_i\choose 2}-\frac{{D_i\choose 2}}{{n-1\choose 2}-{D_i\choose 2}}\left({k\choose 2}-{R_i\choose 2}\right).$$

For $i\in S^c$, the distribution of $D_i$ is the same as under $H_0$. So, from \eqref{di_bound_h0_er}, we have that with probability at least $1-1/n^c$,
\begin{equation}
\left|\frac{{D_i\choose 2}}{{n-1\choose 2}-{D_i\choose 2}}-\frac{p^2}{1-p^2}\right|=O\left(p^{\frac{3}{2}}\sqrt{\frac{\log n}{n}}\right).
\label{erh1p3ns}
\end{equation}

\begin{align*}
     &\,\left|{R_i\choose 2}-\frac{{D_i\choose 2}}{{n-1\choose 2}-{D_i\choose 2}}\left({k\choose 2}-{R_i\choose 2}\right)\right|\\[.1cm]
        =&\,\left|{R_i\choose 2}-\left(\frac{{D_i\choose 2}}{{n-1\choose 2}-{D_i\choose 2}}-\frac{p^2}{1-p^2}\right)\left({k\choose 2}-{R_i\choose 2}\right) -\frac{p^2}{1-p^2}\left({k\choose 2}-{R_i\choose 2}\right)\right|\\[.1cm]
        =&\,\left|\frac{1}{1-p^2}\left({R_i\choose 2}-{k\choose 2}p^2\right)-\left(\frac{{D_i\choose 2}}{{n-1\choose 2}-{D_i\choose 2}}-\frac{p^2}{1-p^2}\right)\left({k\choose 2}-{R_i\choose 2}\right)\right|\\[.1cm]
        \leq&\, \frac{1}{1-p^2}\left|{R_i\choose 2}-{k\choose 2}p^2\right|+O\left(p^{\frac{3}{2}}\sqrt{\frac{\log n}{n}}\right){k\choose 2}= \frac{1}{1-p^2}\left|{R_i\choose 2}-{k\choose 2}p^2\right|\\[.1cm]
        +&\, O((kp)^{\frac{3}{2}}\sqrt{\log n}).
\end{align*}

Observe that in the proof of Theorem \ref{knownp_ranking_alt}, $S_{2i}=\left({R_i\choose 2}-{k\choose 2}p^2\right)(1-p)$ under the ER model. Therefore, from \eqref{s2i}, we have that with probability at least $1-(k+1)/n^c$,
\begin{equation*}
    \begin{split}
    \left|{R_i\choose 2}-{k\choose 2}p^2\right|(1-p)=O(k\sqrt{\log n}\max(\sqrt{kp^3},\sqrt{\log n})).
\end{split}
\end{equation*}

Therefore, with probability at least $1-(k+2)/n^c$,
\begin{equation*}
    \left|{R_i\choose 2}-\frac{{D_i\choose 2}}{{n-1\choose 2}-{D_i\choose 2}}\left({k\choose 2}-{R_i\choose 2}\right)\right|=O(k\sqrt{\log n}\max(\sqrt{kp^3},\sqrt{\log n})).
    \label{erh1p4ns}
\end{equation*}

Consider $i\in S, i'\in S^c$.
\begin{align*}
    &\,T_{i,ER}-T_{i',ER}\\
    =&\,S_{1i} + {k-1\choose 2}(1-p) -\frac{{D_i\choose 2}}{{n-1\choose 2}-{D_i\choose 2}}\widetilde{S}_{1i} - S_{1i'}  +\frac{{D_{i'}\choose 2}}{{n-1\choose 2}-{D_{i'}\choose 2}}\widetilde{S}_{1i'} - (1-p)\widetilde{R}_{i'}\\[.1cm]
    \geq &\,{k-1\choose 2}(1-p) -|S_{1i}| - |S_{1i'}| - \frac{{D_i\choose 2}}{{n-1\choose 2}-{D_i\choose 2}}|\widetilde{S}_{1i}| - \frac{{D_{i'}\choose 2}}{{n-1\choose 2}-{D_{i'}\choose 2}}|\widetilde{S}_{1i'}| -|\widetilde{R}_{i'}|\\[.1cm]
    = &\,{k-1\choose 2}(1-p) -O\left(np^{\frac{3}{2}}\max\left(\sqrt{\frac{2k}{np}},1,\frac{p}{1-p^2}\right)\sqrt{\log n}\right) -O\left(n\gamma^1_{n,k,p} \sqrt{p\log n}\right)\\[.1cm]
    -&\,O\left(k\sqrt{\log n}\max(\sqrt{kp^3},\sqrt{\log n})\right)\\[.1cm]
    >&\, 0,\ \text{by Eq. \eqref{er_detectcond}}.
\end{align*}

Hence,
\begin{equation*}
    \PP_{H_1}\left(\min_{i\in S,\,i'\in S^c}(T_{i,ER}-T_{i',ER})>0\right) > 1-O(n^{-(c-1)}).
\end{equation*}

\subsection{Proof of Theorem \ref{unknownph0cl}}
Let us define the matrix $\widecheck{P}$ such that 
$$\widecheck{P}_{jk} \coloneqq \frac{\EE(D_j)\EE(D_k)}{2\,\EE(M)},\ 1\leq j,k\leq n.$$
The matrix $\widecheck{P}$ should be close to $\widehat{P}$ under both $H_0$ and $H_1$.
\subsubsection*{Under $H_0$}
\begin{align*}
    T_{i,CL}=&\, \underset{j<k}{\sum\sum} A_{ij}A_{ik}(A_{jk}-\widehat{P}_{jk})= \underset{j<k}{\sum\sum} A_{ij}A_{ik}(A_{jk}-\widecheck{P}_{jk})+ \underset{j<k}{\sum\sum} A_{ij}A_{ik}(\widecheck{P}_{jk}- \widehat{P}_{jk})\\
    =&\, \underset{j<k}{\sum\sum} A_{ij}A_{ik}(A_{jk}-P_{jk}) + \underset{j<k}{\sum\sum} A_{ij}A_{ik}(P_{jk}-\widecheck{P}_{jk})+ \underset{j<k}{\sum\sum} A_{ij}A_{ik}(\widecheck{P}_{jk}- \widehat{P}_{jk})\\
    =&\, S_{1i} + \Delta_i + S_{3i},\ \text{say.}
\end{align*}

In the proof of Theorem \ref{knownp_h0_alt}, we showed that for any $c>1$,
\begin{equation}
    \PP_{H_0}\left(|S_{1i}| \leq C\,\delta(P)\sqrt{\log n}\right)\geq 1-\frac{2}{n^c}.
    \label{clh0p1}
\end{equation}

\textbf{Bounding $\Delta_i$:}
$$|\Delta_i| =\left|\underset{j<k}{\sum\sum} A_{ij}A_{ik}(P_{jk}-\widecheck{P}_{jk})\right| \leq \underset{j<k}{\sum\sum} |P_{jk}-\widecheck{P}_{jk}|.$$

We derive
\begin{align*}
    &\, |P_{jk}-\widecheck{P}_{jk}| =  \left|\theta_j\theta_k - \frac{\EE(D_j)\EE(D_k)}{2\,\EE(M)}\right|=  \left|\theta_j\theta_k - \frac{\theta_j\theta_k\sum_{\ell\neq j} \theta_\ell  \sum_{\ell\neq k} \theta_\ell }{\sum\sum_{j\neq k} \theta_j\theta_k}\right|\\[.1cm]
    =&\,\, \left|\frac{\theta_j\theta_k(\left(\sum_{\ell} \theta_\ell \right)^2 - \sum_{\ell} \theta_\ell ^2) -\theta_j\theta_k\left(\sum_{\ell} \theta_\ell -\theta_j\right)\left(\sum_{\ell} \theta_\ell -\theta_k\right)}{\sum\sum_{j\neq k} \theta_j\theta_k}\right|\\[.1cm]
    =&\,\, \frac{\left|-\theta_j\theta_k \sum_{\ell} \theta_\ell ^2+\theta_j\theta_k(\theta_j+\theta_k)\sum_{\ell} \theta_\ell  - (\theta_j\theta_k)^2\right|}{\sum\sum_{j\neq k} \theta_j\theta_k}
    \leq \frac{\theta_j\theta_k(3\,\delta(P) + 1)}{\sum\sum_{j\neq k} \theta_j\theta_k}.
\end{align*}

Therefore,
\begin{equation}
    \label{clh0p3}
    |\Delta_i| \leq \frac{3\,\delta(P) + 1}{2}.
\end{equation}

\textbf{Bounding $S_{3i}$:}
\begin{align*}
   &|S_{3i}| =\left|\underset{j<k}{\sum\sum} A_{ij}A_{ik}(\widehat{P}_{jk} - \widecheck{P}_{j,k})\right|
        =\frac{1}{2}\left|\underset{j\neq k}{\sum\sum} A_{ij}A_{ik}\left(\frac{D_jD_k}{2M} - \frac{\EE(D_j)\EE(D_k)}{2\,\EE(M)}\right)\right|\\[.1cm]
        \leq&\,\frac{1}{2}\left|\underset{j, k}{\sum\sum} A_{ij}A_{ik}\left(\frac{D_jD_k}{2M} - \frac{\EE(D_j)\EE(D_k)}{2\,\EE(M)}\right)\right| + \frac{1}{2}\left|\sum_j A_{ij}\left(\frac{D_j^2}{2M} - \frac{(\EE(D_j))^2}{2\,\EE(M)}\right)\right|.
\end{align*}

\begin{align*}
    &\, \left|\underset{j, k}{\sum\sum} A_{ij}A_{ik}\left(\frac{D_jD_k}{2M} - \frac{\EE(D_j)\EE(D_k)}{2\,\EE(M)}\right)\right|\\[.1cm]
    =&\,\left|\underset{j, k}{\sum\sum} A_{ij}A_{ik}\left(\frac{(D_j-\EE(D_j))D_k}{2M} + \frac{\EE(D_j)D_k}{2}\left(\frac{1}{M}-\frac{1}{\EE(M)}\right) + \frac{\EE(D_j)(D_k - \EE(D_k))}{2\,\EE(M)}\right)\right|\\[.1cm]
    \leq&\, 2\left|\sum_jA_{ij}(D_j-\EE(D_j))\right| + \frac{D_i\,\delta(P)|M-\EE(M)|}{2\,\EE(M)},
\end{align*}
since $\sum_jA_{ij}D_j\leq 2M,\ \sum_jA_{ij}\EE(D_j)\leq 2\,\EE(M)$ and $\EE(D_j)\leq \delta(P)$.

\begin{align*}
    &\left|\sum_j A_{ij}\left(\frac{D_j^2}{2M} - \frac{(\EE(D_j))^2}{2\,\EE(M)}\right)\right|\\[.1cm]
    = &\sum_j A_{ij}\left(\frac{D_j(D_j-\EE(D_j))}{2M} + \frac{D_j\EE(D_j)}{2}\left(\frac{1}{M}-\frac{1}{\EE(M)}\right) + \frac{\EE(D_j)(D_j-\EE(D_j))}{2\,\EE(M)}\right)\\[.1cm]
    \leq&\, 2\,\max_j|D_j-\EE(D_j)| + \frac{\delta(P)|M - \EE(M)|}{2\,\EE(M)},
\end{align*}
since $\sum_jA_{ij}D_j\leq 2M,\ \sum_jA_{ij}\EE(D_j)\leq 2\,\EE(M)$ and $\EE(D_j)\leq \delta(P)$.

To establish a bound on $|S_{3i}|$, we first derive bound for each of the three terms: $\max\limits_j|D_j - \EE(D_j)|$, $|M - \EE(M)|$ and  $|\sum_jA_{ij}(D_j-\EE(D_j))|$.\\

\textbf{Bounding $\max\limits_j|D_j - \EE(D_j)|$:}
Using Bernstein's inequality and Lemma \ref{lem:irl}, we have that for all $j\in[n]$,
\begin{equation*}
    \PP_{H_0}\left(|D_j - \EE(D_j)|\leq \sqrt{2(c+1)\,\delta(P)\log n} + \frac{2(c+1)}{3}\log n\right) \geq 1-\frac{2}{n^{c+1}}.
\end{equation*}
Since $\delta(P)\geq c_0\log n$, we have
\begin{equation}
    \PP_{H_0}\left(|D_j - \EE(D_j)|\leq C\sqrt{\delta(P)\log n}\right) \geq 1-\frac{2}{n^{c+1}}.
    \label{dih0}
\end{equation}

\textbf{Bounding $|M-\EE(M)|$:}
$M-\EE(M)=\sum_{j<k}(A_{jk}-P_{jk})$. Applying Bernstein's inequality and Lemma \ref{lem:irl}, we have with probability at least $1-2/n^c$,
\begin{equation}
    |M-\EE(M)|\leq \sqrt{2c\,\EE(M)\log n} + \frac{2c}{3}\log n \leq C\sqrt{\EE(M)\log n},
    \label{m_bound}
\end{equation} 
since $\EE(M)\geq \delta(P)\geq c_0\log n$.

\textbf{Bounding $|\sum_j A_{ij}(D_j-\EE(D_j))|$:}
\begin{align*}
    &\left|\sum_{j\neq i} A_{ij}(D_j-\EE(D_j))\right|=\left|\sum_{j\neq i}\sum_\ell A_{ij}(A_{j\ell} -\EE(A_{j\ell}))\right|\\[.1cm]
    \leq&\,\left|\sum_{j\neq i} A_{ij}(A_{ji} -\EE(A_{ji}))\right| + \left|\underset{j,\ell \neq i}{\sum\sum} A_{ij}(A_{j\ell}-\EE(A_{j\ell}))\right|
    \leq D_i + 2\left|\underset{j<\ell:\, j,\ell\neq i}{\sum\sum} A_{ij}(A_{j\ell}-P_{j\ell})\right|.
\end{align*}

Since $\EE(D_i)\leq \delta(P)$, Eq. \eqref{dih0} implies
\begin{equation}
    \PP_{H_0}\left(D_i \leq C\delta(P)\right) \geq 1-\frac{2}{n^c}.
    \label{di_bound_cl}
\end{equation}

Given $a_i$, $\underset{j<\ell:\, j,\ell\neq i}{\sum\sum} A_{ij}(A_{j\ell}-P_{j\ell})$ is a sum of independent centered Bernoulli random variables. Applying Bernstein's inequality and Lemma \ref{lem:irl}, we have with probability at least $1-2/n^c$,
\begin{equation}
    \begin{split}
        \left|\underset{j<\ell:\, j,\ell\neq i}{\sum\sum} A_{ij}(A_{j\ell}-P_{j\ell})\right|\leq&\, \sqrt{2c\underset{j<\ell:\, j,\ell\neq i}{\sum\sum} A_{ij}P_{j\ell}(1-P_{j\ell})\log n} + \frac{2c}{3}\log n\\[.1cm]
        \leq&\,\sqrt{c\, D_i\,\delta(P)\log n} + \frac{2c}{3}\log n.
    \end{split}
    \label{aij_ajl_bound}
\end{equation}

From Equations \eqref{di_bound_cl}, \eqref{aij_ajl_bound}, and the expansion of $|\sum_{j\neq i} A_{ij}(D_j-\EE(D_j))|$, we obtain with probability at least $1-4/n^c$,
\begin{equation}
    \left|\sum_{j\neq i} A_{ij}(D_j-\EE(D_j))\right|\leq C\delta(P)\sqrt{\log n}.
    \label{aij_dj_bound}
\end{equation}

Combining Equations \eqref{dih0}, \eqref{m_bound}, \eqref{aij_dj_bound}, and the expansion of $|S_{3i}|$, we obtain
\begin{equation*}
    \PP_{H_0}\left(|S_{3i}| \leq C\delta(P)\sqrt{\log n} + C\frac{\delta(P)^2\sqrt{\log n}}{\sqrt{\EE(M)}}\right) \geq 1-O(n^{-c}).
\end{equation*}
Note that, $\delta(P)\leq \sum_\ell \theta_\ell$, and $2\,\EE(M)=(\sum_\ell\theta_\ell)^2 - \sum_\ell\theta_\ell^2 \geq (1/2) (\sum_\ell\theta_\ell)^2$ by Condition \eqref{degree_cond}.
Therefore,
\begin{equation}
    \PP_{H_0}(|S_{3i}| \leq C\delta(P)\sqrt{\log n}) \geq 1-O(n^{-c}).
    \label{clh0p2}
\end{equation}

\textbf{Conclusion:} From Equations \eqref{clh0p1},\eqref{clh0p3} and \eqref{clh0p2}, we conclude
\begin{equation*}
    \PP_{H_0}(T_{i,CL} \leq C\delta(P)\sqrt{\log n}) \geq 1-O(n^{-c}).
\end{equation*}

\subsubsection*{Under $H_1$}

Consider $i\in S$.
\begin{align*}
    T_{i,CL}=&\, \underset{j<k}{\sum\sum} A_{ij}A_{ik}(A_{jk}-\widehat{P}_{jk})\\
    =&\, \underset{j<k}{\sum\sum} A_{ij}A_{ik}(A_{jk}-Q_{jk}) + \underset{j<k}{\sum\sum} A_{ij}A_{ik}(Q_{jk}-\widecheck{P}_{jk}) +
    \underset{j<k}{\sum\sum} A_{ij}A_{ik}(\widecheck{P}_{jk}- \widehat{P}_{jk})\\[.1cm]
    =&\, \underset{j<k}{\sum\sum} A_{ij}A_{ik}(A_{jk}-Q_{jk}) + \underset{j<k}{\sum\sum} A_{ij}A_{ik}(P_{jk}-\widecheck{P}_{jk})\,\mathbb{I}(j\in S^c\ \text{or}\ k\in S^c)\\[.1cm]
    +&\, \underset{j<k}{\sum\sum} A_{ij}A_{ik}(\widecheck{P}_{jk}- \widehat{P}_{jk}) + \underset{j<k:\,j,k\neq i}{\sum\sum} (1-\widecheck{P}_{jk})\,\mathbb{I}(j,k\in S)\\[.1cm]
    =&\, S_{1i} + \Delta_i + S_{3i} + \underset{j<k:\,j,k\neq i}{\sum\sum} (1-\widecheck{P}_{jk})\,\mathbb{I}(j,k\in S),\ \text{say.}
\end{align*}

From the  proof of Theorem \ref{knownp_h0_alt}, we have that for any $c>1$,
\begin{equation}
    \PP_{H_0}\left(|S_{1i}| \leq C\,\delta(Q)\sqrt{\log n}\right)\geq 1-\frac{2}{n^c}.
    \label{clh1p1s}
\end{equation}

\textbf{Bounding $\Delta_i$:}
\begin{equation}
    \begin{split}
    &\,|\Delta_i| =\left|\underset{j<k}{\sum\sum} A_{ij}A_{ik}(P_{jk}-\widecheck{P}_{jk})\,\mathbb{I}(j\in S^c\ \text{or}\ k\in S^c)\right|\\[.1cm]
    \leq&\, \underset{j<k}{\sum\sum} A_{ij} A_{ik} |P_{jk}-\widecheck{P}_{jk}|\,\mathbb{I}(j,k\in S^c) + \underset{j,k}{\sum\sum} A_{ik}|P_{jk}-\widecheck{P}_{jk}|\,\mathbb{I}(j\in S,\,k\in S^c)
\end{split}
\label{deltai_break}
\end{equation}

For $j,k\in S^c$,
\begin{align*}
    &\,|P_{jk} - \widecheck{P}_{jk}| = \left|\theta_j\theta_k - \frac{\EE(D_j)\EE(D_k)}{2\,\EE(M)}\right| = \left|\theta_j\theta_k - \frac{\theta_j\theta_k\sum_{\ell\neq j} \theta_\ell  \sum_{\ell\neq k} \theta_\ell }{\sum\sum_{j\neq k} \theta_j\theta_k+\umod{B_S}_{1,1}}\right| \\[.1cm]
    =&\, \left|\frac{\theta_j\theta_k\left(\left(\sum_{\ell} \theta_\ell \right)^2 - \sum_{\ell} \theta_\ell ^2\right)+\theta_j\theta_k\umod{B_S}_{1,1}-\theta_j\theta_k\left(\sum_{\ell} \theta_\ell -\theta_j\right)\left(\sum_{\ell} \theta_\ell -\theta_k\right)}{\sum\sum_{j\neq k} \theta_j\theta_k+\umod{B_S}_{1,1}}\right|\\[.1cm]
    =&\, \left|\frac{\theta_j\theta_k\umod{B_S}_{1,1} -\theta_j\theta_k \sum_{\ell} \theta_\ell ^2+\theta_j\theta_k(\theta_j+\theta_k)\sum_{\ell} \theta_\ell  - (\theta_j\theta_k)^2}{\sum\sum_{j\neq k} \theta_j\theta_k+\umod{B_S}_{1,1}}\right| \leq (1+o(1))\,\theta_j\theta_k \frac{\umod{B_S}_{1,1}}{(\sum_\ell \theta_\ell)^2},
\end{align*}
where the last step follows from Conditions \eqref{cl_detectcond1} and \eqref{cl_detectcond2}, noting that the second, third and fourth term in the numerator of the above expression are, in absolute value, $O(\delta(P))$ and $\umod{B_S}_{1,1}\gg \delta(P)$.
Hence,
\begin{equation}
    \underset{j<k:\,j,k\neq i}{\sum\sum} A_{ij}A_{ik}|P_{jk}-\widecheck{P}_{jk}|\,\mathbb{I}(j, k\in S^c)\leq \left(\frac{1}{2}+o(1)\right)\umod{B_S}_{1,1} \frac{(\sum_{\ell\in S^c} A_{i\ell}\theta_\ell)^2}{(\sum_\ell \theta_\ell)^2}.
    \label{deltai_pt1}
\end{equation}

For $j\in S,k\in S^c$,
\begin{align*}
     &\,|P_{jk} - \widecheck{P}_{jk}| = \left|\theta_j\theta_k - \frac{\EE(D_j)\EE(D_k)}{2\,\EE(M)}\right|\\[.1cm]
     =&\,\left|\theta_j\theta_k - \frac{(\sum_{\ell\neq j} \theta_j\theta_\ell  +\sum_{\ell\in S:\, \ell\neq j}(1-\theta_j\theta_\ell))\, (\sum_{\ell\neq k} \theta_k\theta_\ell )}{\sum\sum_{j\neq k} \theta_j\theta_k+\umod{B_S}_{1,1}}\right|\\
     \leq&\,\left|\theta_j\theta_k - \frac{\theta_j\theta_k\sum_{\ell\neq j} \theta_\ell  \sum_{\ell\neq k} \theta_\ell}{\sum\sum_{j\neq k} \theta_j\theta_k+\umod{B_S}_{1,1}}\right| + \left|\frac{\theta_k(\sum_{\ell\neq k} \theta_\ell) \sum_{\ell\in S:\, \ell\neq j}(1-\theta_j\theta_\ell )}{\sum\sum_{j\neq k} \theta_j\theta_k+\umod{B_S}_{1,1}}\right|\\[.1cm]
    \leq&\, (1+o(1))\, \theta_j\theta_k \frac{\umod{B_S}_{1,1}}{(\sum_\ell \theta_\ell)^2} + (1+o(1))\,\theta_k\frac{\sum_{\ell\in S:\, \ell\neq j}(1-\theta_j\theta_\ell)}{\sum_\ell \theta_\ell},
\end{align*}
by Conditions Conditions \eqref{cl_detectcond1} and \eqref{cl_detectcond2}.
Hence,
\begin{equation}
    \underset{j,k\neq i}{\sum\sum} A_{ik}|P_{jk}-\widecheck{P}_{jk}|\,\mathbb{I}(j\in S, k\in S^c)\leq (2+o(1))\umod{B_S}_{1,1} \frac{\sum_{\ell\in S^c} A_{i\ell}\theta_\ell}{\sum_\ell \theta_\ell}.
    \label{deltai_pt2}
\end{equation}

Using Bernstein's inequality and Lemma \ref{lem:irl}, we have that for all $j\in[n]$,
\begin{equation*}
    \PP_{H_1}\left(\sum_{\ell\in S^c} A_{i\ell}\theta_\ell\leq \sum_{\ell\in S^c} P_{i\ell}\theta_\ell + \sqrt{2c\,\sum_{\ell\in S^c} P_{i\ell}\theta_\ell^2\log n} + \frac{2c}{3}\log n\right) \geq 1-\frac{2}{n^{c+1}}.
\end{equation*}
since $\log n\ll \delta(P)\leq \sum_\ell\theta_\ell$, we have
\begin{equation}
    \PP_{H_1}\left(\sum_{\ell\in S^c} A_{i\ell}\theta_\ell\leq \left(\max_{j\in S^c} P_{ij} + o(1)\right) \sum_\ell\theta_\ell\right) \geq 1-\frac{2}{n^{c+1}}.
    \label{deltai_pt3}
\end{equation}

Combining Equations \eqref{deltai_break}, \eqref{deltai_pt1}, \eqref{deltai_pt2}, and \eqref{deltai_pt3} yields
\begin{equation}
    \PP_{H_1}\left(|\Delta_i|\leq \left(\frac{1}{2}\max_{j\in S^c} (P_{ij}^2 + 4 P_{ij}) + o(1)\right) \umod{B_S}_{1,1}\right) \geq 1-O(n^{-c}).
    \label{clh1p2s}
\end{equation}

\textbf{Bounding $S_{3i}$:}
Following the calculations for $S_{3i}$ under $H_0$, we have
\begin{align*}
    |S_{3i}| \leq \left|\sum_{j\neq i} A_{ij}(D_j-\EE(D_j))\right| + \max_j|D_j-\EE(D_j)| +\frac{(D_i+1)\,\delta(Q)\,|M-\EE(M)|}{4\,\EE(M)}.
\end{align*}

\textbf{Bounding $\max\limits_j|D_j - \EE(D_j)|$:}
Using Bernstein's inequality and Lemma \ref{lem:irl}, we have that for all $j\in[n]$,
\begin{equation*}
    \PP_{H_0}\left(|D_j - \EE(D_j)|\leq \sqrt{2(c+1)\,\delta(Q)\log n} + \frac{2(c+1)}{3}\log n\right) \geq 1-\frac{2}{n^{c+1}}.
\end{equation*}
Since $\delta(Q)\geq \delta(P)\geq c_0\log n$, we have
\begin{equation}
    \PP_{H_0}\left(|D_j - \EE(D_j)|\leq C\sqrt{\delta(Q)\log n}\right) \geq 1-\frac{2}{n^{c+1}}.
    \label{dih1s}
\end{equation}

\textbf{Bounding $|M-\EE(M)|$:}
$M-\EE(M)=\sum_{j<k}(A_{jk}-Q_{jk})$. Applying Bernstein's inequality and Lemma \ref{lem:irl}, we have with probability at least $1-2/n^c$,
\begin{equation}
    |M-\EE(M)|\leq \sqrt{2c\,\EE(M)\log n} + \frac{2c}{3}\log n \leq C\sqrt{\EE(M)\log n},
    \label{m_bound_h1s}
\end{equation} 
since $\EE(M)\geq \delta(P)\geq c_0\log n$.

\textbf{Bounding $|\sum_j A_{ij}(D_j-\EE(D_j))|$:}
\begin{align*}
    &\left|\sum_{j\neq i} A_{ij}(D_j-\EE(D_j))\right|=\left|\sum_{j\neq i}\sum_\ell A_{ij}(A_{j\ell} -\EE(A_{j\ell}))\right|\\[.1cm]
    \leq&\,\left|\sum_{j\neq i} A_{ij}(A_{ji} -\EE(A_{ji}))\right| + \left|\underset{j,\ell \neq i}{\sum\sum} A_{ij}(A_{j\ell}-\EE(A_{j\ell}))\right|
    \leq D_i + 2\left|\underset{j<\ell:\, j,\ell\neq i}{\sum\sum} A_{ij}(A_{j\ell}-Q_{j\ell})\right|.
\end{align*}

Since $\EE(D_i)\leq \delta(Q)$, Eq. \eqref{dih1s} implies
\begin{equation}
    \PP_{H_0}\left(D_i \leq C\delta(Q)\right) \geq 1-\frac{2}{n^c}.
    \label{di_bound_cl_h1s}
\end{equation}

Given $a_i$, $\underset{j<\ell:\, j,\ell\neq i}{\sum\sum} A_{ij}(A_{j\ell}-P_{j\ell})$ is a sum of independent centered Bernoulli random variables. Applying Bernstein's inequality and Lemma \ref{lem:irl}, we have with probability at least $1-2/n^c$,
\begin{equation}
    \begin{split}
        \left|\underset{j<\ell:\, j,\ell\neq i}{\sum\sum} A_{ij}(A_{j\ell}-P_{j\ell})\right|\leq&\, \sqrt{2c\underset{j<\ell:\, j,\ell\neq i}{\sum\sum} A_{ij}Q_{j\ell}(1-Q_{j\ell})\log n} + \frac{2c}{3}\log n\\[.1cm]
        \leq&\,\sqrt{2c\, D_i\,\delta(Q)\log n} + \frac{2c}{3}\log n.
    \end{split}
    \label{aij_ajl_bound_h1s}
\end{equation}

From Equations \eqref{di_bound_cl_h1s}, \eqref{aij_ajl_bound_h1s}, and the expansion of $|\sum_{j\neq i} A_{ij}(D_j-\EE(D_j))|$, we obtain with probability at least $1-4/n^c$,
\begin{equation}
    \left|\sum_{j\neq i} A_{ij}(D_j-\EE(D_j))\right|\leq C\delta(Q)\sqrt{\log n}.
    \label{aij_dj_bound_h1s}
\end{equation}

Combining Equations \eqref{dih1s}, \eqref{m_bound_h1s}, \eqref{aij_dj_bound_h1s}, and the expansion of $|S_{3i}|$, we obtain
\begin{equation*}
    \PP_{H_0}\left(|S_{3i}| \leq C\delta(Q)\sqrt{\log n} + C\frac{\delta(Q)^2\sqrt{\log n}}{\sqrt{\EE(M)}}\right) \geq 1-O(n^{-c}).
\end{equation*}
Note that, $\delta(Q) = \delta(P) + \delta(B_S)\leq 2\,\delta(P) \leq \sum_\ell \theta_\ell$ since $\delta(B_S)\ll \delta(P)$ by Conidition \eqref{cl_detectcond2}, and $2\,\EE(M)=(\sum_\ell\theta_\ell)^2 - \sum_\ell\theta_\ell^2 + \umod{B_S}_{1,1} \geq (1/2) (\sum_\ell\theta_\ell)^2$ by Condition \eqref{degree_cond}.
Therefore,
\begin{equation}
    \PP_{H_0}(|S_{3i}| \leq C\delta(P)\sqrt{\log n}) \geq 1-O(n^{-c}).
    \label{clh1p3s}
\end{equation}

\textbf{Lower bound on $\underset{j<k:\,j,k\neq i}{\sum\sum} (1-\widecheck{P}_{jk})\,\mathbb{I}(j,k\in S)$:}
Note that $\EE(M)=\sum\sum_{j\neq k} \theta_j\theta_k + \sum\sum_{j,k\in S:\,j\neq k}(1-\theta_j\theta_k)= \left(\sum_{\ell} \theta_\ell \right)^2 - \sum_{\ell} \theta_\ell ^2+\umod{B_S}_{1,1}.$
Hence, for $j,k\in S$,
\begin{align*}
    \widecheck{P}_{jk}=&\, \frac{\EE(D_j)\EE(D_k)}{2\,\EE(M)}
    = \frac{(\sum_{\ell\neq j} \theta_j\theta_\ell  +\sum_{\ell\in S:\, \ell\neq j}(1-\theta_j\theta_\ell )) (\sum_{\ell\neq k} \theta_k\theta_\ell + \sum_{\ell\in S:\,\ell\neq k}(1-\theta_k\theta_\ell ))}{\left(\sum_{\ell} \theta_\ell \right)^2 - \sum_{\ell} \theta_\ell ^2+\umod{B_S}_{1,1}}\\[.1cm]
    \leq&\, \frac{\theta_j\theta_k(\sum_{\ell} \theta_\ell)^2  + \theta_k(\sum_{\ell} \theta_\ell)\sum_{\ell\in S:\, \ell\neq j}(1-\theta_j\theta_\ell) + \theta_j(\sum_{\ell} \theta_\ell)\sum_{\ell\in S:\,\ell\neq k}(1-\theta_k\theta_\ell) }{\left(\sum_{\ell} \theta_\ell \right)^2 - \sum_{\ell} \theta_\ell ^2+\umod{B_S}_{1,1}}\\[.1cm]
    +&\, \frac{\sum_{\ell\in S:\, \ell\neq j}(1-\theta_j\theta_\ell)\sum_{\ell\in S:\,\ell\neq k}(1-\theta_k\theta_\ell)}{\left(\sum_{\ell} \theta_\ell \right)^2 - \sum_{\ell} \theta_\ell ^2+\umod{B_S}_{1,1}} \\[.1cm]
    =&\, \theta_j\theta_k + \frac{\theta_j\theta_k(\sum_{\ell} \theta_\ell ^2 - \umod{B_S}_{1,1})}{\left(\sum_{\ell} \theta_\ell \right)^2 - \sum_{\ell} \theta_\ell ^2+\umod{B_S}_{1,1}} + \frac{\sum_{\ell\in S:\, \ell\neq j}(1-\theta_j\theta_\ell)\sum_{\ell\in S:\,\ell\neq k}(1-\theta_k\theta_\ell)}{\left(\sum_{\ell} \theta_\ell \right)^2 - \sum_{\ell} \theta_\ell ^2+\umod{B_S}_{1,1}}\\[.1cm]
    +&\, \frac{\theta_k(\sum_{\ell} \theta_\ell)\sum_{\ell\in S:\, \ell\neq j}(1-\theta_j\theta_\ell) + \theta_j(\sum_{\ell} \theta_\ell)\sum_{\ell\in S:\,\ell\neq k}(1-\theta_k\theta_\ell) }{\left(\sum_{\ell} \theta_\ell \right)^2 - \sum_{\ell} \theta_\ell^2 + \umod{B_S}_{1,1}},\\[.1cm]
    \leq&\, \theta_j\theta_k+ \theta_j\theta_k\,\frac{4\umod{B_S}_{1,1}}{\left(\sum_{\ell} \theta_\ell \right)^2} + \frac{2\,\sum_{\ell\in S:\, \ell\neq j}(1-\theta_j\theta_\ell)\sum_{\ell\in S:\,\ell\neq k}(1-\theta_k\theta_\ell)}{\left(\sum_{\ell} \theta_\ell \right)^2}\\[.1cm]
    +&\, \frac{2\,\theta_k(\sum_{\ell} \theta_\ell)\sum_{\ell\in S:\, \ell\neq j}(1-\theta_j\theta_\ell) + 2\,\theta_j(\sum_{\ell} \theta_\ell)\sum_{\ell\in S:\,\ell\neq k}(1-\theta_k\theta_\ell)}{\left(\sum_{\ell} \theta_\ell \right)^2},
\end{align*}
for all sufficiently large $n$, where the last inequality obtained via Conditions \eqref{degree_cond}, \eqref{cl_detectcond1} and \eqref{cl_detectcond2}.
In particular, we use the fact that $\sum_{\ell} \theta_\ell ^2 \leq \delta(P)\leq \umod{B_S}_{1,1}$ and $\left(\sum_{\ell} \theta_\ell \right)^2 - \sum_{\ell} \theta_\ell^2\geq (1/2)\left(\sum_{\ell} \theta_\ell \right)^2$ for all sufficiently large $n$.

Therefore,
\begin{align*}
    &\,\underset{j<k:\,j,k\neq i}{\sum\sum} (1-\widecheck{P}_{jk})\,\mathbb{I}(j,k\in S)=\frac{1}{2} \underset{j\neq k:\,j,k\neq i}{\sum\sum} (1-\widecheck{P}_{jk})\,\mathbb{I}(j,k\in S)\\[.1cm]
    \geq&\, \frac{1}{2}\underset{j\neq k:\,j,k\neq i}{\sum\sum} (1-P_{jk})\,\mathbb{I}(j,k\in S) - 2\umod{B_S}_{1,1}\frac{(\sum_{\ell\in S}\theta_\ell)^2}{\left(\sum_{\ell} \theta_\ell \right)^2} - \frac{\umod{B_S}_{1,1}^2}{\left(\sum_{\ell} \theta_\ell \right)^2} - 2\umod{B_S}_{1,1}\frac{\sum_{\ell\in S}\theta_\ell}{\sum_{\ell} \theta_\ell}\\[.1cm]
    \geq&\, \frac{1}{2}\umod{B_S}_{1,1} - o(\umod{B_S}_{1,1}),
\end{align*}
where the last inequality follows from Conditions \eqref{cl_detectcond1} and \eqref{cl_detectcond2}.

\textbf{Conclusion:} Combining the lower bound derived above with Equations \eqref{clh1p1s}, \eqref{clh1p2s}, \eqref{clh1p3s}, we obtain with probability at least $1-O(n^{-c})$,
\begin{align*}
    T_i \geq&\, \left(\frac{1}{2} -\frac{1}{2}\max_{j\in S^c} (P_{ij}^2 + 4 P_{ij})-o(1)\right)\umod{B_S}_{1,1} \geq \gamma_1 \umod{B_S}_{1,1},
\end{align*}
for some constant $\gamma_1 > 0$, provided that $\max_{j\in S^c} (P_{ij}^2 + 4 P_{ij}) < 1$, which is satisfied if $P_{ij}< \sqrt{5}-2$.

\subsection{Proof of Theorem \ref{unknownph0rdpg}}
\subsubsection*{Under $H_0$}
\begin{equation*}
    2\,T_{i,RDPG}=\underset{j,k}{\sum\sum} A_{ij}A_{ik}(A_{jk}-\widehat{P}_{jk})=a_i^\top (A-\widehat{P})a_i
    =a_i^\top (A-P)a_i + a_i^\top (P-\widehat{P})a_i.
\end{equation*}

In the proof of Theorem \ref{knownp_h0_alt}, we showed that for any $c>1$,
\begin{equation}
    \PP_{H_0}\left(|a_i^\top (A-{P})a_i| \leq \sqrt{2c\log n}\,\sigma_{i,A} + \frac{2c}{3}\log n\right)\geq 1-\frac{2}{n^c}.
\end{equation}

Let $\mathcal{O}_{d\times d}$ be the set of all $d\times d$ orthogonal matrices.
For $W\in\mathcal{O}_{d\times d}$,
\begin{equation}
    \label{ai_phat_minus_p_ai_expand}
\begin{split}
     &\, a_i^\top (\widehat{P}-P)a_i= a_i^\top (\widehat{X}\widehat{X}^\top -XX^\top)a_i\\[.2cm]
     =&\, a_i^\top ((\widehat{X}-XW +XW)(\widehat{X}-XW+XW)^\top -XX^\top )a_i\\[.2cm]
     =&\, a_i^\top (\widehat{X}-XW)(\widehat{X}-XW)^\top a_i + 2\,a_i^\top (\widehat{X}-XW)W^\top X^\top a_i.
\end{split}
\end{equation}

Therefore,
$$|a_i^\top (\widehat{P}-P)a_i|\leq \|a_i^\top (\widehat{X}-XW)\|^2 + 2\,\|a_i^\top (\widehat{X}-XW)\| \|X^\top a_i\|.$$

Here,
$$\|X^\top a_i\| \leq \umod{X}\umod{a_i}\leq \sqrt{\lambda_1(P)}\sqrt{D_i}.$$
Using Bernstein's inequality and Lemma \ref{lem:irl}, we have that with probability at least $1-2/n^c$,
$$|D_i - \EE(D_i)|\leq \sqrt{2c\,\delta(P)\log n} + \frac{2c}{3}\log n.$$
Since $\EE(D_i)\leq \delta(P)$ and $\delta(P)\gg \log n$, we have that for any $\epsilon>0$,
\begin{equation}
    \PP_{H_0}(D_i\leq (1+\epsilon)\,\delta(P))\geq 1-O(n^{-c}),
    \label{di_bound}
\end{equation}
for all sufficiently large $n$.
Also, $\lambda_1(P)=O(\delta(P))$ by the Gershgorin circle theorem. 
Hence, 
\begin{equation*}
    \PP_{H_0}(\|X^\top a_i\| \leq C\delta(P))\geq 1-O(n^{-c}).
\end{equation*}

From \citet[][Theorem 4.4]{xie2021entrywise}, we have that under the condition \eqref{rdpg_cond}, there exists $W^{\star}\in\mathcal{O}_{d\times d}$ such that,
\begin{equation*}
    \widehat{X}-XW^{\star} = (A-P)U_PS_P^{-\frac{1}{2}} +E_0, \text{ such that }\umod{{E_0}}_{2\rightarrow\infty}\leq C\,\dfrac{\log n}{\sqrt{n\,\delta(P)}},
\end{equation*}
with probability at least $1-O(n^{-c})$, for all sufficiently large $n$.
Then,
\begin{equation*}
    \|a_i^\top (\widehat{X}-XW^{\star})\|\leq \|a_i^\top (A-P)U_PS_P^{-\frac{1}{2}}\|+\|a_i^\top E_0\|.
\end{equation*}

Applying Bernstein's inequality, we can show that $\umod{a_i^\top (A-P)U_P}=O(\sqrt{D_i})=O(\sqrt{\delta(P)})$ with probability at least $1-O(n^{-c})$.
Therefore, the first term,
\begin{equation*}
    \|a_i^\top (A-P)U_PS_P^{-\frac{1}{2}}\| \leq \|a_i^\top (A-P)U_P\| \|S_P^{-\frac{1}{2}}\|=O(1),
\end{equation*}
with probability at least $1-O(n^{-c})$, since $\|S_P^{-\frac{1}{2}}\|=\dfrac{1}{\sqrt{\lambda_d(P)}}=O(\delta^{-\frac{1}{2}}(P))$, by assumption.

Now, the second term,
\begin{equation*}
    \|a_i^\top E_0\| \,\leq\, D_i\,\|E_0\|_{2\rightarrow\infty} \,\leq\, C D_i\,\dfrac{\log n}{\sqrt{n\,\delta(P)}}\leq C\,\dfrac{\sqrt{\delta(P)}\log n}{\sqrt{n}},
\end{equation*}
with probability at least $1-O(n^{-c})$, invoking \eqref{di_bound} in the final step.
Therefore,
\begin{align*}
    &\,|a_i^\top (\widehat{P}-P)a_i|\leq\|a_i^\top (\widehat{X}W^{\star}-X)\|^2 + 2\|a_i^\top (\widehat{X}W^{\star}-X)\| \|X^\top a_i\| \\[.2cm]
    \leq&\,C\,\dfrac{\delta(P)(\log n)^2}{n} + C\,\dfrac{\sqrt{\delta(P)}\log n}{\sqrt{n}}\cdot \delta(P)\leq C\,\dfrac{(\delta(P))^{\frac{3}{2}}\log n}{\sqrt{n}},
\end{align*}
with probability at least $1-O(n^{-c})$.
Hence,
\begin{equation*}
    \PP_{H_0}\left(|T_{i,RDPG}|\leq C\,\delta(P)\sqrt{\log n}\max\left\{1,\sqrt{\dfrac{\delta(P)\log n}{n}}\right\}\right)\geq 1-O(n^{-c}),
\end{equation*}
for all suficiently large $n$.

\subsubsection*{Under $H_1$}
\begin{equation}
    \label{ti_expand_rdpg}
    \begin{split}
        &\,2\,T_{i,RDPG}=\underset{j,k}{\sum\sum} A_{ij}A_{ik}(A_{jk}-\widehat{P}_{jk})=a_i^\top (A-\widehat{P})a_i\\
        =&\,a_i^\top (A-Q)a_i+a_i^\top (Q-\widecheck{P})a_i + a_i^\top (\widecheck{P}-\widehat{P})a_i.
    \end{split}
\end{equation}

Following the proof of Theorem \ref{knownp_h0_alt}, we have
\begin{equation}
    \PP_{H_1}\left(|a_i^\top (A-Q)a_i| \leq C\,\delta(P)\sqrt{\log n}\right)\geq 1-\frac{2}{n^c}.
\end{equation}

Following the expansion of $a_i^\top (\widehat{P}-P)a_i$ in \eqref{ai_phat_minus_p_ai_expand}, we can expand $a_i^\top (\widehat{P}-\widecheck{P})a_i$ such that for $W\in\mathcal{O}_{d\times d}$,
\begin{equation*}
    |a_i^\top (\widehat{P}-\widecheck{P})a_i| \,\leq\, \| a_i^\top (\widehat{X}W-\widecheck{X})\|^2 + 2\,\|a_i^\top (\widehat{X}W-\widecheck{X})\| \|\widecheck{X}^\top a_i\|.
\end{equation*}

Here,
$$\|\widecheck{X}^\top a_i\|\leq \|\widecheck{X}\|\umod{a_i}\leq \sqrt{\lambda_1(Q)}\sqrt{D_i}.$$
We can bound $D_i$ using Bernstein's inequality as before, so that for any $\epsilon>0$,
\begin{equation}
    \PP_{H_1}(D_i\leq (1+\epsilon)\,\delta(Q))\geq 1-O(n^{-c}),
    \label{di_bound_q}
\end{equation}
for all sufficiently large $n$.
Now$\lambda_1(Q)\leq \lambda_1(P)+\umod{B_S}=O(\delta(P))$,  using Weyl's inequality and Condition \eqref{rdpg_detectcond1}.
Also, $\delta(Q))\leq \delta(P)+\delta(B_S)=O(\delta(P))$ by Condition \eqref{rdpg_detectcond1}.
Hence,
\begin{equation}
    \PP_{H_1}(\|\widecheck{X}^\top a_i\|\leq C\,\delta(P)) \geq 1-O(n^{-c}).
\end{equation}

We are going to show that 
$$\|a_i^\top (\widehat{X}-\widecheck{X}W)\|=O(\max\{\log n,\umod{B_S}\delta^{-\frac{1}{2}}(P)\}),$$
with probability at least $1-O(n^{-c})$.
We first show that for some $W\in\mathcal{O}_{d\times d}$,
$$\widehat{X} - \widecheck{X}W = (A-Q)U_QS_Q^{-\frac{1}{2}} + E_1, \text{ such that }$$
$$\|E_1\| = O(\max\{\sqrt{\log n}\ \delta^{-\frac{1}{2}}(P),\umod{B_S}\delta^{-1}(P)\}),$$
with probability at least $1-O(n^{-c})$.

We adapt the proof technique from the proof of Theorem 50 in \cite{athreya2017statistical}, with the necessary changes required since $Q$ does not represent a rank-$d$ RDPG model. 
Let $U_Q^\top U_A=W_1\Sigma W_2^\top $ be the singular value decomposition of $U_Q^\top U_A$, where
$\Sigma \coloneqq diag(\sigma_1, \ldots, \sigma_d),\sigma_1\geq\sigma_2\geq\ldots\geq \sigma_d$. Let $\theta_i=\cos^{-1}\sigma_i$ and define the matrix $\sin\Theta=diag(\sin\theta_1,\ldots,\sin\theta_d)$. 
From the Davis-Kahan Theorem \cite{yu2015useful}, 
we have
\begin{align}
    &\umod{U_AU_A^\top - U_QU_Q^\top } = \umod{\sin\Theta}\leq O\left(\frac{\umod{A-Q}}{\lambda_d(Q) - \lambda_{d+1}(Q)}\right),\label{dk1}\\[.2cm]
    &\umod{U_A-U_QW}_F = O\left(\frac{C\sqrt{d}\umod{A-Q}}{\lambda_d(Q)-\lambda_{d+1}(Q)}\right),\ \text{for some }W\in\mathcal{O}_{d\times d}. \label{dk3}
\end{align}

From Theorem 20 of \cite{athreya2017statistical}, we have
\begin{equation}
    \PP_{H_1}(\umod{A-Q}\leq C\sqrt{\delta(Q)\log n}) \geq 1-O(n^{-c}).
    \label{aminusq}
\end{equation}
By Weyl's inequality, we have that $\lambda_d(Q)-\lambda_{d+1}(Q)\geq (\lambda_d(P)-\umod{B_S})-(\lambda_{d+1}(P)+\umod{B_S}) = \lambda_d(P) - 2\umod{B_S}$. 
Since $\lambda_d(P)\geq c_1\,\delta(P)\gg \umod{B_S}$, we have that for any $\epsilon>0$, $\lambda_d(Q)-\lambda_{d+1}(Q)\geq c_1\,(1-\epsilon)\,\delta(P)$ for all sufficiently large $n$.
Hence, 
\begin{equation}
    \PP_{H_1}\left(\frac{\umod{A-Q}}{\lambda_d(Q) - \lambda_{d+1}(Q)} \leq C\,\sqrt{\frac{\log n}{\delta(P)}}\right) \geq 1-O(n^{-c}).
    \label{dk4}
\end{equation}

Define $W^{\star}\coloneqq W_1W_2^\top$. Then, we have 
$$\|U_Q^\top U_A-W^\star\| =\umod{\Sigma-I}=\max_i|1-\sigma_i|\leq \max_i(1-\sigma_i^2)=\umod{\sin\Theta}^2\leq C\,\frac{\log n}{\delta(P)},$$
with probability at least $1-O(n^{-c})$, where the last inequality follows from \eqref{dk1} and \eqref{dk4}.

Let us define
$$R_1 \coloneqq U_A-U_QU_Q^\top U_A, R_2 \coloneqq U_QU_Q^\top U_A-U_QW^{\star},R_3 \coloneqq W^{\star}S_A^{\frac{1}{2}}-S_Q^{\frac{1}{2}}W^{\star}, R_4 \coloneqq U_A-U_QW^{\star}.$$

\textbf{Bounding $R_1$:} Note that $R_1$ is the residual after projecting $U_A$ orthogonally onto the column space of $U_Q$, and hence
\begin{equation}
    \umod{R_1}_F\leq \min_{W\in\mathcal{O}_{d\times d}}\umod{U_A-U_QW}_F \leq C\,\sqrt{\frac{\log n}{\delta(P)}},
    \label{r1_bound}
\end{equation}
with probability at least $1-O(n^{-c})$, from the Davis-Kahan Theorem. 

\textbf{Bounding $R_2$:}
\begin{equation}
    \umod{R_2} = \|U_QU_Q^\top U_A-U_QW^{\star}\| \leq \|U_Q^\top U_A-W^{\star}\| \leq C\,\frac{\log n}{\delta(P)},
    \label{r2_bound}
\end{equation}
with probability at least $1-O(n^{-c})$.

\textbf{Bounding $R_3$:}
Note that 
$$\widehat{P}U_A=U_AS_A=AU_A, \widecheck{P}U_Q=U_QS_Q=QU_Q.$$

We derive
\begin{align*}
    &\, W^{\star} S_A\\[.2cm]
    =&\, (W^{\star} - U_Q^\top U_A)S_A + U_Q^\top U_AS_A=(W^{\star} - U_Q^\top U_A)S_A + U_Q^\top AU_A\\[.2cm]
    =&\, (W^{\star} - U_Q^\top U_A)S_A + U_Q^\top (A-Q)U_A+U_Q^\top QU_A\\[.2cm]
    =&\, (W^{\star} - U_Q^\top U_A)S_A + U_Q^\top (A-Q)R_1 + U_Q^\top (A-Q)U_QU_Q^\top U_A + S_QU_Q^\top U_A\\[.2cm]
    =&\, (W^{\star} - U_Q^\top U_A)S_A + U_Q^\top (A-Q)R_1 + U_Q^\top (A-Q)U_QU_Q^\top U_A + S_Q(U_Q^\top U_A-W^{\star}) + S_QW^{\star}.
\end{align*}

Therefore,
\begin{align*}
    &\,\umod{W^{\star} S_A-S_QW^{\star}}\\[.2cm]
    \leq&\, \|W^{\star} - U_Q^\top U_A\|(\umod{S_A}+\umod{S_Q}) + \|U_Q^\top (A-Q)R_1\| + \|U_Q^\top (A-Q)U_Q\| \|U_Q^\top U_A\|\\[.2cm]
    =&\, O(\log n) + O(\log n) + \|U_Q^\top (A-Q)U_Q\|,
\end{align*}
with probability at least $1-O(n^{-c})$, because $\|W^{\star} - U_Q^\top U_A\| = O(\log n\,\delta^{-1}(P))$, both $\umod{S_A}$ and $\umod{S_Q}$ are $O(\delta(P))$, $\umod{R_1}_F = O(\sqrt{\log n}\delta^{-\frac{1}{2}}(P)) $, and $\|U_Q^\top U_A\|\leq 1$, with probability at least $1-O(n^{-c})$.
Applying Hoeffding's inequality, we can show that
\begin{equation}
    \PP_{H_1}(\umod{U_Q^\top (A-Q)U_Q}_F \leq C\sqrt{\log n}) \geq 1-O(n^{-c}).
    \label{uq_aminusq_uq}
\end{equation}
Therefore,
\begin{equation}
    \PP_{H_1}(\umod{W^{\star} S_A-S_QW^{\star}} \leq C\log n) \geq 1-O(n^{-c}).
\end{equation}

Now, we can establish a bound on $\umod{R_3}_F$ 
by noting that the $(i,j)$-th entry of $R_3=W^{\star}S_A^{\frac{1}{2}}-S_Q^{\frac{1}{2}}W^{\star}$ can be written as
$$W^{\star}_{ij}(\lambda_i^{\frac{1}{2}}(A)-\lambda_j^{\frac{1}{2}}(Q))=\frac{W^{\star}_{ij}(\lambda_i(A)-\lambda_j(Q))}{\lambda_i^{\frac{1}{2}}(A)+\lambda_j^{\frac{1}{2}}(Q)}.$$

Then,
\begin{equation}
    \label{r3_bound}
    \umod{R_3}_F\leq \frac{\umod{W^{\star} S_A-S_QW^{\star}}_F}{\lambda_d^{\frac{1}{2}}(A)+\lambda_d^{\frac{1}{2}}(Q)} = O(\log n\, \delta^{-\frac{1}{2}}(P)),
\end{equation}
with probability at least $1-O(n^{-c})$,
since 
\begin{gather}
    \lambda_d(Q)\geq \lambda_d(P)-\umod{B_S} =\Omega(\delta(P)), \label{ldq_bound}\\[.2cm] 
    \lambda_d(A)\geq \lambda_d(Q)-\umod{A-Q}\geq \lambda_d(Q)-O(\sqrt{\delta(P)\log n})=\Omega(\delta(P)), \label{lda_bound}
\end{gather}
with probability at least $1-O(n^{-c})$, by Conditions \eqref{rdpg_cond} and \eqref{rdpg_detectcond1}, and the result \eqref{aminusq}.

\textbf{Bounding $\|\widehat{X}-\widecheck{X}W^{\star}\|$:}
We derive
\begin{align*}
    \widehat{X}-\widecheck{X}W^{\star}=&\, U_AS_A^{\frac{1}{2}}-U_QS_Q^{\frac{1}{2}}W^{\star}=U_AS_A^{\frac{1}{2}}-U_QW^{\star}S_A^{\frac{1}{2}} + U_QR_3\\[.2cm]
    =&\, U_AS_A^{\frac{1}{2}}-U_QU_Q^\top U_AS_A^{\frac{1}{2}} + R_2S_A^{\frac{1}{2}} + U_QR_3.
\end{align*}
Observe that $U_QU_Q^\top \widecheck{P}=\widecheck{P}$ and $U_AS_A^{\frac{1}{2}}=AU_AS_A^{-\frac{1}{2}}$. Thus,
\begin{align*}
    \widehat{X}-\widecheck{X}W^{\star}=&\,(A-\widecheck{P})U_AS_A^{-\frac{1}{2}}-U_QU_Q^\top (A-\widecheck{P})U_AS_A^{-\frac{1}{2}} + R_2S_A^{\frac{1}{2}} + U_QR_3\\[.2cm]
    =&\,(A-\widecheck{P})U_QW^{\star}S_A^{-\frac{1}{2}}-U_QU_Q^\top (A-\widecheck{P})U_QW^{\star}S_A^{-\frac{1}{2}} + R_2S_A^{\frac{1}{2}} + U_QR_3\\[.2cm]
    +&\,(I-U_QU_Q^\top)(A-\widecheck{P})R_4S_A^{-\frac{1}{2}}\\[.2cm]
    =&\,(A-Q)U_QW^{\star}S_A^{-\frac{1}{2}}-U_QU_Q^\top (A-Q)U_QW^{\star}S_A^{-\frac{1}{2}} + R_2S_A^{\frac{1}{2}} + U_QR_3\\[.2cm]
    +&\,(I-U_QU_Q^\top )(A-\widecheck{P})R_4S_A^{-\frac{1}{2}},\ \text{since }\widecheck{P}U_Q=QU_Q\\[.2cm]
    =&\,(A-\widecheck{P})U_QW^{\star}S_A^{-\frac{1}{2}} + E_1,\ \text{say.}
\end{align*}

\textbf{Bounding $\|E_1\|$:}
We show that $\|E_1\|=O(\sqrt{\log n}\max\{\sqrt{\log n}\ \delta^{-\frac{1}{2}}(P),\|B_S\|\delta^{-1}(P)\})$ with probability at least $1-O(n^{-c})$. First,
$$\|U_QU_Q^\top (A-Q)U_QW^{\star}S_A^{-\frac{1}{2}}\|\leq \umod{U_Q^\top (A-Q)U_Q}_F\|S_A^{-\frac{1}{2}}\|=O(\sqrt{\log  n}\,\delta^{-\frac{1}{2}}(P)),$$
with probability at least $1-O(n^{-c})$, by \eqref{uq_aminusq_uq} and \eqref{lda_bound}.

Next,
$$\|R_2 S_A^{\frac{1}{2}}\|\leq \|R_2\| \|S_A^{\frac{1}{2}}\| = O(\delta^{-\frac{1}{2}}(P)),\umod{U_Q R_3}\leq \umod{R_3}= O(\sqrt{\log n}\ \delta^{-\frac{1}{2}}(P)),$$
with probability at least $1-O(n^{-c})$, invoking \eqref{r2_bound} and \eqref{r3_bound}.

The last term in $E_1$,
\begin{align*}
    &\,\|(I-U_QU_Q^\top )(A-\widecheck{P})R_4 S_A^{-\frac{1}{2}}\| \leq \|(I-U_QU_Q^\top )\| \|A-\widecheck{P}\| \umod{R_4}\|S_A^{-\frac{1}{2}}\|\\[.2cm]
    \leq&\, (\|A-Q\|+\|Q-\widecheck{P}\|)\cdot (\umod{R_1}+\umod{R_2})\cdot \umod{S_A}^{-\frac{1}{2}}\\[.2cm]
    =&\, (O(\sqrt{\delta(P)\log n}) + \lambda_{d+1}(Q))\cdot O(\sqrt{\log n}\,\delta^{-\frac{1}{2}}(P))\cdot O(\delta^{-\frac{1}{2}}(P))\\[.2cm]
    \leq&\, (O(\sqrt{\delta(P)\log n}) + \umod{B_S})\cdot O(\sqrt{\log n}\,\delta^{-1}(P))\\[.2cm] 
    =&\, O(\sqrt{\log n}\max\{\sqrt{\log n}\, \delta^{-\frac{1}{2}}(P),\umod{B_S}\delta^{-1}(P)\}).
\end{align*}

Hence, 
\begin{equation}
    \PP_{H_1}(\umod{E_1}\leq C\sqrt{\log n}\max\{\sqrt{\log n}\ \delta^{-\frac{1}{2}}(P),\umod{B_S}\delta^{-1}(P)\}) \geq 1-O(n^{-c}).
    \label{e1_bound}
\end{equation}

\textbf{Bounding $\|a_i^\top (\widehat{X}-\widecheck{X}W^{\star})\|$:}
Now, we derive the bound for $\|a_i^\top (\widehat{X}-\widecheck{X}W^{\star})\|$.
\begin{equation*}
    \|a_i^\top (\widehat{X}-\widecheck{X}W^{\star})\|\leq \|a_i^\top (A-Q)U_QW^{\star}S_A^{-\frac{1}{2}}\| + \|a_i^\top E_1\|.
\end{equation*}
Applying Bernstein's inequality, we can show that $\|a_i^\top (A-Q)U_Q\| = O(\sqrt{D_i})=O(\sqrt{\delta(P)})$ with probability at least $1-O(n^{-c})$. Therefore, the first term,
$$\|a_i^\top (A-Q)U_QW^{\star}S_A^{-\frac{1}{2}}\| \leq \|a_i^\top (A-Q)U_Q\| \|S_A^{-\frac{1}{2}}\|=O(1),$$
with probability at least $1-O(n^{-c})$.
And the second term,
\begin{align*}
    \|a_i^\top E_1\|\leq&\, \|a_i^\top \| \|E_1\| = \sqrt{D_i}\cdot O(\sqrt{\log n} \max\{\sqrt{\log n}\ \delta^{-\frac{1}{2}}(P),\umod{B_S}\delta^{-1}(P)\})\\[.2cm]
    =&\, O(\max\{\log n, \sqrt{\log n}\umod{B_S}\delta^{-\frac{1}{2}}(P)\}),
\end{align*}
with probability at least $1-O(n^{-c})$, leveraging \eqref{di_bound_q} and \eqref{e1_bound}.

Therefore,
\begin{align*}
    |a_i^\top (\widehat{P}-\widecheck{P})a_i|\leq&\, \|a_i^\top (\widehat{X}-\widecheck{X}W^{\star})\|^2 + 2\|a_i^\top (\widehat{X}-\widecheck{X}W^{\star})\| \|\widecheck{X}^\top a_i\|\\[.2cm]
    =&\, (O(\max\{\log n,\sqrt{\log n}\umod{B_S}\delta^{-\frac{1}{2}}(P))\})^2 \\[.2cm]
    +&\, O(\max\{\log n, \sqrt{\log n}\umod{B_S}\delta^{-\frac{1}{2}}(P)\})\cdot O(\delta(P))\\[.2cm]
    =&\, O(\max\{\delta(P)\log  n,\umod{B_S}\sqrt{\delta(P)\log n}\}).
\end{align*}

Let us expand the remaining term $a_i^\top (Q-\widecheck{P})a_i$.
\begin{align*}
    &\, a_i^\top (Q-\widecheck{P})a_i = \underset{j,k}{\sum\sum}A_{ij}A_{ik}(Q_{jk}-\widecheck{P}_{jk})\\[.2cm]
    =&\, \underset{j,k}{\sum\sum}(A_{ij}A_{ik} -Q_{ij}Q_{ik})(Q_{jk}-\widecheck{P}_{jk}) + q_i^\top (Q-\widecheck{P})q_i.
\end{align*}

The sum $\underset{j,k}{\sum\sum}(A_{ij}A_{ik} -Q_{ij}Q_{ik})(Q_{jk}-\widecheck{P}_{jk})$ is $O(\max\{\delta(P)\log  n,\umod{B_S}\sqrt{\delta(P)\log n}\})$ with probability at least $1-O(n^{-c})$.

From \eqref{ti_expand_rdpg}, we obtain that for all $i\in[n]$,
\begin{equation*}
  2T_{i,RDPG}\geq q_i^\top (Q-\widecheck{P})q_i - O(\max\{\delta(P)\log  n,\umod{B_S}\sqrt{\delta(P)\log n}\}),
\end{equation*}
with probability at least $1-O(n^{-c})$. 
Therefore, Eq. \eqref{unknownp_power_rdpg} is proved due to Condition \eqref{rdpg_detectcond2}.

\section{Proof of Theorem \ref{unknownprankingrdpg}}
Recall the definitions and results from the proof of Theorem \ref{unknownph0rdpg}.
Consider $i\in S, i'\in S^c$.
\begin{align*}
    &\,2\,(T_{i,RDPG}-T_{i',RDPG})\\[.2cm]
    \geq&\, {q_i^{(1)}}^\top (Q-\widecheck{P})q_i^{(1)}-{q_{i'}^{(1)}}^\top (Q-\widecheck{P})q_{i'}^{(1)} -2|{q_i^{(1)}}^\top (Q-\widecheck{P})q_i^{(2)}-{q_{i'}^{(1)}}^\top (Q-\widecheck{P})q_{i'}^{(2)}|\\[.2cm]
    -&\, |{q_i^{(2)}}^\top (Q-\widecheck{P})q_i^{(2)} -{q_{i'}^{(2)}}^\top (Q-\widecheck{P})q_{i'}^{(2)}|-O(\max\{\delta(P)\log  n,\umod{B_S}\sqrt{\delta(P)\log n}\})\\[.2cm]
    \geq&\, \gamma_2\umod{B_S}^2 - 7\gamma_2\epsilon \umod{B_S}^2 - O(\max\{\delta(P)\log  n,\umod{B_S}\sqrt{\delta(P)\log n}\}) 
\end{align*}
with probability at least $1-O(n^{-c})$, provided that Conditions \eqref{rdpg_localizationcond1} and \eqref{rdpg_localizationcond2} holds.
Since $\umod{B_S}^2\gg \delta(P)\log n$ by Condition \eqref{rdpg_detectcond1}, we have
\begin{equation*}
    \PP_{H_1}\left(\min_{i\in S, i'\in S^c}(T_{i,RDPG}-T_{i',RDPG}) > 0\right) \geq 1-O(n^{-(c-1)}).
\end{equation*}

\bibliographystyle{chicagoa}
\bibliography{ref, ref2}

@InProceedings{adamic2005political,
  Title                    = {The political blogosphere and the 2004 {US} election: divided they blog},
  Author                   = {Adamic, Lada A and Glance, Natalie},
  Booktitle                = {Proceedings of the 3rd International Workshop on Link Discovery},
  Year                     = {2005},
  Organization             = {ACM},
  Pages                    = {36--43}
}

@InProceedings{greene2013producing,
  Title                    = {Producing a unified graph representation from multiple social network views},
  Author                   = {Greene, Derek and Cunningham, P{\'a}draig},
  Booktitle                = {Proceedings of the 5th Annual ACM Web Science Conference},
  Year                     = {2013},
  Organization             = {ACM},
  Pages                    = {118--121}
}

@Article{karrer2011stochastic,
  Title                    = {Stochastic blockmodels and community structure in networks},
  Author                   = {Karrer, B. and Newman, Mark E. J.},
  Journal                  = {Physical Review E},
  Year                     = {2011},
  Pages                    = {016107},
  Volume                   = {83},

  Publisher                = {APS}
}

@Article{lorrain1971structural,
  Title                    = {Structural equivalence of individuals in social networks},
  Author                   = {Lorrain, Francois and White, Harrison C},
  Journal                  = {The Journal of Mathematical Sociology},
  Year                     = {1971},
  Pages                    = {49--80},
  Volume                   = {1},

  Publisher                = {Taylor \& Francis}
}

@Article{zhao2012consistency,
  Title                    = {Consistency of community detection in networks under degree-corrected stochastic block models},
  Author                   = {Zhao, Yunpeng and Levina, Elizaveta and Zhu, Ji},
  Journal                  = {The Annals of Statistics},
  Year                     = {2012},
  Pages                    = {2266--2292},
  Volume                   = {40},

  Publisher                = {Institute of Mathematical Statistics}
}

@article {Bickel2016estimating,
author = {Bickel, Peter J. and Sarkar, Purnamrita},
title = {Hypothesis testing for automated community detection in networks},
journal = {Journal of the Royal Statistical Society: Series B (Statistical Methodology)},
volume = {78},
number = {1},
issn = {1467-9868},
url = {http://dx.doi.org/10.1111/rssb.12117},
doi = {10.1111/rssb.12117},
pages = {253--273},
year = {2016},
}

@article {senguptapabm,
author = {Sengupta, Srijan and Chen, Yuguo},
title = {A block model for node popularity in networks with community structure},
journal = {Journal of the Royal Statistical Society: Series B (Statistical Methodology)},
 volume={80},
  number={2},
  pages={365--386},
  year={2018},
}

@article{alon1998finding,
  title={Finding a large hidden clique in a random graph},
  author={Alon, Noga and Krivelevich, Michael and Sudakov, Benny},
  journal={Random Structures \& Algorithms},
  volume={13},
  number={3-4},
  pages={457--466},
  year={1998},
  publisher={Wiley Online Library}
}

@article{dekel2014finding,
  title={Finding hidden cliques in linear time with high probability},
  author={Dekel, Yael and Gurel-Gurevich, Ori and Peres, Yuval},
  journal={Combinatorics, Probability and Computing},
  volume={23},
  number={1},
  pages={29--49},
  year={2014},
  publisher={Cambridge University Press}
}

@inproceedings{feige2010finding,
  title={Finding hidden cliques in linear time},
  author={Feige, Uriel and Ron, Dorit},
  booktitle={Discrete Mathematics and Theoretical Computer Science},
  pages={189--204},
  year={2010},
  organization={Discrete Mathematics and Theoretical Computer Science}
  }

@article{verzelen2015community,
  title={Community detection in sparse random networks},
  author={Verzelen, Nicolas and Arias-Castro, Ery},
  journal={The Annals of Applied Probability},
  volume={25},
  number={6},
  pages={3465--3510},
  year={2015},
  publisher={Institute of Mathematical Statistics}
}

@article{arias2014community,
  title={Community detection in dense random networks},
  author={Arias-Castro, Ery and Verzelen, Nicolas},
  journal={The Annals of Statistics},
  volume={42},
  number={3},
  pages={940--969},
  year={2014},
  publisher={Institute of Mathematical Statistics}
}

@article{chung2002average,
  title={The average distances in random graphs with given expected degrees},
  author={Chung, Fan and Lu, Linyuan},
  journal={Proceedings of the National Academy of Sciences},
  volume={99},
  number={25},
  pages={15879--15882},
  year={2002},
  publisher={National Acad Sciences}
}

@inproceedings{young2007random,
  title={Random dot product graph models for social networks},
  author={Young, Stephen J and Scheinerman, Edward R},
  booktitle={International Workshop on Algorithms and Models for the Web-Graph},
  pages={138--149},
  year={2007},
  organization={Springer}
}

@article{sussman2012consistent,
  title={A consistent adjacency spectral embedding for stochastic blockmodel graphs},
  author={Sussman, Daniel L and Tang, Minh and Fishkind, Donniell E and Priebe, Carey E},
  journal={Journal of the American Statistical Association},
  volume={107},
  number={499},
  pages={1119--1128},
  year={2012},
  publisher={Taylor \& Francis Group}
}

@book{crossley2015social,
  title={Social network analysis for ego-nets: Social network analysis for actor-centred networks},
  author={Crossley, Nick and Bellotti, Elisa and Edwards, Gemma and Everett, Martin G and Koskinen, Johan and Tranmer, Mark},
  year={2015},
  publisher={Sage}
}

@article{xie2021entrywise,
  title={Entrywise limit theorems for eigenvectors of signal-plus-noise matrix models with weak signals},
  author={Xie, Fangzheng},
  journal={Bernoulli},
  volume = {30},
  pages = {388--418},
  year={2024}
}

@Article{amini2013,
  Title                    = {Pseudo-likelihood methods for community detection in large sparse networks},
  Author                   = {Amini, Arash A. and Chen, Aiyou and Bickel, Peter J. and Levina, Elizaveta},
  Journal                  = {Ann. Statist.},
  Year                     = {2013},

  Month                    = {08},
  Number                   = {4},
  Pages                    = {2097--2122},
  Volume                   = {41},

  Doi                      = {10.1214/13-AOS1138},
  Fjournal                 = {The Annals of Statistics},
  Publisher                = {The Institute of Mathematical Statistics},
  Url                      = {http://dx.doi.org/10.1214/13-AOS1138}
}

@Manual{R,
    title = {R: A Language and Environment for Statistical Computing},
    author = {{R Core Team}},
    organization = {R Foundation for Statistical Computing},
    address = {Vienna, Austria},
    year = {2016},
    url = {https://www.R-project.org/},
  }

@Article{Erdoes1959,
  author  = {Erd{\"o}s, Paul and R{\'e}nyi, Alfr{\'e}d},
  title   = {On random graphs},
  journal = {Publicationes Mathematicae Debrecen},
  year    = {1959},
  volume  = {6},
  pages   = {290--297},
}

@inproceedings{li2010detecting,
  title={Detecting blackhole and volcano patterns in directed networks},
  author={Li, Zhongmou and Xiong, Hui and Liu, Yanchi and Zhou, Aoying},
  booktitle={IEEE 10th International Conference on Data Mining (ICDM)},
  pages={294--303},
  year={2010},
  organization={IEEE}
}

@inproceedings{pan2012decoding,
  title={Decoding social influence and the wisdom of the crowd in financial trading network},
  author={Pan, Wei and Altshuler, Yaniv and Pentland, Alex},
  booktitle={2012 International Confernece on Social Computing (SocialCom)},
  pages={203--209},
  year={2012},
  organization={IEEE}
}

@article{lin2006organizing,
  title={Organizing principles of real-time memory encoding: neural clique assemblies and universal neural codes},
  author={Lin, Longnian and Osan, Remus and Tsien, Joe Z},
  journal={TRENDS in Neurosciences},
  volume={29},
  number={1},
  pages={48--57},
  year={2006},
  publisher={Elsevier}
}

@incollection{bollobas1998random,
  title={Random graphs},
  author={Bollob{\'a}s, B{\'e}la},
  booktitle={Modern graph theory},
  pages={215--252},
  year={1998},
  publisher={Springer}
}

@article{kuvcera1993expected,
  title={Expected complexity of graph partitioning problems},
  author={Ku{\v{c}}era, Lud{\v{e}}k},
  year={1993},
  publisher={Max-Planck-Institut f{\"u}r Informatik}
}

@article{athreya2017statistical,
  title={Statistical inference on random dot product graphs: a survey},
  author={Athreya, Avanti and Fishkind, Donniell E and Tang, Minh and Priebe, Carey E and Park, Youngser and Vogelstein, Joshua T and Levin, Keith and Lyzinski, Vince and Qin, Yichen},
  journal={The Journal of Machine Learning Research},
  volume={18},
  number={1},
  pages={8393--8484},
  year={2017},
  publisher={JMLR. org}
}

@article{bogerd2021detecting,
  title={Detecting a planted community in an inhomogeneous random graph},
  author={Bogerd, Kay and Castro, Rui M and van der Hofstad, Remco and Verzelen, Nicolas},
  journal={Bernoulli},
  volume={27},
  number={2},
  pages={1159--1188},
  year={2021},
  publisher={Bernoulli Society for Mathematical Statistics and Probability}
}

@article{deshpande2015finding,
  title={Finding hidden cliques of size $\sqrt (N/e)$ in nearly linear time},
  author={Deshpande, Yash and Montanari, Andrea},
  journal={Foundations of Computational Mathematics},
  volume={15},
  number={4},
  pages={1069--1128},
  year={2015},
  publisher={Springer}
}

@InProceedings{Hajek2015computational,
  title = {Computational Lower Bounds for Community Detection on Random Graphs},
  author = {Hajek, Bruce and Wu, Yihong and Xu, Jiaming},
  booktitle = {Proceedings of The 28th Conference on Learning Theory},
  pages = {899--928},
  year = {2015},
  editor = {Grünwald, Peter and Hazan, Elad and Kale, Satyen},
  volume = {40},
  series = {Proceedings of Machine Learning Research},
  address = {Paris, France},
  month = {03--06 Jul},
  publisher =    {PMLR}
}

@article{chen2016statistical,
  title={Statistical-computational tradeoffs in planted problems and submatrix localization with a growing number of clusters and submatrices},
  author={Chen, Yudong and Xu, Jiaming},
  journal={The Journal of Machine Learning Research},
  volume={17},
  number={1},
  pages={882--938},
  year={2016},
  publisher={JMLR.org}
}

@article{rudelson2013hanson,
author = {Mark Rudelson and Roman Vershynin},
title = {{Hanson-Wright inequality and sub-gaussian concentration}},
volume = {18},
journal = {Electronic Communications in Probability},
number = {none},
publisher = {Institute of Mathematical Statistics and Bernoulli Society},
pages = {1 -- 9},
year = {2013},
doi = {10.1214/ECP.v18-2865},
URL = {https://doi.org/10.1214/ECP.v18-2865}
}

@article{yu2015useful,
  title={A useful variant of the Davis--Kahan theorem for statisticians},
  author={Yu, Yi and Wang, Tengyao and Samworth, Richard J},
  journal={Biometrika},
  volume={102},
  number={2},
  pages={315--323},
  year={2015},
  publisher={Oxford University Press}
}

@article{cape2019signal,
author = {J.~Cape and M.~Tang and C.~E.~Priebe},
title = {Signal-plus-noise matrix models: eigenvector deviations and fluctuations},
journal = {Biometrika},
volume = {106},
pages = {243--250},
year = {2019}
}

@book{matula1976largest,
  title={The largest clique size in a random graph},
  author={Matula, David W},
  year={1976},
  publisher={Department of Computer Science, Southern Methodist University Dallas, Texas~…}
}
\end{document}